\documentclass[a4paper,10pt]{article}
\pdfoutput=1
\usepackage{jcappub}
\usepackage{amsmath}
\usepackage{mathtools}
\usepackage{graphicx}
\usepackage{xcolor}
\usepackage{footmisc}

\usepackage[normalem]{ulem}

\usepackage{float}

\usepackage{braket}
\usepackage{comment}

\newcommand{\be}{\begin{equation}}
	\newcommand{\ee}{\end{equation}}

\begin{document}
	\title{Neutrino winds on the sky}
	\author{Caio Nascimento}
	\author{and Marilena Loverde}
	\affiliation{Department of Physics, University of Washington, Seattle, WA, USA}
	\emailAdd{caiobsn@uw.edu}
	\emailAdd{mloverde@uw.edu}

\abstract{
We develop a first-principles formalism to compute the distortion to the relic neutrino density field caused by the peculiar motions of large-scale structures. This distortion slows halos down due to dynamical friction, causes a local anisotropy in the neutrino-CDM cross-correlation, and reduces the global cross-correlation between neutrinos and CDM. The local anisotropy in the neutrino-CDM cross-spectrum is imprinted in the three point cross-correlations of matter and galaxies, or the bispectrum in Fourier space, producing a signal peaking at squeezed triangle configurations. This bispectrum signature of neutrino masses is not limited by cosmic variance or potential inaccuracies in the modeling of complicated nonlinear and galaxy formation physics, and it is not degenerate with the optical depth to reionization. We show that future surveys have the potential to detect the distortion bispectrum. }	
\maketitle

	

\section{Introduction}
\label{sec:int}

Neutrino oscillation experiments have established that at least two of the neutrino mass eigenstates have a non-zero mass. Two squared mass differences are known from atmospheric and solar neutrino oscillations, which leads to a lower bound on the sum of the masses of $\sum_{\nu} m_{\nu} \gtrsim 0.06,0.1$eV for the normal and inverted hierarchies respectively \cite{de2020global, esteban2020fate, capozzi2017global}. Additionally, terrestrial beta decay experiments set an upper bound to a weighted sum of the masses of $m_{\nu, \beta} < 0.8$eV at $90\%$ confidence level \cite{aker2022katrin}. Tighter constraints are obtained from the neutrino-induced suppression of cosmological structure formation at small scales, leading to the bound $\sum_{\nu} m_{\nu} \lesssim 0.12$eV at $95\%$ confidence \cite{collaboration2020planck}, depending on the dataset. Future cosmological surveys will reach the required sensitivity to detect the neutrino mass scale at the lower limits imposed by oscillation data. This highly anticipated measurement will place tight constraints on sub-percent components of the matter density that can alter the power spectrum, and put the synergy between particle physics and cosmology to the test \cite{green2021cosmological}.

Relic neutrinos are nonrelativistic at late times but even today still have a nonnegligible (thermal) velocity dispersion of $ \sigma_{\nu} \sim 3T_{\nu,0}/m_{\nu} \approx 1500$ km/s for a neutrino mass $m_{\nu}=0.1$eV, and temperature $T_{\nu,0} \approx 1.95$K $\approx 1.7 \times 10^{-4}$eV. We can define the free-streaming scale as the distance traveled by neutrinos over the course of one expansion time, i.e.,  $\lambda_{\textrm{fs}} \sim \sigma_{\nu}/H_{0} \sim 20$ Mpc when evaluated at the present time, where $H_{0} \approx 70$ km/s/Mpc is the Hubble expansion rate. At scales below the free-streaming length ($k > k_{\textrm{fs}}$) pressure dominates over the gravitational potential and neutrino anisotropies are washed out creating a homogeneous sea of neutrinos with a characteristic size $L \sim \lambda_{\textrm{fs}}$. At the present time cold dark matter (CDM) halos have a typical peculiar velocity $v_{\textrm{H}} \sim 500$ km/s. We will show that the peculiar motion of halos causes neutrino particles to accumulate behind the moving halo, generating wakes that slow the halo down due to dynamical friction, and reducing the cross-correlation between neutrinos and CDM. Some previous studies of neutrino wake effects can be found in \cite{Fry:1980wf, okoli2017dynamical, zhu2016probing} (also see \cite{stebbins1987cosmic} for cosmic string wakes), we will comment on similarities and differences between our approaches and those throughout this text. As we will see, the neutrino wake effects are also directly related to the CDM-$\nu$ dipole distortion produced by the relative velocity between CDM and neutrinos \cite{zhu2014measurement, zhu2020measuring, inman2015precision}.

In this work we develop an analytic approach to offsets in neutrino and CDM clustering on small scales. To do this, we leverage the fact that the physics of neutrino clustering simplifies in the limit that the neutrino free-streaming scale is large in comparison with the scale of nonlinearities. The CDM bulk flow produces a distortion in the neutrino density field that naturally emerges in nonlinear structure formation with massive neutrinos. This distortion can be explicitly extracted from the solution to the Boltzmann equation for the neutrino distribution function in the background of a nonlinear dark matter structure, while remaining completely agnostic about the nonlinear dynamics of CDM. In contrast, previous approaches to neutrino wakes are anchored on the introduction of an ad-hoc effective relative displacement between CDM and neutrino fluids, from which the observational signals are obtained.\footnote{This is analogous to the moving background perturbation theory approach to modeling the effects of the relative bulk motion between CDM and baryons on the formation of the first structures \cite{tseliakhovich2010relative}.} This framework is effective on scales below a relative velocity coherence length but it breaks down at sufficiently large scales. Our approach improves on the analytical understanding of neutrino wakes in an expanding universe and allows for a more complete treatment of observables.

In general, the problem of including massive neutrinos in nonlinear structure formation is rather challenging and has been investigated in the literature using different approaches.  That includes simulations that directly implement the neutrinos as N-body particles \cite{bird2012massive, villaescusa2014cosmology, villaescusa2018imprint, rossi2020sejong, adamek2017relativistic, brandbyge2008effect, villaescusa2013non, castorina2015demnuni, emberson2017cosmological, Zimmer:2023jbb}, but also (the computationally less expensive) simulations that assume massive neutrinos remain linearly clustered up to late times \cite{brando2021relativistic, tram2019fully, chiang2019first, brandbyge2009grid, archidiacono2015efficient, ali2013efficient, Chiang:2017vuk}, and analytic approaches as well \cite{lesgourgues2009non, dupuy2014describing, fuhrer2015higher, blas2014structure, levi2016massive, senatore2017effective}. The peculiar motion of halos and the associated distortion to the neutrino density field should, in principle, be included in some of these methods and indeed the existence of a CDM-$\nu$ dipole distortion has been explicitly verified in the TianNu simulations using a real space three point function correlating the CDM and neutrino densities with their relative velocity field \cite{inman2017simulating}. Additionally,  signatures of neutrino wakes were also observed in simulations that directly integrate the six-dimensional Boltzmann (Vlasov)-Poisson equations in phase space \cite{yoshikawa2020cosmological}. 

Our novel theoretical framework enables a throughout investigation of potential observational signatures of this dipole distortion effect; it slows halos down due to dynamical friction, decreases the cross correlation between CDM and neutrinos, and leaves an imprint in three point cross correlations of matter and galaxies, or the bispectrum in Fourier space, peaking at squeezed triangle configurations. As we shall see, this observable benefits from cosmic variance cancellation: it is proportional to the particular realization of the CDM power spectrum on small scales. Furthermore, it is not limited by the (lack of) knowledge of optical depth to reionization,\footnote{Indeed, both of these qualities are common to multi-tracer approaches to measuring the neutrino mass with large-scale structure \cite{loverde2016neutrino, yu2018towards, ballardini2022constraining, Schmittfull:2017ffw,tanseri2022updated}.} and only has a small degeneracy with additional contributions to the bispectrum from standard nonlinear structure formation. This is then a promising additional independent signature of neutrino masses in the large-scale structure that can be probed in future surveys and complement the standard method based on the small-scale matter power suppression in the presence of massive neutrinos. While the suppression in the matter power spectrum from neutrinos is expected to provide the most stringent limits on the neutrino mass, the power spectrum method is limited by the knowledge of optical depth to reionization as probed by the cosmic microwave background (CMB), uncertainties associated with modeling complex nonlinear structure and galaxy formation physics, and degeneracies with other new physics such as a dynamical dark energy component (see e.g. \cite{qu2022probing,brinckmann2019promising, allison2015towards, mishra2018neutrino}). The neutrino wake effects discussed in this paper are free from some of these limitations. 

Our analysis of wakes only assumes generic features of a thermally produced collisionless non-cold dark species and could therefore apply to other sources of hot dark matter. Nevertheless, we work with neutrinos for definiteness. In numerical calculations throughout the paper we assume a reference flat $\Lambda$CDM cosmology, with $\Omega_{\textrm{m},0} = \Omega_{\textrm{c},0}  + \Omega_{\textrm{b},0} + \Omega_{\nu,0} = 0.32$, $\Omega_{\textrm{b},0} = 0.05$, $\Omega_{\Lambda,0} = 0.68$  and $h=0.67$. We emphasize that the neutrino effects considered in this paper scale with powers of the individual neutrino masses, rather than the total energy density in neutrinos. This is in contrast to the linear-theory suppression of the matter power spectrum, which scales with the total energy density in neutrinos as $\sum_i m_{\nu i}$. For simplicity we will work with a single massive species with mass $m_\nu$ and consider the dynamical friction and neutrino distortions produced by that mass state. The net effects of three massive neutrinos can be computed straightforwardly. For a given neutrino mass $m_{\nu}$ we adjust the fractional contribution of CDM to the energy density $\Omega_{\textrm{c},0}$ in such a way as to keep the total matter fraction $\Omega_{\textrm{m},0}$ fixed. Linear transfer functions are obtained with the Cosmic Linear Anisotropy Solving System (CLASS) \cite{blas2011cosmic}, and we compute power spectra assuming an almost scale invariant primordial power defined by a tilt $n_{s} = 0.96$ and amplitude $A_s=2.2 \times 10^{-9}$ at the pivot scale $k_{\textrm{piv}}=0.05$Mpc$^{-1}$. Throughout this paper we treat baryons and CDM identically and henceforth CDM refers both. 

The paper is organized as follows: In Section \ref{sec:form} we review the formalism necessary to study the distortion in the neutrino density field produced by the peculiar motion of halos. In Section \ref{sec:1h} we apply it to the simple case of a moving point mass halo that approximates the distribution of CDM structure at sufficiently small scales. We also derive the 1-halo contribution to the dynamical friction effect, and discuss the limitations associated to neglecting the large-scale gravitational potential.  In Section \ref{sec:dfolss} we apply our formalism to a general nonlinear CDM distribution where we only impose kinematic relations and hence remain agnostic about the nonlinear dynamics of CDM. We then revisit the dynamical friction effect and comment on large-scale structure observables. In Section \ref{sec:calc} we carry out some numerical calculations using the halo model to derive the 2-halo contribution to the dynamical friction effect, and standard perturbation theory (SPT) to compute two and three point cross correlations of neutrinos and CDM. The latter are used to forecast the detectability of neutrino wakes effect in future surveys combining two tracers of the matter distribution. In \ref{sec:conc} we summarize our findings. Finally, in Appendix \ref{sec:stn} we present the details of signal-to-noise calculations.

\section{Basic formalism}
\label{sec:form}

We follow the standard approach to study how neutrinos are distributed around CDM, i.e., to solve for the neutrino phase space distribution function $f_{\nu}(t,\vec{x},\vec{q})$, where $t$ is the Friedmann-Lemaître-Robertson-Walker (FLRW) cosmic time, $\vec{x}$ the comoving position, and $\vec{q} = m_{\nu}a^{2}(t) d\vec{x}/dt$ the comoving momentum in the cosmic frame. Here $m_{\nu}$ is the neutrino mass and $a(t)$ is the scale factor. We work with one neutrino mass eigenstate at a time. The collisionless Boltzmann equation reads \cite{ma1995cosmological},
\begin{equation}
\label{eq:boltzmanneq}
	\frac{\partial f_{\nu}}{\partial{t}} + \frac{d\vec{x}}{dt} \cdot \frac{\partial f_{\nu}}{\partial \vec{x}} + \frac{d \vec{q}}{dt} \cdot \frac{\partial f_{\nu}}{\partial \vec{q}} = 0 \,.
\end{equation}
The neutrinos move in the gravitational potential $\phi(t, \vec{x})$ of CDM, from where it follows that,  
\begin{equation}
\label{eq:dynamical}
	\frac{d\vec{q}}{dt} = -m_{\nu}\frac{\partial \phi}{\partial \vec{x}} \,. 
\end{equation}
Three approximations are being made here. First, we consider the motion of neutrinos in the external gravitational potential of CDM, ignoring the back reaction of neutrinos on the gravitational potential. This is a good approximation as long as the fractional contribution of neutrinos to the matter density is small. We also neglect large-scale general relativistic corrections, such as dilation effects via $\phi'$ terms. These do not contribute to the physics on the sub-horizon scales we consider. Finally, we assume that neutrinos are nonrelativistic, so our equations apply at sufficiently late times. Note that for a neutrino mass $m_{\nu} \sim 0.1$eV, the transition between relativistic and nonrelativistic regimes happens at a redshift $z_{\textrm{tr}} \sim m_{\nu}/3T_{\nu,0} \sim 200$, much earlier than the onset of nonlinear structure formation.

The Boltzmann Eq.~(\ref{eq:boltzmanneq}) can then be written as:
\begin{equation}
\label{eq:boltzmanneq2}
	\frac{\partial f_{\nu}}{\partial{t}} + \frac{\vec{q}}{m_{\nu}a^2} \cdot \frac{\partial f_{\nu}}{\partial \vec{x}} -m_{\nu}\frac{\partial \phi}{\partial \vec{x}} \cdot \frac{\partial f_{\nu}}{\partial \vec{q}} = 0\,.
\end{equation}
We split the neutrino phase-space distribution function as follows: $f_{\nu}=f_{\nu,0}+f_{\nu,1}$, where $f_{\nu,0}$ is a solution to the case $\phi =0$, and $f_{\nu,1}$ represents a response to the gravitational field of dark matter. Setting $\phi=0$ in Eq.~(\ref{eq:boltzmanneq2}) gives,
\begin{equation}
\label{eq:boltzmanneqback}
	\frac{\partial f_{\nu,0}}{\partial{t}} + \frac{\vec{q}}{m_{\nu}a^2} \cdot \frac{\partial f_{\nu,0}}{\partial \vec{x}} = 0 \,.
\end{equation}
A general solution to Eq.~(\ref{eq:boltzmanneqback}), which is also isotropic in the cosmic frame, is an arbitrary function of the magnitude of the comoving momentum $f_{\nu,0}=f_{\nu,0}(q)$. Since we are considering the distribution of a single relic neutrino species, we set this to be the relativistic Fermi-Dirac distribution $f_{\nu,0}(q) = f_{\textrm{FD}}(q/T_{\nu,0})$, with $T_{\nu,0} \approx 1.95$K the  neutrino temperature today, and:
\begin{equation}
\label{eq:fd}
	f_{\textrm{FD}}(x) = \frac{g_{\nu}}{e^{x}+1} \, ,
\end{equation}
with $g_{\nu}=2$ to account for both left-handed neutrinos and right-handed antineutrinos. Going back to the general case of a nonvanishing CDM potential, we arrive at
\begin{equation}
\label{eq:boltzmanneqpert}
	\frac{\partial f_{\nu,1}}{\partial{t}} + \frac{\vec{q}}{m_{\nu}a^2} \cdot \frac{\partial f_{\nu,1}}{\partial \vec{x}} = m_{\nu}\frac{\partial \phi}{\partial \vec{x}} \cdot \frac{\partial f_{\nu,0}}{\partial \vec{q}} + m_{\nu}\frac{\partial \phi}{\partial \vec{x}} \cdot \frac{\partial f_{\nu,1}}{\partial \vec{q}} \,.
\end{equation}
We now make the Brandenberger-Kaiser-Turok (BKT) approximation which consists in dropping the second term in the right-hand side of Eq.~(\ref{eq:boltzmanneqpert}) \cite{Brandenberger:1987kf}. This can be phrased as the statement that we compute the linear response of the neutrino distribution to the (nonlinear) gravitational potential of CDM. We expect the approximation to break down at sufficiently small scales, or in close encounters of neutrinos with very massive halos, in which case one can proceed to solve the Eq.~(\ref{eq:boltzmanneqpert}) iteratively \cite{senatore2017effective}. We are interested in the perturbative quasi-linear regime where the BKT approximation should be adequate.

The Boltzmann Eq.~(\ref{eq:boltzmanneqpert}) then  greatly simplifies to:
\begin{equation}
\label{eq:beqbkt}
	\frac{\partial f_{\nu,1}}{\partial{\eta}} + \frac{\vec{q}}{ m_{\nu}} \cdot \frac{\partial f_{\nu,1}}{\partial \vec{x}} =  m_{\nu}a^2(\eta)\frac{\partial \phi}{\partial \vec{x}} \cdot \frac{\partial f_{\nu,0}}{\partial \vec{q}} \,,
\end{equation}
where we also introduced a new time coordinate $\eta$ via $dt/d\eta = a^{2}(\eta)$, proportional to the comoving distance traveled by a nonrelativistic particle throughout cosmic history. In the remainder of this manuscript we will study how the solution to Eq.~(\ref{eq:beqbkt}) captures the distortion in the neutrino density field induced by the peculiar motion of halos. As a warm-up, in Sec. \ref{sec:1h} we will first solve it for the simple case that $\phi$ is the potential for a single moving point mass halo that works as a toy model for the distribution of CDM at sufficiently small scales. The single-halo approach neglects the clustering of nearby halos and the overall incoherence of the velocity field of nearby structures. In Sec. \ref{sec:dfolss} we will consider Eq. (\ref{eq:beqbkt}) in the context of the full large-scale CDM density field. Our approach will be to use the Poisson equation and the fully nonlinear continuity equation to develop a systematic expansion for perturbations to the neutrino density in terms of derivatives of the CDM density field. In this sense, our predictions for the neutrino wakes do not rely on any particular implementation of non-linear gravitational evolution.

\section{A warm-up exercise: Neutrino wakes around a single moving halo}
\label{sec:1h}

Our goal in this section is to determine the late-time clustering of relic neutrinos around a moving point mass halo with a given mass $M$, following the comoving trajectory $\vec{x}_{\textrm{H}}(\eta)$. We first compute the neutrino energy density and then proceed to determine the resulting dynamical friction force. The section ends with a discussion on the limitations associated to neglecting the large-scale gravitational potential which provides intuition for some of the calculations that are carried out in Section \ref{sec:dfolss}. 

\subsection{Anisotropic neutrino density profile}

As a first step we solve Eq. (\ref{eq:beqbkt}) for the gravitational potential of a moving point mass halo. In general, the solution to Eq.~(\ref{eq:beqbkt}) reads
\begin{equation}
	\label{eq:solf1}
	f_{\nu,1}(\eta, \vec{x}, \vec{q}) = \int d\eta' \int d^3 \vec{x}' \  G(\eta-\eta', \vec{x}-\vec{x}') m_{\nu}a^2(\eta')\frac{\partial \phi}{\partial \vec{x}'} \cdot \frac{\partial f_{\nu,0}}{\partial \vec{q}} \,,
\end{equation}
where:
\begin{equation}
	\label{eq:gsf}
	G(\eta, \vec{x}) = \Theta(\eta)\delta^{(3)}\left(\vec{x}-\frac{\vec{q}}{m_{\nu}} \eta \right) \,,
\end{equation}
is the causal Green's function of Eq.~(\ref{eq:beqbkt}), with $\Theta(\eta)$ the Heaviside step function. The substitution of Eq.~(\ref{eq:gsf}) into Eq.~(\ref{eq:solf1}) gives, after some algebra,
\begin{equation}
	\label{eq:solf1final}
	f_{\nu,1}(\eta, \vec{x}, \vec{q}) = m_{\nu} \int_{0}^{\eta} d\eta' a^2(\eta') \frac{\partial \phi}{\partial \vec{y}} \cdot \frac{\partial f_{\nu,0}}{\partial \vec{q}} \,,
\end{equation}
where we set $\eta = 0$ corresponding to the arbitrary initial integration time and
\begin{equation}
	\label{eq:y} 
	\vec{y} = \vec{x} - \frac{\vec{q}}{m_{\nu}} (\eta-\eta') \,.
\end{equation} 
It is now straightforward to compute the perturbation to the neutrino energy density as follows:
\begin{equation}
	\label{eq:massdensity}
	\begin{split}
		\delta \rho_{\nu} (\eta, \vec{x}) &= m_{\nu} a^{-3}(\eta) \int \frac{d^3 \vec{q}}{(2\pi)^{3}} \ f_{\nu,1}(\eta, \vec{x}, \vec{q}) \,, \\
		&= m_{\nu}^2 a^{-3}(\eta) \int_{0}^{\eta} d\eta' a^2(\eta') \int \frac{d^3 \vec{q}}{(2\pi)^{3}} \ \frac{\partial \phi}{\partial \vec{y}} \cdot \frac{\partial f_{\nu,0}}{\partial \vec{q}} \,.
	\end{split}
\end{equation}
After integration by parts, and using Eq.~(\ref{eq:y}), this becomes
\begin{equation}
	\label{eq:massdensityfinal}
	\delta \rho_{\nu} (\eta, \vec{x}) = m_{\nu} a^{-3}(\eta) \int_{0}^{\eta} d \eta' (\eta-\eta') a^2(\eta') \int \frac{d^3 \vec{q}}{(2\pi)^{3}} \ f_{\nu,0}(q) \nabla^{2}_{y} \phi(\eta', \vec{y}) \,.
\end{equation}
For a point mass halo with a given mass $M$ and following the comoving trajectory $\vec{x}_{\textrm{H}}(\eta)$,
\begin{equation}
	\label{eq:pmh}
	\nabla^{2}_{y} \phi(\eta', \vec{y}) = \frac{4\pi GM}{a(\eta')} \delta^{(3)}(\vec{y}-\vec{x}_{\textrm{H}}(\eta')) \,,
\end{equation}
with Newton's gravitational constant $G$. The substitution of Eq.~(\ref{eq:pmh}) into Eq.~(\ref{eq:massdensityfinal}) yields, after integrating over the comoving momentum to eliminate the delta function,
\begin{equation}
	\label{eq:massdensitypmh}
	\delta \rho_{\nu} (\eta, \vec{x}) = \frac{GMm_{\nu}^4}{2\pi^2} a^{-3}(\eta) \int_{0}^{\eta} d\eta' \frac{a(\eta')}{(\eta-\eta')^2} f_{\textrm{FD}}\left(\frac{m_{\nu}}{T_{\nu,0}} \frac{|\vec{x}-\vec{x}_{\textrm{H}}(\eta')|}{\eta-\eta'}\right) \,,
\end{equation}
where we used the relation $f_{\nu,0}(q) = f_{\textrm{FD}}(q/T_{\nu,0})$, along with Eq.~(\ref{eq:fd}) (with $g_{\nu}=2$). This is the expression for the mass density perturbation of relic neutrinos in the presence of a moving point mass halo, that we will apply in the next section in order to study the dynamical friction effect. 

\begin{figure}
	\centering
	\includegraphics[width=0.75\textwidth]{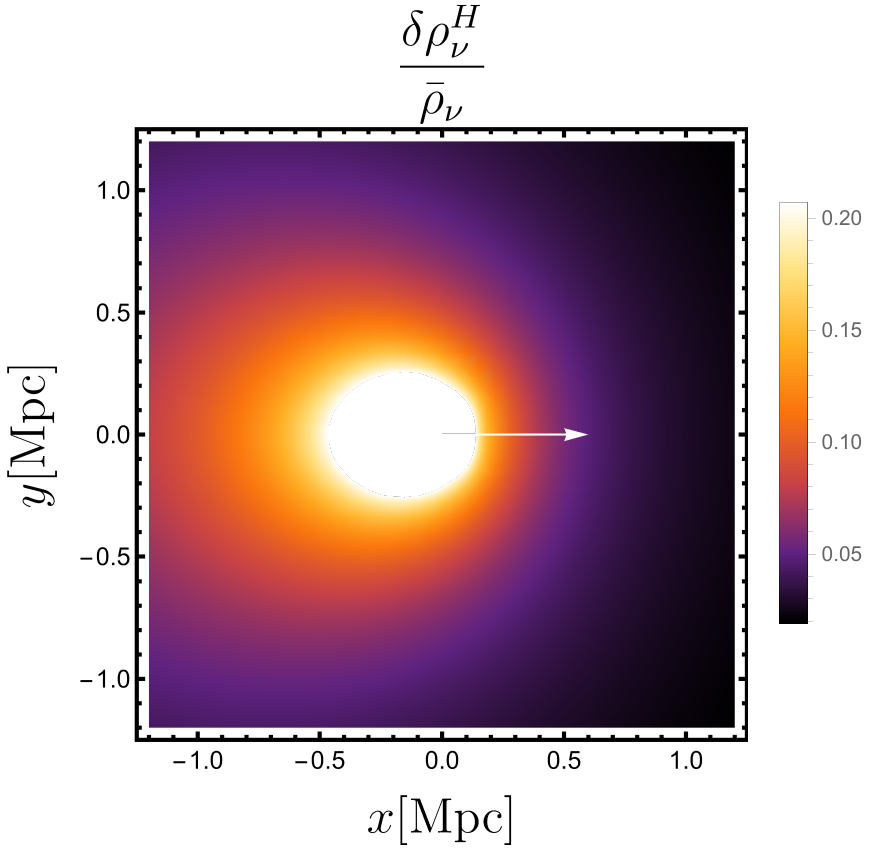}
	\caption{Anisotropic clustering of massive neutrinos behind a point mass halo with a constant velocity in a static universe. Here $m_{\nu}=0.1$eV is the neutrino mass, $M=10^{13}M_{\odot}$ is the halo mass and $v_{\textrm{H}} = 500$km/s is the halo velocity along the x-axis as illustrated by the white arrow. The plot shows the neutrino density contrast in a frame of reference comoving with the halo, which diverges at the origin (the point mass halo position) and hence we introduce a cutoff at an arbitrary maximum density contrast which here corresponds to the value of $\delta_{\nu, \textrm{max}}^{H} = 0.2$.}
	\label{fig:2dplot}
\end{figure}

For now, let us build some intuition on the physics at hand by considering the case of a halo moving with constant velocity $\vec{v}_{\textrm{H}}$ in a static universe, i.e., $a(t)=1$ and $\vec{x}_{\textrm{H}}(t) = \vec{x}_{\textrm{H}}(0)+\vec{v}_{\textrm{H}}t$ with $\eta=t$. In this case, the Eq.~(\ref{eq:massdensitypmh}) reads in the halo frame 
\begin{equation}
\label{eq:haloframe}
	\delta \rho_{\nu}^{\textrm{H}} (t, \vec{x}) = \delta \rho_{\nu} (t, \vec{x} + \vec{x}_{\textrm{H}}(t)) \underset{t\to \infty}{=} \frac{GMm_{\nu}^3T_{\nu,0}}{2\pi^2} \int_{0}^{\infty} du \  f_{\textrm{FD}}(|u\vec{x}+\vec{\alpha}|) \,,
\end{equation}
where we introduce the vector
\begin{equation}
\label{eq:alpha}
	\vec{\alpha} = \frac{m_{\nu}a}{T_{\nu,0}} \vec{v}_{\textrm{H}} \,,
\end{equation}
which points in the direction of the halo motion and whose magnitude is proportional to the ratio of halo to thermal neutrino velocities, and also changed the integration variable to $u=(m_{\nu}/T_{\nu,0})(1/t-t')$ before taking the static limit $t \to \infty$.  Equation (\ref{eq:haloframe}) is plotted in Fig.~\ref{fig:2dplot} for a neutrino mass $m_{\nu} = 0.1$eV, and a halo with mass $M=10^{13}M_{\odot}$ moving with a velocity $v_{\textrm{H}} = 500$km/s along the x-axis. Note that the halo peculiar velocity creates a distortion in the neutrino density field, which clusters anisotropically behind the moving halo.

\subsection{Dynamical friction on a single halo}

\begin{figure}
	\centering
	\includegraphics[width=0.75\textwidth]{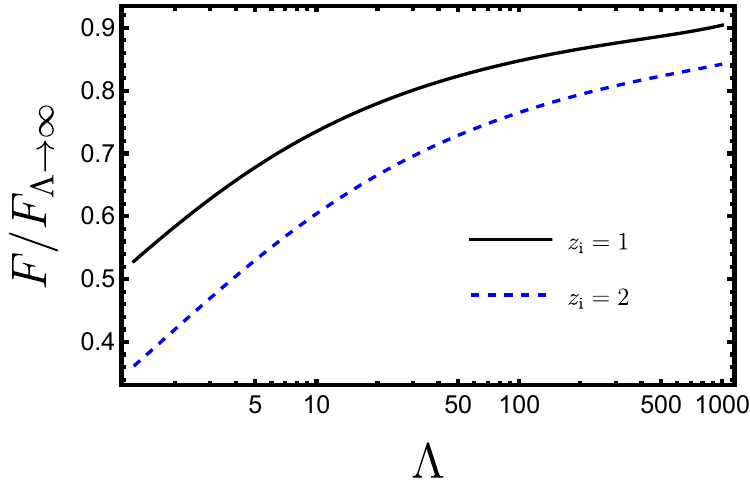}
	\caption{Suppression of Chandrasekhar's formula for the dynamical friction force $F/F_{\Lambda \to \infty}$ for finite values of the cutoff $\Lambda$ in the simple case of a halo subject to Hubble drag. The solid black curve corresponds to a choice of halo formation redshift of $z_{\textrm{i}}=1$, and the dashed blue curve to $z_{\textrm{i}}=2$. In realistic applications we might expect $\Lambda \approx \lambda_{\rm{coh}}/R_{\rm halo} \approx 10-1000$ as we will see in Sec.~\ref{sec:calc}.}
	\label{fig:ckcutoff}
\end{figure}

We now have an expression for the anisotropic clustering of neutrinos around moving point mass halos in an expanding universe, i.e., Eq.~(\ref{eq:massdensitypmh}). We expect this to generate dynamical friction that slows the halo down and this is what we compute next. Using Newton's third law, the force acting on the halo can be obtained by integrating the neutrino mass density over the gravitational field of the halo
\begin{equation}
	\label{eq:dff}
	\vec{F}(\eta) = - a^{3}(\eta) \int d^3 \vec{x} \  \delta \rho_{\nu} (\eta, \vec{x}) \vec{g}_{\textrm{H}}(\eta, \vec{x}) \,,
\end{equation}
where for a moving point mass halo in an expanding universe the gravitational field reads
\begin{equation}
	\label{eq:gfpmh}
	\vec{g}_{\textrm{H}}(\eta, \vec{x}) = - \frac{\vec{\nabla}_{x} \phi(\eta, \vec{x})}{a(\eta)} = - \frac{GM}{a^2(\eta)}\frac{\vec{x}- \vec{x}_{\textrm{H}}(\eta)}{|\vec{x}- \vec{x}_{\textrm{H}}(\eta)|^3}\,.
\end{equation}
Substituting Eqs.~(\ref{eq:massdensitypmh}) and (\ref{eq:gfpmh}) into Eq.~(\ref{eq:dff}) gives
\begin{equation}
	\label{eq:dff2}
	\vec{F}(\eta) = \frac{G^2M^2m_{\nu}^4}{2\pi^2} \int_{0}^{\eta} \frac{d\eta'}{(\eta-\eta')^2} \frac{a(\eta')}{a^2(\eta)} \int d^3 \vec{x} \  f_{\textrm{FD}}\left(\frac{m_{\nu}}{T_{\nu,0}} \frac{|\vec{x}-\vec{x}_{\textrm{H}}(\eta')|}{\eta-\eta'}\right) \frac{\vec{x}- \vec{x}_{\textrm{H}}(\eta)}{|\vec{x}- \vec{x}_{\textrm{H}}(\eta)|^3}\,,
\end{equation}
we now change the integration variable in the second integral in the right hand side of Eq.~(\ref{eq:dff2}). In terms of $\vec{y}=(m_{\nu}/T_{\nu,0})(\vec{x}-\vec{x}_H(\eta'))/(\eta-\eta')$ the integral is formally equivalent to the problem of calculating the gravitational field generated by the isotropic mass density $f_{\textrm{FD}}(y)$. Applying Gauss's law we find 
\begin{equation}
	\label{eq:gausslaw}
	\vec{F}(\eta) = - \frac{2}{\pi} G^2M^2m_{\nu}^4 \left(\frac{T_{\nu,0}}{m_{\nu}}\right)^3 \int_{0}^{\eta} \frac{d\eta'}{\eta-\eta'} \frac{a(\eta')}{a^2(\eta)} \frac{\frac{\vec{x}_{\textrm{H}}(\eta)-\vec{x}_{\textrm{H}}(\eta')}{\eta-\eta'}}{|\frac{\vec{x}_{\textrm{H}}(\eta)-\vec{x}_{\textrm{H}}(\eta')}{\eta-\eta'}|^3} \int_{0}^{\frac{m_{\nu}}{T_{\nu,0}} |\frac{\vec{x}_{\textrm{H}}(\eta)-\vec{x}_{\textrm{H}}(\eta')}{\eta-\eta'}|} dy \ y^2 f_{\textrm{FD}}(y)\,.
\end{equation}
In the limit $\eta' \to \eta$ we have that $(\vec{x}_{\textrm{H}}(\eta)-\vec{x}_{\textrm{H}}(\eta'))/(\eta-\eta')\rightarrow a(\eta)\vec{v}_H(\eta)$, where $\vec{v}_{\textrm{H}} = ad\vec{x}_{\textrm{H}}/dt$ is the halo peculiar velocity, and the integral is singular due to the factor of $(\eta-\eta')$ in the denominator. To regularize it we introduce the dimensionless time integration variable $\Delta=\eta/(\eta-\eta')$ in terms of which the Eq.~(\ref{eq:gausslaw}) becomes
\begin{equation}
	\label{eq:dff4}
	\vec{F}(\eta) = -\frac{2}{\pi}G^2M^2m_{\nu}^4 \left(\frac{T_{\nu,0}}{m_{\nu}}\right)^3  \int_{1}^{\infty} \frac{d\Delta}{\Delta} \frac{a(\eta')}{a^2(\eta)} \frac{\frac{\vec{x}_{\textrm{H}}(\eta)-\vec{x}_{\textrm{H}}(\eta')}{\eta-\eta'}}{|\frac{\vec{x}_{\textrm{H}}(\eta)-\vec{x}_{\textrm{H}}(\eta')}{\eta-\eta'}|^3} \int_{0}^{\frac{m_{\nu}}{T_{\nu,0}} |\frac{\vec{x}_{\textrm{H}}(\eta)-\vec{x}_{\textrm{H}}(\eta')}{\eta-\eta'}|} dy \ y^2 f_{\textrm{FD}}(y) \,,
\end{equation}
with $\eta' = \eta(1-1/\Delta)$ implicit on Eq.~(\ref{eq:dff4}). The limit $\eta' \to \eta$ is mapped into $\Delta \to \infty$ where the integrand approaches $\sim d\Delta/\Delta$, i.e., we encounter a logarithmic divergence. This is to be expected in our simple model of a point mass halo since gravity is a long range force and hence neutrinos coming towards the halo with arbitrarily large impact parameter give a nonvanishing contribution \cite{10.2307/j.ctvc778ff}. However, in realistic scenarios one has to account for the large-scale gravitational potential which sets a cutoff $\Lambda$ for the integral. We postpone a discussion of this to the next subsection and for now simply introduce the cutoff by hand. The expression in Eq.~(\ref{eq:dff4}) greatly simplifies in the limit $\Lambda \gg 1$, which motivates the following regularization scheme:
\begin{equation}
	\label{eq:dff5}
	\vec{F}(\eta) = -\frac{2}{\pi} \log\Lambda G^2M^2m_{\nu}^4 \left(\frac{T_{\nu,0}}{m_{\nu}}\right)^3 \lim_{\Lambda \to \infty} \frac{1}{\log \Lambda}  \int_{1}^{\Lambda} \frac{d\Delta}{\Delta} \frac{a(\eta')}{a^2(\eta)} \frac{\frac{\vec{x}_{\textrm{H}}(\eta)-\vec{x}_{\textrm{H}}(\eta')}{\eta-\eta'}}{|\frac{\vec{x}_{\textrm{H}}(\eta)-\vec{x}_{\textrm{H}}(\eta')}{\eta-\eta'}|^3} \int_{0}^{\frac{m_{\nu}}{T_{\nu,0}} |\frac{\vec{x}_{\textrm{H}}(\eta)-\vec{x}_{\textrm{H}}(\eta')}{\eta-\eta'}|} dy \ y^2 f_{\textrm{FD}}(y) \,.
\end{equation}
the integral is now dominated by its contribution from $\Delta \gg 1$ so we arrive at,
\begin{equation}
	\label{eq:cdff}
	\vec{F} = -\frac{2}{\pi} \log \Lambda G^2 M^2 m_{\nu}^4 \left(\frac{T_{\nu,0}}{m_{\nu}a}\right)^3 \frac{\vec{v}_{\textrm{H}}}{v_{\textrm{H}}^3} \int_{0}^{\alpha} dy \ y^2 f_{\textrm{FD}}(y) \,,
\end{equation}
where the time dependence is now implicit in Eq.~(\ref{eq:cdff}), and $\alpha = |\vec{\alpha}|$ is the magnitude of the vector defined in Eq.~(\ref{eq:alpha}). This is the Chandrasekhar dynamical friction formula \cite{chandrasekhar1943dynamical} as used, for instance, in the pioneering paper \cite{Fry:1980wf}. Before moving forward, let us stop for a moment to investigate the explicit dependence of Eq.~(\ref{eq:dff4}) on the cutoff $\Lambda$ in the simple scenario where the halo is only subject to Hubble drag, that is, assuming $v_{\textrm{H}} \sim a^{-1}$ and hence $d\vec{x}/d\eta = a\vec{v}_{\textrm{H}}$ is constant. In this case when evaluating Eq.~(\ref{eq:dff4}) at redshift $z=0$ (or $a=1$) we reproduce Eq.~(\ref{eq:cdff}) but now with an additional factor of:
\begin{equation}
\label{eq:supp}
	\frac{F}{F_{\Lambda \to \infty}} = \frac{1}{\log \Lambda} \int_{1}^{\Lambda} \frac{d\Delta}{\Delta} a\left(\eta\left(1-\frac{1}{\Delta}\right)\right) \,.
\end{equation}
This can be calculated in our reference  $\Lambda$CDM cosmology for any given initial redshift $z_{i}$ associated to the halo formation time (where we set $\eta_{i}=0$) and is plotted in Fig. \ref{fig:ckcutoff}. One can see the trend of $F/F_{\Lambda \to \infty} \to 1$ in the limit $\Lambda \to \infty$, with Chandrasekhar's formula in Eq.~(\ref{eq:cdff}) being suppressed for any finite value of the cutoff $\Lambda$ (also note that the suppression is larger for earlier forming halos which move faster at higher redshift and hence probe larger distance scales). We will give a physical interpretation for this result in the next subsection where we discuss the limitations associated to ignoring the large-scale gravitational potential in our simple single halo model. 
 
We are primarily interested in the small halo velocity regime $\alpha \ll 1$ of Eq.~(\ref{eq:cdff}), suitable for realistic neutrino masses as constrained by cosmology. In this case the dynamical friction force is proportional to the halo velocity,
\begin{equation}
\label{eq:cdflv}
	\vec{F} \approx -\frac{2}{3\pi} \log \Lambda G^2M^2 m_{\nu}^4 \vec{v}_{\textrm{H}} \,,
\end{equation}
and scales like $m_{\nu}^4$ such that the dynamical friction effect is dominated by the most massive neutrino eigenstate. We may then assume a single neutrino species for simplicity, or explicitly write the total dynamical friction force as a sum of individual contribution from different eigenstates.

In order to connect Eq.~(\ref{eq:cdflv}) with the results of future sections, it is convenient to rewrite it in terms of a quantity with dimension of inverse time. Since $F = M dv_{\textrm{H}}/dt$, we can define:
\begin{equation}
\label{eq:timescale}
	\tau^{-1} = -\frac{\vec{F} \cdot \vec{v}_{\textrm{H}}}{Mv_{\textrm{H}}^2} = \frac{2}{3\pi} \log \Lambda G^2M m_{\nu}^4 = 3.4 \times 10^{-5} \frac{\log \Lambda}{\log 100} \frac{M}{10^{13} M_{\bigodot}} \left(\frac{m_{\nu}}{0.1\textrm{eV}}\right)^4 H_{0}   \,,
\end{equation}
which is the characteristic time scale for an order one fractional decrease in the halo velocity due to the dynamical friction effect. Note that $1/\tau H_{0} = \Delta v/v$ is the overall relative decrease in the halo velocity over the age of the Universe $t \sim 1/H_{0}$. We obtain a numerical value of $\Delta v/v = 3.4 \times 10^{-5}$ for a halo mass $M = 10^{13} M_{\bigodot}$ and individual neutrino mass $m_{\nu}=0.1$eV, when also assuming $\Lambda = 100$. This already suggests that the dynamical friction effect is quite small, although it can pick up some significant contributions from the clustering of nearby halos as we will see in Sec.~\ref{sec:calc}.

\subsection{Limitations to the 1-halo approach}
\label{sec:b1a}

Thus far we have determined the anisotropic clustering of massive neutrinos behind moving point mass halos and the corresponding dynamical friction force. A more realistic calculation would have to account for both the finite extent of the halo and the presence of large-scale structure. Indeed, the Eq.~(\ref{eq:timescale}) involves an unknown Coulomb logarithm, $\log \Lambda$, where in typical applications of the dynamical friction formula the cutoff $\Lambda$ can be estimated as the ratio of maximum and minimum impact parameters, $\Lambda \sim b_{\textrm{max}}/b_{\textrm{min}}$ \cite{10.2307/j.ctvc778ff}. Here $b_{\textrm{min}} \sim R_{\rm halo}$ is the halo radius, and $b_{\textrm{max}} \sim \lambda_{\textrm{coh}} \sim 0.1$ Mpc$^{-1}$ is the CDM velocity coherence scale. The CDM bulk flow is only coherent over sufficiently small scales and hence our analysis based on a single moving halo is expected to break down at scales $\lambda \gtrsim \lambda_{\textrm{coh}}$.\footnote{We should also impose a cutoff corresponding to the distance traveled by free-streaming neutrinos, which sets the scale where neutrino inhomogeneities are coherent with CDM. As we shall see the neutrino free-streaming scale is much larger than the CDM velocity coherence length $\lambda_{\textrm{coh}}$. Relatedly, since our formalism assumes neutrinos are non-relativistic, one might also expect that we should impose a cutoff corresponding to the distance neutrinos have traveled while non-relativistic. Yet this scale is much larger than the free-streaming scale and, of course, the CDM velocity coherence length $\lambda_{\textrm{coh}}$ and is therefore irrelevant.} This point will be made more clear in the next section, where we also provide a precise definition for the velocity coherence scale. 

\begin{figure}
	\centering
	\includegraphics[width=0.75\textwidth]{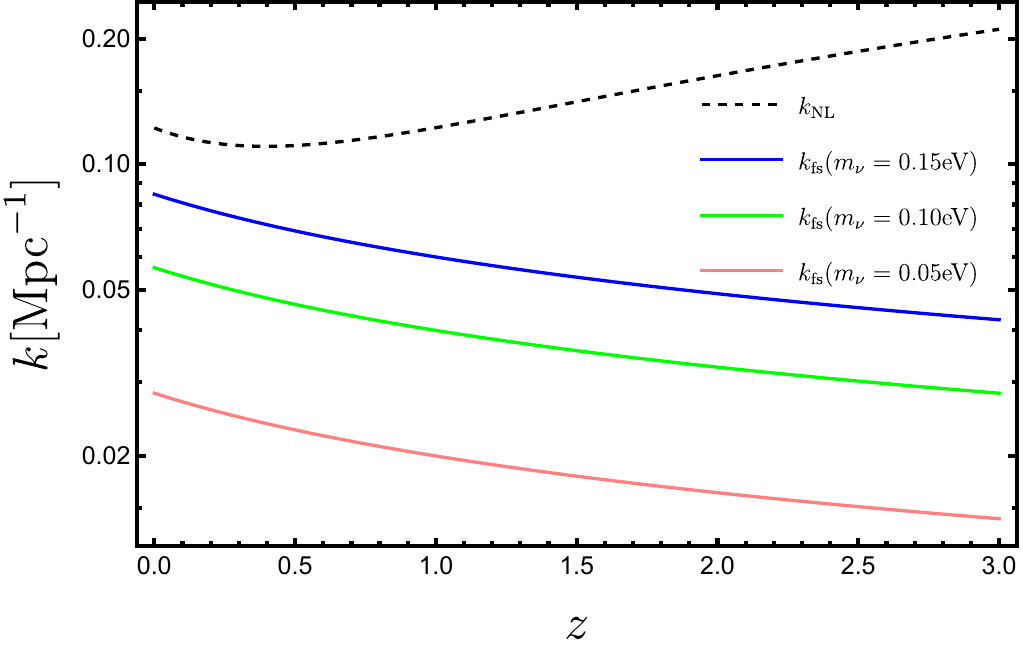}
	\caption{Scale of nonlinearities, $k_{\textrm{NL}}$, and neutrino free-streaming scale, $k_{\textrm{fs}}$, as a function of redshift and for varying neutrino mass. The black dashed curve is the scale of nonlinearities and blue, green and pink solid curves correspond to the free-streaming scale for neutrino masses of $m_{\nu}=0.05$eV,$m_{\nu}=0.1$eV and $m_{\nu}=0.15$eV respectively.} 
	\label{fig:scales}
\end{figure}

When accounting for the large-scale structure, the halo velocity $v_{\textrm{H}}$ is replaced by the root mean square CDM velocity dispersion $\sigma$ given by,
\begin{equation}
	\label{eq:cdmdispersion}
	\sigma^2 =\  \langle v^2 \rangle \ = \int \frac{dk}{2\pi^2} P_{\theta \theta}(z,k) \,,
\end{equation}
with $P_{\theta \theta}(z,k)$ the power spectrum of the divergence of the CDM velocity field $\theta = \vec{\nabla} \cdot \vec{v}$ (note that the velocity dispersion depends on redshift). This can be used to define the scale associated to the CDM bulk flow, which is also known as the scale of nonlinearities\footnote{In the literature one often finds alternative definitions for the scale of nonlinearities, such as the scale at which the dimensionless power is unity, which we will here denote by $\tilde{k}_{\textrm{NL}}$. In our reference $\Lambda$CDM cosmology one finds $\tilde{k}_{\textrm{NL}} \sim 0.15$Mpc$^{-1} \sim k_{\textrm{NL}}$ at $z=0$ so the two definitions produce numerically equivalent results. However, their time evolution is different and this can produce significant differences at higher redshift. For example, $\tilde{k}_{\textrm{NL}} \approx 0.5$Mpc$^{-1}$ at $z=1$, while $k_{\textrm{NL}}$ changes only slightly when compared to its $z=0$ value, as depicted in Fig.~\ref{fig:scales}.}
\begin{equation}
\label{eq:scalenl}
	k_{\textrm{NL}} = \frac{aH}{\sigma} \,,
\end{equation}
where $a$ is the scale factor and $H=\dot{a}/a$ is the Hubble rate, with dot denoting a derivative with respect to cosmic time $t$. We evaluate the Eq.~(\ref{eq:cdmdispersion}) numerically in our reference $\Lambda$CDM cosmology and in linear perturbation theory, using $P_{\theta \theta}(z,k) = (faH)^2 [D_{\textrm{L}}(z)/D_{\textrm{L}}(z=0)]^2 P_{\delta \delta}(z=0,k)$, where $P_{\delta \delta}(z=0,k)$ is the CDM power spectrum evaluated at $z=0$. The linear growth factor $D_{\textrm{L}}$ and linear growth rate $f = d\log D_{\textrm{L}}/d\log a$ are obtained with the approximate formulas in \cite{hamilton2001formulae}.

In our calculations thus far we have implicitly assumed a large-scale homogeneous distribution of neutrinos. In general, neutrinos have a large velocity dispersion\footnote{In later sections we provide the motivation for this particular definition of the neutrino velocity dispersion.},
\begin{equation}
\label{eq:nudispersion}
	\sigma_{\nu} = \sqrt{\frac{3\zeta(3)}{\log 4}} \frac{T_{\nu,0}}{m_{\nu}a} \,,
\end{equation}
and hence free-stream over cosmological distances with a characteristic scale $k_{\textrm{fs}}$ of,
\begin{equation}
\label{eq:fstreaming}
	 k_{\textrm{fs}}^2 = \frac{4\pi Ga^2\bar{\rho}(a)}{\sigma_{\nu}^2} =  \frac{3}{2} \Omega_{\textrm{m}}(a) \left(\frac{aH}{\sigma_{\nu}}\right)^2 \,,
\end{equation}
where $\bar{\rho}(a)$ is the background matter energy density and we used the Friedmann equation. In Fig.~\ref{fig:scales} we plot the scale of nonlinearities, $k_{\textrm{NL}}$, and the neutrino free-streaming scale, $k_{\textrm{fs}}$, as a function of redshift in our reference $\Lambda$CDM cosmology for three values of the neutrino mass: $m_{\nu}=0.05$eV, $m_{\nu}=0.1$eV and $m_{\nu}=0.15$eV.

At scales smaller than the free-streaming scale ($k \gtrsim k_{\textrm{fs}}$), pressure dominates over the gravitational field and neutrino anisotropies are washed-out. On the other hand, at larger scales ($k \ll k_{\textrm{fs}}$) neutrinos cluster like cold matter and there are significant anisotropies in the neutrino fluid.\footnote{In practice the transition at the free-streaming scale is rather smooth, with the neutrino species fully behaving like CDM only at scales above the neutrino horizon scale which is approximately set by the free-streaming scale when evaluated at the moment when neutrinos first became nonrelativistic and lies at much larger scales. See Fig.1 in \cite{nascimento2023accurate} for more details.} As a consequence we expect our previous analysis, which assumes a homogeneous distribution of neutrinos, to break down when the CDM flows over scales that are larger than $\lambda_{\textrm{fs}} \sim 1/k_{\textrm{fs}}$, i.e., when $k_{\textrm{fs}} \gtrsim k_{\textrm{NL}}$. From Fig.~\ref{fig:scales} we conclude that for the small neutrino masses as constrained by cosmology, $k_{\textrm{fs}}<k_{\textrm{NL}}$ and hence our assumptions are consistent. 

Note that since $k_{\textrm{fs}} \sim aH/\sigma_{\nu}$ and $k_{\textrm{NL}} \sim aH/\sigma$, we have that $k_{\textrm{fs}}/k_{\textrm{NL}} \sim \sigma/\sigma_{\nu}$ is analogous to the parameter $\alpha$ introduced in Eq.~(\ref{eq:alpha}) and this has two important consequences. First, as we showed before, the limit $\alpha \ll 1$ allowed for a significant simplification of the dynamical friction effect, leading to the simple formula in Eq.~(\ref{eq:timescale}). In an approach that accounts for the large-scale structure, we expect a similar simplification to happen in the limit $k_{\textrm{fs}}/k_{\textrm{NL}} \ll 1$, and indeed we will use this as a guiding principle in future sections. Second, due to the breakdown of our framework in the regime $k_{\textrm{fs}}/k_{\textrm{NL}} \sim \sigma/\sigma_{\nu} \gtrsim 1$, the Chandrasekhar formula in Eq.~(\ref{eq:cdff}) can no longer be applied when $\alpha \gtrsim 1$. If we try to ignore this and proceed to extract the $\alpha \gg 1$ limit in Eq.~(\ref{eq:cdff}) we would (erroneously) arrive at,
\begin{equation}
\label{eq:cdflhv}
\vec{F} \approx -\frac{6}{\pi} \zeta(3) \log \Lambda G^2M^2m_{\nu}^4\left(\frac{T_{\nu,0}}{m_{\nu}a}\right)^3\frac{\vec{v}_{\textrm{H}}}{v_{\textrm{H}}^3} \,,
\end{equation} 
where we used Eq.~(\ref{eq:fd}) with $g_{\nu}=2$. For a given fixed neutrino temperature today $T_{\nu,0}$, one would then conclude that $F \sim m_{\nu}$ and hence the dynamical friction effect can be made arbitrarily large by increasing $m_{\nu}$, which in turn would enable us to place an upper bound on the neutrino mass scale based solely on the observation that galaxies have some nonzero peculiar velocities today and hence the dynamical friction effect cannot be too large \cite{Fry:1980wf}.  This argument completely ignores the cutoff at the free-streaming scale and is hence incorrect. Indeed, we can get some valuable intuition from considering the limit $k_{\textrm{fs}}/k_{\textrm{NL}} \sim \sigma/\sigma_{\nu} \gg 1$ (analogous to $\alpha \gg 1$), which corresponds to a neutrino mass so large that the neutrinos are no longer streaming over cosmological distances and hence effectively behave just like a cold matter component. Then neutrinos will cluster along with the CDM and there no longer exists a homogeneous sea of neutrinos around the halo. Another way to think about it is that in this limit, neutrinos are comoving with CDM and then, in the frame of the halo, the situation is equivalent to a halo at rest in a homogeneous sea of neutrinos. In that case, neutrino clustering is isotropic in the halo frame and there are no wakes and no dynamical friction.

In later sections we will see precisely how the free-streaming scale comes into play when generalizing the approach of this section to a more realistic dark matter distribution. Another limitation to Eq.~(\ref{eq:timescale}) is that it neglects the effects of clustering between nearby halos. In studying neutrino wakes and dynamical friction in large-scale structure, we then expect to see some corrections accounting for the fact that nearby halos will share neutrino wakes. The two-halo term was studied in detail in \cite{okoli2017dynamical} under the assumption of a static universe, as seen from the frame of reference of a moving halo. This approach implicitly assumes that all of the CDM moves coherently, and hence breaks down at scales above the CDM velocity coherence scale. In the next section, we will develop a framework that incorporates the expansion of the Universe and also allows us to work with CDM structures that have a velocity dispersion, allowing us to push the regime of validity of the calculations all the way up to the free-streaming scale.

\section{Neutrino wakes in large-scale structure}
\label{sec:dfolss}

In the previous section we determined that the peculiar motion of halos generates anisotropic neutrino wakes that slow halos down due to dynamical friction.  Our goal now is to extended the analysis of neutrino wakes to a general CDM distribution (in an expanding universe) in order to account for the large scale structure. This improves on several limitations of the previous approach: The cutoffs at both the free-streaming and CDM coherence scale will naturally emerge from the formalism, as opposed to being introduced by hand to regularize divergent integrals. We will also derive a 2-halo term to the dynamical friction formula and naturally account for the CDM velocity dispersion, as opposed to introducing an ad-hoc halo peculiar velocity. Finally, this approach allows us to derive corrections to the matter power spectrum from neutrino wakes and compute the matter-neutrino cross-bispectrum generated by the distortions to the neutrino field.

Based on our intuition from the previous section, we expect the problem to significantly simplify in the limit $k_{\textrm{fs}}/k_{\textrm{NL}}\ll 1$. As we will see, by working in this regime we are able to address the problem analytically. Our strategy is to solve the Boltzmann Eq.~(\ref{eq:beqbkt}) in Fourier space, integrate the solution over comoving momenta to obtain the neutrino density perturbation, and then consider a Taylor expansion in the small parameter $k_{\textrm{fs}}/k_{\textrm{NL}}\ll 1$. We will then calculate both the zeroth and first order terms. The latter will correspond to the distortion in the neutrino density field that we are looking for. In our calculations we remain agnostic about the nonlinear evolution of CDM perturbations, imposing only the continuity and Poisson equations. Our final expressions in this section will be written in terms of CDM power spectra and bispectra without assuming a particular model for their computation. 

\subsection{Solution to the Boltzmann equation for an arbitrary CDM density field}

Going back to the problem at hand, we need to consider Eq.~(\ref{eq:beqbkt}) in its full generality. In Fourier space (let $\vec{\nabla} \to i\vec{k}$), it reads:
\begin{equation}
	\label{eq:befs}
	\frac{\partial f_{\nu,1}}{\partial \eta} + i \frac{\vec{k} \cdot \vec{q}}{m_{\nu}} f_{\nu,1} = im_{\nu}a^2(\eta)\phi(\eta, \vec{k}) \vec{k} \cdot \frac{\partial f_{\nu,0}}{\partial \vec{q}} \,,
\end{equation}
where we remind the reader that  $f_{\nu,0}(q)$ corresponds to the background neutrino distribution function, while $f_{\nu,1}(\eta,\vec{k},\vec{q})$ is the linear response to the CDM gravitational potential $\phi(\eta, \vec{k})$. Furthermore, $\vec{q}$ is the comoving momentum, and we are working in the time coordinate $\eta$ defined by $dt/d\eta = a^2(\eta)$. This is a first-order ordinary differential equation whose solution is straightforward to write down \footnote{The distribution function quickly loses sensitivity to the initial conditions at sub-horizon scales, hence we set $\phi(0,\vec{k})=0$.}
\begin{equation}
	\label{eq:solbefs}
	f_{\nu,1}(\eta, \vec{k}, \vec{q}) = im_{\nu}\vec{k} \cdot \frac{\partial 	f_{\nu,0}}{\partial \vec{q}} \int_{0}^{\eta} d\eta' a^2(\eta') \phi(\eta',\vec{k}) e^{-i\frac{\vec{k} \cdot \vec{q}}{m_{\nu}} (\eta-\eta')} \,,
\end{equation}
and hence the perturbation to the neutrino mass density is,
\begin{equation}
	\label{eq:neupert}
	\begin{split}
		\delta \rho_{\nu}(\eta, \vec{k}) &= m_{\nu}a^{-3}(\eta) \int \frac{d^3 \vec{q}}{(2\pi)^3} f_{\nu,1}(\eta, \vec{k}, \vec{q}) \,, \\ &= im_{\nu}^2 a^{-3} \vec{k} \cdot \int_{0}^{\eta} d\eta' a^2(\eta') \phi(\eta', \vec{k}) \int \frac{d^3 \vec{q}}{(2\pi)^3} \frac{\partial f_{\nu,0}}{\partial \vec{q}} e^{-i\frac{\vec{k} \cdot \vec{q}}{m_{\nu}} (\eta-\eta')} \,,
	\end{split}
\end{equation}
now integrate the second integral by parts, followed by integration over the solid angle, to arrive at:
\begin{equation}
	\label{eq:nuepert2}
	\delta \rho_{\nu}(\eta,\vec{k}) = -\frac{m_{\nu}^2 k}{2\pi^2} a^{-3}(\eta) \int_{0}^{\eta} d\eta' a^2(\eta') \phi(\eta', \vec{k}) \int_{0}^{\infty} dq \ q f_{\nu,0}(q) \sin \left[\frac{kq}{m_{\nu}}(\eta-\eta')\right] \,.
\end{equation}
To proceed use Poisson's equation $k^2 \phi(\eta, \vec{k}) = -4\pi Ga^2\delta \rho(\eta, \vec{k})$ with $\delta \rho(\eta, \vec{k})$ the CDM density in excess of the average $\bar{\rho}(\eta)$, i.e., $\rho(\eta, \vec{x}) = \bar{\rho}(\eta) + \delta \rho(\eta, \vec{x}) \equiv \bar{\rho}(\eta)[1+\delta(\eta,\vec{x})]$ in real space, with $\delta(\eta,\vec{x})$ the CDM density contrast. We also change the integration variable in the integral over comoving momenta to $x=q/T_{\nu,0}$, using the relation $f_{\nu,0}(q) = f_{\textrm{FD}}(q/T_{\nu,0}) = f_{\textrm{FD}}(x)$. Equation~(\ref{eq:nuepert2}) now reads,
\begin{equation}
	\label{eq:nuepert3}
	\delta_{\nu}(\eta,\vec{k}) = \frac{\delta \rho_{\nu}(\eta,\vec{k})}{\bar{\rho}_{\nu}(\eta)} = \frac{4\pi G\bar{\rho}_{0}}{3\zeta(3)} \int_{0}^{\infty} dx \, x^2 f_{\textrm{FD}}(x) \int_{0}^{\eta} d\eta' (\eta-\eta') a(\eta') \delta(\eta',\vec{k}) j_{0}\left[k\frac{T_{\nu,0}}{m_{\nu}}x(\eta-\eta')\right] \,,
\end{equation}
with $\bar{\rho}_{0}$ the comoving background matter energy density (or the matter density when evaluated at the present time), $\delta_{\nu}(\eta,\vec{k})$ the neutrino density contrast, $j_{0}(x)=\sin x/x$ the spherical Bessel function of zeroth order and the background neutrino energy density $\bar{\rho}_{\nu}(\eta)$ is given by
\begin{equation}
\label{eq:nudensity}
	\bar{\rho}_{\nu}(\eta) = m_{\nu}a^{-3}(\eta) \int \frac{d^3 \vec{q}}{(2\pi)^3} f_{\nu,0}(\vec{q}) =  \frac{3\zeta(3)}{2\pi^2} a(\eta)^{-3}m_{\nu}T_{\nu,0}^3 \,,
\end{equation}
where we used $f_{\nu,0}(q) = f_{\textrm{FD}}(q/T_{\nu,0})$ and Eq.~(\ref{eq:fd}) with $g_{\nu}=2$. Also note that $3\zeta(3) = \int_{0}^{\infty} dx \, x^2 f_{\textrm{FD}}(x)$ sets the numerical factor in the denominator of Eq.~(\ref{eq:nuepert3}). Equation~\ref{eq:nuepert3} is a key equation from which we will derive several observables. We emphasize that when referring to the CDM density or gravitational potential we add no subscripts, while the corresponding neutrino quantities always carry the subscript $\nu$. In this work we always assume a small fractional contribution of neutrinos to the total matter, $f_{\nu} = \bar{\rho}_{\nu}/\bar{\rho} \ll 1$.

We first consider scales that are above the Jeans length, $kT_{\nu,0}\eta/m_{\nu} \ll 1$. In this case we can work under the approximation $j_{0}(x) \approx 1$ for $x\ll 1$, and assume a linear evolution for the CDM field (since we are interested in neutrino masses for which $k_{\textrm{NL}}>k_{\textrm{fs}}$)
\begin{equation}
\label{eq:lineargrowth}
	\delta(\eta',\vec{k})=\frac{D_{\textrm{L}}(\eta')}{D_{\textrm{L}}(\eta)} \delta(\eta,\vec{k}) \,,
\end{equation}
with $D_{\textrm{L}}(\eta)$ the linear growth factor, which satisfies the following differential equation:
\begin{equation}
\label{eq:eqlingro}
	\frac{d^2}{d\eta^2} D_{\textrm{L}}(\eta) = 4\pi G \bar{\rho}_{0} a(\eta) D_{\textrm{L}}(\eta) \,.
\end{equation}  
It then follows from Eq.~(\ref{eq:nuepert3}) that, for $kT_{\nu,0}\eta/m_{\nu} \ll 1$,
\begin{equation}
\label{eq:largescales}
	\delta_{\nu}(\eta,\vec{k})  \underset{k\ll k_{\textrm{fs}}}{=}  4\pi G \bar{\rho}_{0} \frac{\delta(\eta,\vec{k})}{D_{\textrm{L}}(\eta)} \int_{0}^{\eta} d\eta' (\eta-\eta')a(\eta')D_{\textrm{L}}(\eta') \,,
\end{equation}
where we used Eq.~(\ref{eq:lineargrowth}). Substituting Eq.~(\ref{eq:eqlingro}) into Eq.~(\ref{eq:largescales}) yields after integrating by parts,
\begin{equation}
\label{eq:largescales2}
	\delta_{\nu}(\eta,\vec{k}) \underset{k\ll k_{\textrm{fs}}}{=} \delta(\eta,\vec{k}) \left[1-\eta \frac{D_{\textrm{L}}'(0)}{D_{\textrm{L}}(\eta)} - \frac{D_{\textrm{L}}(0)}{D_{\textrm{L}}(\eta)} \right] \approx \delta(\eta,\vec{k}) \,,
\end{equation}
where in the second equality we use the fact that the additional terms in the square brackets can be made negligible by setting the initial conditions at sufficiently early times (which is equivalent to just keeping the growing mode). This is the standard result that at scales above the neutrino horizon the neutrino species exactly traces the cold dark matter. In our derivation we had to assume a linear evolution for the CDM component at scales above the neutrino horizon, with the linear growth factor satisfying Eq.~(\ref{eq:eqlingro}), in order to be consistent within our framework of a linear neutrino response. However, the final result is general as it is essentially the statement of causality and hence holds in any nonlinear approach, as long as one consistently accounts for the nonlinearities in both neutrinos and CDM. 

Let us now investigate the opposite regime, $kT_{\nu,0}\eta/m_{\nu} \gg 1$, in Eq.~(\ref{eq:nuepert3}). The spherical Bessel function peaks at $\Delta \eta = \eta-\eta'=0$ and becomes vanishing when,
\begin{equation}
\label{eq:spherical}
	k \frac{T_{\nu,0}}{m_{\nu}} \eta \frac{\Delta \eta}{\eta} \gg 1 \iff \frac{\Delta \eta}{\eta} \gg \frac{1}{k \frac{T_{\nu,0}}{m_{\nu}}\eta} \,,
\end{equation}
so the time integral in Eq.~(\ref{eq:nuepert3}) becomes dominated by its contribution from $\Delta \eta/\eta \ll 1$, which suggests a Taylor series expansion of the integrand around $\eta'=\eta$. To leading order we can simply evaluate the integrand at $\eta'=\eta$, which implies that the clustering of neutrinos only depends on a snapshot of the CDM distribution at the final time. This can be interpreted as follows: Neutrino fluctuations with a (comoving) wavenumber $k$ have a characteristic time scale, $t \sim a/k\sigma_{\nu} \sim (k_{\textrm{fs}}/k)H^{-1}$, which is much smaller than a Hubble time at sub-free streaming scales. We can then effectively consider a quasi-static space in terms of comoving quantities. However, since our goal in this paper is to study the effects of halo peculiar velocities, we need to go beyond the leading order term. Indeed, as long as $k_{\textrm{fs}}/k_{\textrm{NL}} \sim \sigma/\sigma_{\nu} \ll 1$, it suffices to consider the next to leading order term in the Taylor series expansion since the CDM bulk flow is slow when compared to the neutrino time scale. In what follows we compute the contributions from zeroth and first order terms in the expansion. 

\subsubsection{Zeroth order: The static limit}

We begin with the zeroth-order term which corresponds to approximating $a(\eta') \delta(\eta',\vec{k}) \approx a(\eta) \delta(\eta,\vec{k})$ when evaluating the time integral in Eq.~(\ref{eq:nuepert3}). As previously discussed, this is justified since the integral is dominated by its contribution from $\eta'=\eta$ in the limit where $kT_{\nu,0}\eta/m_{\nu} \gg 1$. This yields
\begin{equation}
	\label{eq:nuepert0th}
	\begin{split}
	\delta^{(0)}_{\nu}(\eta,\vec{k}) &= \frac{4\pi G \bar{\rho}_{0}}{3\zeta(3)} a(\eta)\delta(\eta,\vec{k}) \int_{0}^{\infty} dx \, x^2  f_{\textrm{FD}}(x) \int_{0}^{\eta} d\eta' (\eta-\eta')  j_{0}\left[k\frac{T_{\nu,0}}{m_{\nu}}x(\eta-\eta')\right] \ \\ & = \frac{4\pi G \bar{\rho}_{0}}{3\zeta(3)} a(\eta)\delta(\eta,\vec{k}) \left(\frac{m_{\nu}}{kT_{\nu,0}}\right)^2 \int_{0}^{\infty} dx f_{\textrm{FD}}(x) \left[1-\cos\left(k\frac{T_{\nu,0}}{m_{\nu}}x\eta\right) \right]  \\ & = \left(\frac{k_{\textrm{fs}}}{k}\right)^2 \delta(\eta,\vec{k}) \,,
	\end{split}
\end{equation}
where we used Eqs.~(\ref{eq:fd}), (\ref{eq:nudispersion}), (\ref{eq:fstreaming}), the Friedmann equation $\Omega_{\textrm{m}}(a) = (8\pi G \bar{\rho}_{0}/3a^3H^2)$, and from the second to the third line we dropped the cosine term since $kT_{\nu,0}\eta/m_{\nu} \gg 1$. Note that our definition of the neutrino velocity dispersion, as given by Eq.~(\ref{eq:nudispersion}), produces the result in the last line of Eq.~(\ref{eq:nuepert0th}) with no extra coefficients. Equation (\ref{eq:nuepert0th}) is a well-known result at sub free-streaming scales, but also under the assumption that halos are static \cite{ringwald2004gravitational}. Ultimately, it is just the statement of neutrino free-streaming: The neutrino density perturbation is suppressed with respect to CDM at sub free-streaming scales.

In this work, however, we are interested in the effect produced by the peculiar motion of halos. That means we need to consider the leading order correction to Eq.~(\ref{eq:nuepert0th}), i.e., the first order term. 

\subsubsection{First order: Accounting for CDM bulk flows}
 
Going back to Eq.~(\ref{eq:nuepert3}), we now need to expand $a(\eta')\delta(\eta',\vec{k})$ to first order in $\Delta \eta=\eta -\eta'$. On one hand,
\begin{equation}
	\label{eq:exp1}
	a(\eta') = a(\eta)\left[1-a^2(\eta)H(\eta)(\eta-\eta') + \dots\right] \,.
\end{equation}
Since only $\Delta \eta \lesssim m_{\nu}/kT_{\nu,0}$ contributes to the integral, and $k_{\textrm{fs}} \sim (m_{\nu}/T_{\nu,0})a^2 H$ from Eqs.~(\ref{eq:nudispersion}) and (\ref{eq:fstreaming}), we find that the first order term in the right hand side of Eq.~(\ref{eq:exp1}) scales like $\sim k_{\textrm{fs}}/k$ with respect to the zeroth order term and can hence be neglected at sufficiently small scales. We also need to consider:
\begin{equation}
	\label{eq:exp2}
	\delta(\eta',\vec{k})  = \delta(\eta,\vec{k}) -a^2(\eta) \dot{\delta}(\eta,\vec{k})(\eta-\eta')+ \dots
\end{equation}
where dot denotes a derivative with respect to cosmic time $t$. To proceed we assume the continuity equation for the CDM fluid in an expanding universe
\begin{equation}
	\label{eq:cont}
		\dot{\delta} = -\frac{1}{a} \vec{\nabla} \cdot \vec{P} \,,
\end{equation}
when written in position space with $\vec{P}=(1+\delta) \vec{v}$ the CDM momentum (density) field, where $\vec{v}(\eta,\vec{x})$ is the CDM velocity. The ratio of first to zeroth order terms in Eq.~(\ref{eq:exp2}) is then, after using Eq.~(\ref{eq:cont}) in Fourier space,
\begin{equation}
\label{eq:ratio}
	\frac{a(i\vec{k} \cdot \vec{P})\Delta \eta}{\delta} \lesssim \frac{1}{\sigma_{\nu}} \frac{P}{\delta} \,,
\end{equation}
where once again only $\Delta \eta \lesssim m_{\nu}/kT_{\nu,0}$ contributes to the integral, $\sigma_{\nu} \sim T_{\nu,0}/m_{\nu}a$ from Eq.~(\ref{eq:nudispersion}), and we introduce the notation $kP=i\vec{k} \cdot \vec{P}$ for the divergence of the momentum field. In the linear regime we have $P=v = (faH)\delta/k$ is the linear CDM velocity field, a tracer of the CDM density field [with $f = d\log D_{\textrm{L}}/d\log a$ the linear growth rate]. The ratio of first to zeroth order terms in Eq.~(\ref{eq:ratio}) then becomes $\sim aH/\sigma_{\nu}k \sim k_{\textrm{fs}}/k$ once again. We reproduce a well-known important result: In linear perturbation theory the leading order correction to Eq.~(\ref{eq:nuepert0th}) is negligible at sufficiently small scales. 

However, we now argue that the nonlinearities in the CDM momentum field, parameterized by $\vec{p}=\vec{P}-\vec{v}$, or $\vec{p}= \delta \vec{v}$ in position space (we also use the notation $kp = i\vec{k}\cdot \vec{p}$ for its divergence) yield a nonnegligible contribution to next to leading order which has to be accounted for. The reason for this is as follows: At small enough scales the CDM  moves coherently, with $v \sim \sigma$ the CDM velocity dispersion as defined in Eq.~(\ref{eq:cdmdispersion}) in which case $p \sim \sigma \delta$. This gives a contribution to the ratio of first to zeroth order terms as given in Eq.~(\ref{eq:ratio}) of size $\sim \sigma/\sigma_{\nu} \sim k_{\textrm{fs}}/k_{\textrm{NL}}$, which is independent of $k$. At scales $k \gtrsim k_{\textrm{NL}}$ this becomes larger than the contributions scaling like $\sim k_{\textrm{fs}}/k$ which we have neglected before, and it persists down to arbitrarily small scales. Such contributions cannot be neglected when studying nonlinear effects which turn on at scales $k\gtrsim k_{\textrm{NL}}$.

The ratio $k_{\textrm{fs}}/k_{\textrm{NL}}$ is analogous to the parameter $\alpha$ we required to be small in Sec. \ref{sec:b1a} as a consistency condition for our framework in terms of halos moving through a homogeneous sea of neutrinos. Here, the smallness of $\sigma/\sigma_{\nu} \sim k_{\textrm{fs}}/k_{\textrm{NL}}$ is required in order to have a well-defined expansion in Eq.~(\ref{eq:exp2}), with the physical interpretation that we can truncate the expansion at next to leading order when the typical halo motions are slow with respect to the characteristic time scale associated to the neutrino dynamics. We reinforce that this is consistent with current bounds on the sum of neutrino mass eigenstates from cosmological observations, as can be seen from Fig.~\ref{fig:scales}.

We are now ready to compute the leading order correction to Eq.~(\ref{eq:nuepert0th}), i.e., to first order in $k_{\textrm{fs}}/k_{\textrm{NL}}$. We substitute Eq.~(\ref{eq:cont}) into the second term in the right hand side of Eq.~(\ref{eq:exp2}). It then follows from Eq.~(\ref{eq:nuepert3}), after dropping contributions that are negligible at sub-free streaming scale as discussed above:
\begin{equation}
	\label{eq:nuepert1th}
	\begin{split}
		\delta_{\nu}^{(1)}(\eta,\vec{k}) & = \frac{4\pi G \bar{\rho}_{0}}{3\zeta(3)} a^2(\eta) kp(\eta,\vec{k}) \int_{0}^{\infty} dx \, x^2 f_{\textrm{FD}}(x) \int_{0}^{\eta} d\eta' (\eta-\eta')^2  j_{0}\left[k\frac{T_{\nu,0}}{m_{\nu}}x(\eta-\eta')\right] \\ & =  \frac{4\pi G \bar{\rho}_{0}}{3\zeta(3)} a^2(\eta) kp(\eta,\vec{k}) \int_{0}^{\infty} dx \, x^2 f_{\textrm{FD}}(x) \left(\frac{m_{\nu}}{kT_{\nu,0}x}\right)^3 \left[\sin\left(k \frac{T_{\nu,0}}{m_{\nu}} x\eta\right) - k \frac{T_{\nu,0}}{m_{\nu}}x\eta \cos\left(k \frac{T_{\nu,0}}{m_{\nu}} x\eta\right) \right] \\ & = \gamma \left(\frac{k_{\textrm{fs}}}{k}\right)^2 \frac{p(\eta,\vec{k})}{\sigma_{\nu}} \,,
	\end{split}
\end{equation}
where we used Eqs.~(\ref{eq:fd}), (\ref{eq:nudispersion}), (\ref{eq:fstreaming}), the Friedmann equation $\Omega_{\textrm{m}}(a) = (8\pi G \bar{\rho}_{0}/3a^3H^2)$, and from the second to the third line we dropped the cosine term since $kT_{\nu,0}\eta/m_{\nu} \gg 1$. We also introduced a numerical factor set by the (relativistic) Fermi-Dirac profile
\begin{equation}
\label{eq:gamma}
	\gamma = \frac{\pi}{6\zeta(3)} \left(\frac{3\zeta(3)}{\log 4}\right)^\frac{3}{2} \approx 1.83 \,.
\end{equation}
We can now combine Eqs.~(\ref{eq:nuepert0th}) and (\ref{eq:nuepert1th}) into an expression for the neutrino density contrast at sub-free streaming scales
\begin{equation}
\label{eq:combine}
	\delta_{\nu}(\vec{k}) \underset{k\gg k_{\textrm{fs}}}{=}  \left(\frac{k_{\textrm{fs}}}{k}\right)^2\left[\delta(\vec{k})+\gamma \frac{p(\vec{k})}{\sigma_{\nu}} + \mathcal{O}\left(\frac{k_{\textrm{fs}}}{k_{\textrm{NL}}}\right)^2 \right] \,.
\end{equation}
Where the reader is reminded that $\vec{p}$ is the non-linear contribution to the CDM momentum, $\vec{p}=\vec{P}-\vec{v}$, or $\vec{p}= \delta \vec{v}$. Henceforth we omit the time dependence for simplicity of notation. At scales much larger than the free-streaming length the neutrinos exactly trace the cold dark matter as given by Eq.~(\ref{eq:largescales2}). On the other hand, at scales below the free-streaming length the neutrino density contrast is mostly proportional to the CDM density contrast with a scale dependent factor, but it also traces the CDM (nonlinear) momentum through a term that scales like $\sim k_{\textrm{fs}}/k_{\textrm{NL}} \ll 1$ which produces a distortion in the neutrino density field due to the CDM bulk flow, i.e., the peculiar motion of halos. This is a nonlinear effect at sub-free streaming scales and hence is not accounted for in linear perturbation theory. However, it is principle accounted for in implementations of massive neutrinos in nonlinear structure formation that consistently evolves neutrino perturbations in the nonlinear gravitational potential of CDM (such as N-body simulations).

It is important to point out that in our calculations we have remained agnostic about the nonlinear gravitational evolution, imposing only kinematic relations. We would need to consider explicit models for the nonlinear dynamics of CDM, such as N-body simulations or perturbative methods, in order to evaluate the higher order terms in Eq.~(\ref{eq:combine}) since they would involve a derivative of the velocity which requires the equations of motion. Additionally, in our approach we do not have a handle on the neutrino density contrast at scales that are comparable to the free-streaming scale as that would require evaluating the time integrals in Eq.~(\ref{eq:nuepert3}), which also requires a model for the nonlinear dynamics of CDM. 

In this work we adopt two strategies to go around those issues. The first is to assume that $k_{\textrm{fs}}/k_{\textrm{NL}} \ll 1$. As depicted in Fig.~\ref{fig:scales}, this a very good approximation for an individual neutrino mass of $m_{\nu}=0.05$eV. However, in the case of a large individual neutrino mass of $m_{\nu}=0.15$eV (corresponding to $\sum m_{\nu i}\approx 0.45$eV), this small parameter can be as large as $k_{\textrm{fs}}/k_{\textrm{NL}} \sim 2/3$ producing $(2/3)^2 \sim 40\%$ level corrections to our calculations so that we can still confidently predict the overall magnitude of distortion effects, even for such high values for the individual neutrino mass. Our second strategy is to introduce a simple interpolation between Eqs.~(\ref{eq:largescales2}) and (\ref{eq:combine}) to ensure that our final formula for the neutrino density contrast can be applied at all scales. For example, under the approximation that the neutrino density contrast is proportional to the CDM density contrast with a scale dependent factor [which is equivalent to dropping the second $p(\vec{k})$ term in the right-hand side of Eq.~(\ref{eq:combine})], the following formula is known to give a good approximation at all scales \cite{ali2013efficient}:
\begin{equation}
\label{eq:glue1}
	\delta_{\nu}(\vec{k}) \approx \frac{1}{(1+\frac{k}{k_{\textrm{fs}}})^2} \delta(\vec{k}) = \frac{\left(\frac{k_{\textrm{fs}}}{k}\right)^2}{(1+\frac{k_{\textrm{fs}}}{k})^2} \delta(\vec{k}) \,.
\end{equation}
Note that this reduces to both Eqs.~(\ref{eq:largescales2}) and (\ref{eq:combine}) in the appropriate limits, as it must. We now slightly generalize this prescription to account for the additional second $p(\vec{k})$ term in the right-hand side of Eq.~(\ref{eq:combine}). Since this contribution must not be present at scales $k \ll k_{\textrm{fs}}$, we simply add an additional power in the denominator,
\begin{equation}
\label{eq:centraleq}
\begin{split}
	\delta_{\nu}(\vec{k}) & \approx \frac{\left(\frac{k_{\textrm{fs}}}{k}\right)^2}{(1+\frac{k_{\textrm{fs}}}{k})^2} \delta(\vec{k})  + \gamma \frac{\left(\frac{k_{\textrm{fs}}}{k}\right)^2}{(1+\frac{k_{\textrm{fs}}}{k})^3} \frac{p(\vec{k})}{\sigma_{\nu}} \\ & = \frac{\left(\frac{k_{\textrm{fs}}}{k}\right)^2}{(1+\frac{k_{\textrm{fs}}}{k})^2} \left[\delta(\vec{k}) + \gamma \frac{p_{\textrm{rel}}(\vec{k})}{\sigma_{\nu}}\right] \,,
\end{split}
\end{equation}
where we introduce the relative (nonlinear) momentum,
\begin{equation}
	\label{eq:relmom}
 	p_{\textrm{rel}}(\vec{k})=\frac{p(\vec{k})}{1+\frac{k_{\textrm{fs}}}{k}} \,,
\end{equation}
which is just a high-pass filtered nonlinear contribution to the CDM momentum field with a cutoff at the free-streaming scale. In that sense the distortion in the neutrino density field is sensitive to the relative bulk flow between CDM and neutrinos and not just the CDM bulk flow, or in other words, we have to remember to impose the cutoff at the free-streaming scale as we did here (also see the discussion in Sec.~\ref{sec:b1a} for a physical interpretation). In the limit $k_{\textrm{fs}}/k_{\textrm{NL}} \ll 1$ any nonlinear physics happens on scales where Eq.~(\ref{eq:combine}) safely applies and hence our interpolation is not really needed. However, as discussed above we will consider applications with $m_\nu$ somewhat larger than the current cosmological bounds where $k_{\textrm{fs}}/k_{\textrm{NL}} \lesssim 2/3$ for which the interpolation can play an important role, i.e., to cutoff the distortion effect at the free-streaming scale. We then expect some inaccuracies in our modeling that nevertheless allow for an estimation of the size of the distortion effect in potential observables.

We can gain some intuition on Eq.~(\ref{eq:centraleq}) by considering what it looks like in case one simply introduces a constant (position independent) relative velocity between the neutrino and CDM fluids $\vec{v}_{\textrm{rel}} = \sigma_{\textrm{rel}} \hat{z}$ (say along the z direction); this is a good approximation at scales below the velocity coherence length (which is also at sub-free streaming scales) and is indeed the approach taken in previous works \cite{zhu2014measurement, zhu2016probing, zhu2020measuring, okoli2017dynamical}. In that case $p_{\textrm{rel}}(\vec{k}) = i(\hat{k} \cdot \hat{z})\sigma_{\textrm{rel}} \delta(\vec{k})$ and Eq.~(\ref{eq:combine}) reads
\begin{equation}
	\label{eq:constantve}
	\delta_{\nu}(\vec{k}) = \left(\frac{k_{\textrm{fs}}}{k}\right)^2\left[1 + i  \gamma (\hat{k} \cdot \hat{z}) \frac{\sigma_{\textrm{rel}}}{\sigma_{\nu}} \right]\delta(\vec{k}) \,.
\end{equation}
An imaginary term appears, which corresponds to a dipole distortion of the neutrino density field along the direction of the relative velocity. Furthermore, we are working in the regime $\sigma_{\textrm{rel}}/\sigma_{\nu} \ll 1$, where
\begin{equation}
\label{eq:displacement}
	e^{i\gamma(\hat{k}\cdot \hat{z}) \frac{\sigma_{\textrm{rel}}}{\sigma_{\nu}}} \approx 1 + i  \gamma (\hat{k} \cdot \hat{z}) \frac{\sigma_{\textrm{rel}}}{\sigma_{\nu}} \,,
\end{equation}
and hence we can absorb the distortion in a phase that corresponds to a shift in coordinates in position space, i.e., going to the neutrino rest frame. To briefly elaborate on this point, if the CDM velocity field is completely coherent, then one can study the problem in the rest frame of CDM, where the neutrino momentum distribution appears to have a dipole component. In this situation the Fourier-space momentum field $p(\vec{k})$ is simply proportional to the density $\delta(\vec{k})$, as opposed to related via a convolution over the velocity field. Our Equation \ref{eq:centraleq} is more general in that it accounts for the fact that the relative bulk flow is not coherent over arbitrarily large scales, so one cannot generally boost to a frame where all the CDM is at rest. This approach therefore allows for a more complete treatment of potential observational signatures which henceforth will be the focus of our attention. 

\subsection{Dynamical friction revisited}

We have investigated the solution to the Boltzmann equation for the neutrino distribution function in the background of a general nonlinear CDM distribution, and we showed how one can extract the distortion in the neutrino density field produced by the CDM bulk flow, building up to our central result in Eq.~(\ref{eq:centraleq}). Our next step is to estimate the impact of this on the CDM dynamics. The motivation for doing so is two fold: First, we want to connect the formalism described in the previous section with our 1-halo results, or more specifically Eq.~(\ref{eq:timescale}). This will enable us to derive the 2-halo term, and to confirm that the second term in the right-hand side of  Eq.~(\ref{eq:centraleq}), the nonlinear momentum term, is indeed producing the dynamical friction effect. Second, if the impact of dynamical friction on the CDM is found to be large, i.e., halos are slowing down significantly due to the anisotropic neutrino wakes then we may use that as an avenue for detection of the dynamical friction effect, through a halo velocity bias or direct measurements of galaxy peculiar velocities, e.g., via the  kinetic Sunyaev Zeldovich (kSZ) effect. Otherwise, we should focus on potential signatures of Eq.~(\ref{eq:centraleq}) on the large-scale structure as observables.

\begin{figure}
	\centering
	\includegraphics[width=0.75\textwidth]{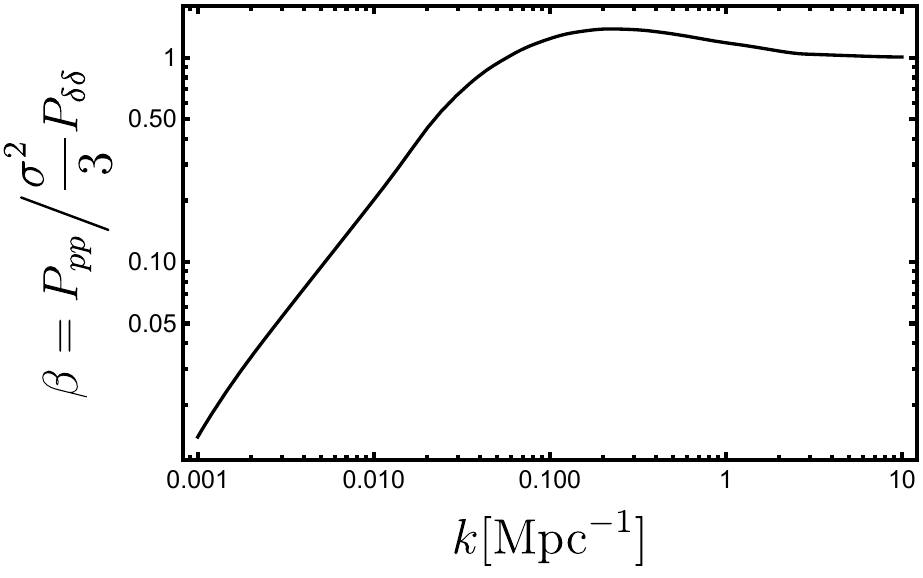}
	\caption{The function $\beta(k)$ as given by Eq. (\ref{eq:parametrization}), which captures the coherence of the CDM velocity field as a function of $k$.  The CDM bulk flow is coherent on scales $k \gg k_{\textrm{coh}}$, with velocity $v=\sigma/\sqrt{3}$. On the other hand, at larger scales this is no longer the case causing a suppression in the momentum power spectrum relative to $\sigma^2 P_{\delta\delta}/3$. From the plot above we read off $k_{\textrm{coh}} \sim 0.1$Mpc$^{-1}$ for our reference $\Lambda$CDM cosmology.}
	\label{fig:f2plot}
\end{figure}

The additional clustering of neutrinos as given by the second term in the right-hand side of Eq.~(\ref{eq:centraleq}) translates into a contribution to the large-scale gravitational field, using Poisson's equation,
\begin{equation}
	\label{eq:gf}
	\vec{g}(\vec{k}) = 4\pi G a \frac{i\hat{k}}{k} \delta \rho_{\nu}^{(1)}(\vec{k}) = \frac{4\pi m_{\nu}^4 G^2 \bar{\rho}_{0}}{(1+\frac{k_{\textrm{fs}}}{k})^3} p(\vec{k}) \frac{i\hat{k}}{k^3} \,,
\end{equation}
where $\delta \rho_{\nu}^{(1)}(\vec{k})= \bar{\rho}_{\nu} \delta^{(1)}_{\nu}(\vec{k})$ is the distortion in the neutrino density perturbation produced by the CDM bulk flow and we used Eqs.~(\ref{eq:nudispersion}), (\ref{eq:fstreaming}), (\ref{eq:nudensity}), (\ref{eq:gamma}) and (\ref{eq:centraleq}). Note that time dependences are implicit in our equations. We wish to compute the large-scale structure analog of the characteristic time scale in Eq.~(\ref{eq:timescale}). In this case the force per unit mass acting on the CDM structure is given by $\delta \vec{g}$, since the integral of the density times the gravitational field produces the overall force on the CDM within a given region of space. Additionally, due to the stochastic nature of cosmological fields we need to take the average such that we arrive at:
\begin{equation}
	\label{eq:dff2nd}
	\begin{split}
		\tau^{-1} &= - \frac{\langle \, \vec{v} \cdot \delta \ \vec{g} \, \rangle}{\langle v^2 \rangle} = -\frac{1}{\sigma^2}  \langle \, \vec{p} \cdot \vec{g} \, \rangle \\ &= \frac{2}{\pi} \frac{G^2 m_{\nu}^4}{\sigma^2} \bar{\rho}_{0} \int d\log k \, \frac{P_{pp}(k)}{(1+\frac{k_{\textrm{fs}}}{k})^3} \,,
	\end{split}
\end{equation}
where in the first line we used Eq.~(\ref{eq:cdmdispersion}) in combination with $\vec{p}=\delta \vec{v}$, and Eq.~(\ref{eq:gf}) to arrive at the second line.\footnote{We assume that the CDM nonlinear momentum can be written as the gradient of a potential. We do not expect vorticity to contribute at the large quasilinear scales of our interest $\sim k_{\textrm{NL}} \sim 0.1$Mpc$^{-1}$, e.g., see Fig.4 in \cite{jelic2018generation}.} In order to make progress in Eq.~(\ref{eq:dff2nd}) we need an expression for the power spectrum of the (nonlinear) momentum field $P_{pp}(k)$. This is what we discuss next, following the work of \cite{ma2002nonlinear}. We are interested in the divergence $p(\vec{k}) = i \hat{k} \cdot \vec{p}(\vec{k})$. In position space $\vec{p}=\delta \vec{v}$ so we need to evaluate the convolution to obtain,
\begin{equation}
	\label{eq:nmf}
	p(\vec{k}) =  \sigma \frac{k_{\textrm{NL}}}{k} \int \frac{d^3\vec{k}_{1}}{(2\pi)^3} \int \frac{d^3\vec{k}_{2}}{(2\pi)^3}(2\pi)^3 \delta^{(3)}(\vec{k}-\vec{k}_{12}) \alpha(\vec{k}_{1},\vec{k}_{2}) \delta(\vec{k}_{1}) \frac{\theta(\vec{k}_{2})}{aH} \,,
\end{equation}
where $\vec{k}_{12}= \vec{k}_{1} + \vec{k}_{2}$, $\alpha(\vec{k}_{1},\vec{k}_{2}) = (\vec{k}_{12} \cdot \vec{k}_{2})/k_{2}^{2}$, $\delta^{(3)}(\vec{k})$ stands for the Dirac delta function and for future convenience we have inserted a factor of $\sigma k_{\textrm{NL}}/aH = 1$. It is now straightforward to compute the two-point function. We find:
\begin{equation}
	\label{eq:qqpsdef}
	\langle p(\vec{k})p(\vec{k}') \rangle =(2\pi)^3 \delta^{(3)}(\vec{k}+\vec{k}') P_{pp}(k) \,,
\end{equation}
where,
\begin{equation}
	\label{eq:qqps}
		P_{pp}(k) = \sigma^2 \int  \frac{d^3\vec{k'}}{(2\pi)^3}  \left(\frac{k_{\textrm{NL}}}{k'}\right)^2  (\hat{k} \cdot  \hat{k}')\Bigg\{ (\hat{k} \cdot \hat{k'}) P_{\delta \delta}(|\vec{k}-\vec{k}'|) \frac{P_{\theta \theta}(k')}{(aH)^2}  + \frac{kk'\left[1-\frac{k'}{k} (\hat{k} \cdot \hat{k'})\right]}{|\vec{k} - \vec{k}'|^2}\frac{P_{\delta \theta}(|\vec{k} - \vec{k}'|)}{aH}\frac{P_{\delta \theta}(k')}{aH}  \Bigg\} \,.
\end{equation}
For wavelengths below the velocity coherence scale, $k \gg k_{\textrm{coh}}$, we can expand the integrand to leading order in $k'/k$ to obtain \cite{hu2000reionization}
\begin{equation}
	\label{eq:momentanl}
	P_{pp}(k) \underset{k \gg k_{\textrm{coh}}}{=} \sigma^2 P_{\delta \delta}(k) \int \frac{d^3\vec{k'}}{(2\pi)^3}  \left(\frac{k_{\textrm{NL}}}{k'}\right)^2  (\hat{k} \cdot \hat{k}')^2 \frac{P_{\theta \theta}(k')}{(aH)^2} = \frac{\sigma^2}{3} P_{\delta \delta}(k) \,,
\end{equation}
where we have used Eqs.~(\ref{eq:cdmdispersion}) and (\ref{eq:scalenl}). In words, the CDM bulk flow is coherent on scales $k \gg k_{\textrm{coh}}$, with velocity $v=\sigma/\sqrt{3}$. On the other hand, at larger scales this is no longer the case causing a suppression in the momentum power spectrum when compared to Eq.~(\ref{eq:momentanl}). This can be parameterized as follows: On large perturbative scales we can set $\theta(\vec{k})=faH\delta(\vec{k})$ into Eq.~(\ref{eq:qqps}), with $f= d\log D_{\textrm{L}}/d\log a$  the linear growth rate, and evaluate the ratio
\begin{equation}
	\label{eq:parametrization}
	\begin{split}
	\beta(k) &= \frac{P_{pp}(k)}{\frac{\sigma^2}{3} P_{\delta \delta}(k)}\Bigg|_{\textrm{linear theory}}   \\ & = 3 f^2 \int \frac{d^3\vec{k'}}{(2\pi)^3} \left(\frac{k_{\textrm{NL}}}{k'}\right)^2 (\hat{k} \cdot \hat{k'})\left\{ (\hat{k} \cdot \hat{k'}) +  \frac{kk'\left[1-\frac{k'}{k} (\hat{k} \cdot \hat{k'})\right]}{|\vec{k} - \vec{k'}|^2} \right\} \frac{P^{\textrm{L}}_{\delta \delta}(|\vec{k} - \vec{k'}|) P^{\textrm{L}}_{\delta \delta}(k')}{P^{\textrm{L}}_{\delta \delta}(k)} \,,
	\end{split}
\end{equation}
with $P^{\textrm{L}}_{\delta \delta}(k)$ the linear power spectrum. Note that $\beta(k) = 1$ for $k\gg k_{\textrm{coh}}$ as we have demonstrated above, and this can be used as a definition of the velocity coherence scale. From $P^{\textrm{L}}_{\delta \delta} \sim D_{\textrm{L}}^2(a)$, Eqs.~(\ref{eq:cdmdispersion}) and  (\ref{eq:scalenl}) it follows that $\beta(k)$ is independent of redshift, and hence so is the velocity coherence scale. This function is plotted in Fig. \ref{fig:f2plot} for our reference $\Lambda$CDM cosmology from where we can read off $k_{\textrm{coh}} \sim 0.1$Mpc$^{-1}$. At scales above the velocity coherence scale, $k<k_{\textrm{coh}}$, the momentum field, and hence the distortion effect, is suppressed. Our model for the momentum power spectrum at all scales now reads,
\begin{equation}
	\label{eq:ppmodel}
	P_{pp}(k) = \frac{\sigma^2}{3} \beta(k) P_{\delta \delta}(k) \,,
\end{equation}
with $P_{\delta \delta}(k)$ the nonlinear power spectrum. Substituting Eq.~(\ref{eq:ppmodel}) into Eq.~(\ref{eq:dff2nd}) yields
\begin{equation}
	\label{eq:invtau}
	\tau^{-1}(a) = \frac{2}{3\pi} G^2 m_{\nu}^4 \bar{\rho}_{0} \int d\log k \, \frac{\beta(k)}{\left[1+\frac{k_{\textrm{fs}}(a)}{k}\right]^3} P_{\delta \delta}(a,k) \,.
\end{equation}
Note that for clarity we insert the time dependences in this final formula, which are the ones implicit through $P_{\delta \delta}(a,k)$ and $k_{\textrm{fs}}(a)$. This can be readily evaluated with a nonlinear power spectrum  for the neutrino masses and redshifts of interest. We use the Boltzmann code CLASS to produce the nonlinear power spectrum with the HMcode prescription \cite{mead2021hmcode}. It turns out that the integral in Eq.~(\ref{eq:invtau}) is dominated by its contribution from large quasilinear scales where the nonlinear corrections are small, and in fact we explicitly verified that simply using the linear power spectrum produces the same numerical results.\footnote{In the next section we will use the halo model to interpret this result as the statement that the 2-halo term dominates over the 1-halo term on average.} We then substitute $P_{\delta \delta}(a,k) = [D_{\textrm{L}}(a)/D_{\textrm{L}}(a=1)]^2P^{\textrm{L}}_{\delta \delta}(a=1,k)$ into Eq.~(\ref{eq:invtau}) to obtain (using the approximate formula in \cite{hamilton2001formulae} for the linear growth factor),
\begin{equation}
	\label{eq:powerlaw}
	(\tau H)^{-1} \approx \sum_{m_{\nu}} 2.2 \times 10^{-4} \left(\frac{m_{\nu}}{0.1\textrm{eV}}\right)^{2.8}(1+z)^{-1.83} \,,
\end{equation}
where we assume our reference $\Lambda$CDM cosmology and add a sum over neutrino mass eigenstates. The reader is reminded that dynamical friction only occurs for neutrino masses small enough that the free-streaming scale is larger than the non-linear scale (roughly $m_\nu \lesssim 0.2$eV). The power law approximation in Eq.~(\ref{eq:powerlaw}) reproduces the numerical results of Eq.~(\ref{eq:invtau}) with errors that can be as large as tens of percents, for neutrino masses in the range $0.05$eV $\leq m_{\nu} \leq 0.15$eV, and redshifts $z \leq 2$. Note that $\Delta v/v \sim 1/\tau H$ is the net average fractional decrease in the velocity of halos due to the dynamical friction over the age of the universe $\sim 1/H$. 

Increasing the neutrino mass causes a dramatic increase in the effect, but there is a significant deviation from the naive scaling from $\sim m_{\nu}^4$ to $\sim m_{\nu}^{2.8}$. Recall that our framework assumes $k_{\textrm{fs}}/ k_{\textrm{NL}} \ll 1$, while $k_{\textrm{fs}}/ k_{\textrm{NL}} \sim 2/3$ already for $m_{\nu} = 0.15$eV. At that point the suppression at the free-streaming scale is playing an important role, with the effect vanishing in the limit $k_{\textrm{fs}}/ k_{\textrm{NL}} \gg 1$. This is relevant since a scenario where $k_{\textrm{fs}}/ k_{\textrm{NL}} > 1$ has been ruled out by cosmological observations as we discussed previously, but laboratory experiments based on measurements of beta decay only set an upper bound to a weighted sum of the masses $m_{\nu, \beta} < 0.8$eV \cite{aker2022katrin}. Furthermore, nonstandard scenarios such as dynamical dark energy and unstable neutrino species allow for less stringent bounds on the neutrino mass scale from cosmological probes \cite{chacko2020cosmological, escudero2020relaxing, joudaki2013constraints, lorenz2017distinguishing, hannestad2005neutrino}. Finally, at higher redshift the effect is smaller due to both linear growth of structure and the smaller age of the universe $\sim 1/H$.

The overall amplitude in Eq.~(\ref{eq:powerlaw}) is quite small, indicating that the decrease in the velocity of halos due to the dynamical friction effect produced by anisotropic neutrino wakes are unlikely to be detected in the foreseeable future. That includes measurements of galaxy peculiar velocities via redshift space distortions, the kSZ effect, or a large-scale velocity bias that deviates from unity. In the absence of massive neutrinos, a large scale velocity bias of $b_{v} \equiv 1$ follows from the equivalence principle \cite{chen2018accurate} (and any initial nonzero $|b_{v}-1|$ decays away \cite{hui2008evolution}). In \cite{villaescusa2018imprint} a $\gtrsim 0.5\%$ deviation from unity in the large scale velocity bias was not found in simulations with massive neutrinos, which is consistent with our results. Previous work has found a more pronounced dynamical friction effect \cite{okoli2017dynamical}, but there the cutoffs at the velocity coherence and free-streaming scales were not consistently accounted for. 

We are still left with the task of connecting Eq.~(\ref{eq:invtau}), our result for the velocity decay time, to the results of Sec.~\ref{sec:1h}, specifically our $1-$halo calculation of that quantity in Eq.~(\ref{eq:timescale}). We will postpone this discussion to Sec.~\ref{sec:calc} where we carry out numerical calculations using the halo model. There we will see that the 1-halo term will naturally reemerge, but now together with a 2-halo term accounting for the clustering of nearby halos.

\subsection{Large-scale structure observables}

We calculated the average fractional decrease in halo velocities due to dynamical friction and found that it is too small to allow for a detection of the anisotropic neutrino wakes. We then take the alternative route of searching for potential observational signatures in the large scale structure. For instance, we can use Eq.~(\ref{eq:centraleq}) to evaluate the CDM-$\nu$ cross power spectrum,
\begin{equation}
\label{eq:crosspower}
	P_{\delta_{\nu} \delta}(k) = \frac{\left(\frac{k_{\textrm{fs}}}{k}\right)^2}{(1+\frac{k_{\textrm{fs}}}{k})^2} \left[ P_{\delta \delta}(k) + \frac{ \gamma}{1+\frac{k_{\textrm{fs}}}{k}} \frac{P_{p\delta}(k)}{\sigma_{\nu}} \right] \,.
\end{equation}
The additional second term in the right-hand side of Eq.~({\ref{eq:crosspower}) is the contribution from the distortion effect that we expect to be small, due to its scaling with $\sim k_{\textrm{fs}}/k_{\textrm{NL}}$, and negative since the relative bulk flow between neutrinos and CDM should reduce their cross correlation. We postpone some numerical calculations, using standard perturbartion theory (SPT) as our model for the nonlinear dynamics of CDM, to Sec.~\ref{sec:calc}. For now we simply point out that the signature of the distortion effect in the two-point function is small and very degenerate with standard nonlinear structure formation. 
	
Instead, we will argue that three-point cross-correlations allow for a separation of the distortion effect from standard gravitational nonlinearities, and that the resulting signal can be potentially measured in upcoming large scale structure surveys. We can use Eq.~(\ref{eq:centraleq}) to compute the neutrino-CDM-CDM bispectrum:
\begin{equation}
	\label{eq:3ptc}
	\langle \delta_{\nu}(\vec{k}_1) \delta(\vec{k}_2) \delta(\vec{k}_3) \rangle = (2\pi)^3 \delta^{(3)}(\vec{k}_{1}+\vec{k}_{2}+\vec{k}_{3}) 	B(k_1,k_2,k_3) \,,
\end{equation}
where\footnote{Note that these are only functions of the magnitude of the wave vectors since the angles between them are constrained by the requirement that they fit into a triangle configuration, i.e., $\vec{k}_1+\vec{k}_2+\vec{k}_3=0$. For example, $\hat{k}_1\cdot \hat{k}_2 = (k_3^2 - k_1^2-k_2^2)/2k_1k_2$.}
\begin{equation}
	\label{eq:bispectrum}
		B(k_1,k_2,k_3) = \frac{\left(\frac{k_{\textrm{fs}}}{k_1}\right)^2}{(1+\frac{k_{\textrm{fs}}}{k_1})^2} \left[ \langle\delta(\vec{k}_1) \delta(\vec{k}_2) \delta(\vec{k}_3) \rangle' + \frac{ \gamma}{1+\frac{k_{\textrm{fs}}}{k_1}} \langle \frac{p(\vec{k}_1)}{\sigma_{\nu}} \delta(\vec{k}_2) \delta(\vec{k}_3) \rangle' \right] \,,
\end{equation}
is written in terms of CDM-only bispectra.\footnote{A prime in correlation functions indicates that the ever present factor of $(2\pi)^3\delta^{(3)}(\sum \vec{k}_{i})$ is to be omitted.} Note that the first term inside the brackets in the right-hand side of Eq.~(\ref{eq:bispectrum}) is symmetric under the exchange $\vec{k}_i \leftrightarrow \vec{k}_j$ for any pair of wavevectors, while the second term is only symmetric under $\vec{k}_2 \leftrightarrow \vec{k}_3$. Because of this it is straightforward to obtain a smoking gun of the distortion effect: Take the antisymmetric combination of Eq.~(\ref{eq:bispectrum}) under $\vec{k}_1 \leftrightarrow \vec{k}_3$ and consider the squeezed limit where $k_1  \approx k_3 \gg k_2$. We postpone explicit calculations using SPT as our model for the nonlinear dynamics of CDM to Section \ref{sec:calc}, and for now we provide some intuition on this observable. In the presence of a long-wavelength velocity potential $\phi(\vec{K})$, the (local) $\nu$-CDM cross correlation picks up a dipole,
\begin{equation}
	\label{eq:dipoleagainpre}
	\langle \delta_{\nu}(\vec{k}_1) \delta(\vec{k}_2)\rangle_{\phi(\vec{K})} \sim \frac{\vec{k}_1\cdot \vec{K}}{k_1}  P_{\delta \delta_{\nu}}^{\textrm{MON}}(k_1) \phi(\vec{K}) \,,
\end{equation}
due to neutrinos clustering behind the moving CDM, where $\vec{K}=\vec{k}_1+\vec{k}_2$.\footnote{Indeed, this is the signature obtained in previous studies that investigated this effect using simplified models at scales below the velocity coherence scale, in two different but related languages: Dynamical friction \cite{okoli2017dynamical} and large-scale relative flow \cite{zhu2014measurement}.} This is qualitatively similar to the picture in Fig.~\ref{fig:2dplot} for the clustering of neutrinos around a moving point mass halo. Due to isotropy of the background, the CDM bulk velocity points in different directions in disparate regions of space so that the dipole averages out to zero, and the net effect in the two-point function is a small suppression in cross power as we pointed out below Eq.~(\ref{eq:crosspower}). On the other hand, the presence of a long-wavelength mode $\vec{K}$ provides a preferred direction which allows the signal to be extracted from three-point correlations. 

Additionally, note from Eq.~(\ref{eq:dipoleagainpre}) that the dipole is proportional to the $\nu$-CDM monopole which is related to the CDM power spectrum by a known scale-dependent transfer function. This implies that a measurement of the distortion effect is not limited by cosmic variance since one can reconstruct the potential $\phi(\vec{K})$ from measurements of both the monopole and dipole, regardless of the specific realization of the small scale CDM power. In the next section we will see that a multi-tracer approach involving measurements of both the matter field and galaxies can be applied to extract this signal since galaxies only trace the CDM, while the matter field traces neutrinos as well. This benefits from cosmic variance cancellation, which makes it possible for this faint signal of neutrino masses to be potentially detectable in upcoming surveys.

\section{Numerical calculations of large-scale structure observables}
\label{sec:calc}

In the previous section we extracted the distortion in the neutrino density field, due to the peculiar motion of halos, from the Boltzmann Eq.~(\ref{eq:beqbkt}). Our approach assumes a rather general CDM distribution on large scales while remaining agnostic about the details of the nonlinear gravitational evolution. We were able to determine the size of the dynamical friction effect and potential observational signatures on the large scale structure. However, it will prove useful to assume specific models for the nonlinear structure formation in order to carry out numerical calculations of interest. Specifically, in Sec. \ref{sec:2ht} we use the halo model to derive how halos of different masses contribute to the average in Eqs.~(\ref{eq:dff2nd}) and (\ref{eq:invtau}). This will enable us to reproduce the results of Sec.~\ref{sec:1h} [specifically Eq.~(\ref{eq:timescale})] but also to derive the 2-halo term. In Sec. \ref{ssec:twopointcc} and Sec. \ref{ssec:threepointcc},  we will apply the standard perturbation theory (SPT) to compute the power spectra and bispectra in Eq.~(\ref{eq:crosspower}) and (\ref{eq:bispectrum}) to study the observational signatures of the distortion effect in the large scale structure, and use it to forecast its observability in future surveys.

\subsection{The 2-halo contribution to dynamical friction}
\label{sec:2ht}

We start with the halo model \cite{cooray2002halo}, which asserts that cold dark matter halos are the basic building blocks of CDM structure. We assume the NFW profile for the halos as suggested by numerical simulations  \cite{navarro1996structure, navarro1997universal}, and we use the expression in \cite{dutton2014cold} for the halo concentration [see Eqs.~(8), (10) and (11) in this reference] which is calibrated from simulations. We adopt the convention that halos are defined as spherical regions with an average density of $\Delta=200$ times the critical density. In order to write down the halo model expression for the power spectrum, we need two additional ingredients: the halo mass function $dn/dM$ that gives the (comoving) number density of halos in a given halo mass window $dM$, and the linear halo bias $b(M)$ that fixes the correlation between nearby halos in terms of matter two-point correlations. As we will see, we will not need their explicit expressions as it will suffice to enforce the usual requirement that the mean bias of halos is unity, that is,
\begin{equation}
	\label{eq:unitymeanbias}
	\frac{1}{\bar{\rho}_{0}}\int dM \frac{dn}{dM} M b(M) = 1 \,,
\end{equation}
where $\bar{\rho}_{0}$ is the comoving matter density (or matter density evaluated at $z=0$) as in previous sections. According to the halo model:
\begin{equation}
	\label{eq:ps}
	\begin{split}
		& P_{\delta \delta}(k)  = \int dM  \frac{dn}{dM} \left(\frac{M}{\bar{\rho}_{0}}\right)^2 |u(k|M)|^2 + \\ & + \int dM \frac{dn}{dM} \int dM' \frac{dn}{dM'} \frac{MM'}{\bar{\rho}_{0}^2} u(k|M) u(k|M') b(M) b(M') P_{\delta \delta}^{L}(k) \,,
	\end{split}
\end{equation}
where $u(k|M)$ is the Fourier transform of the normalized NFW profile, $P_{\delta \delta}^{L}(k)$ is the linear theory power spectrum and time dependences are implicit. Now substituting Eq.~(\ref{eq:ps}) into Eq.~(\ref{eq:invtau}) gives
\begin{equation}
	\label{eq:dfhm1}
	\begin{split}
		\tau^{-1} =    \frac{1}{\bar{\rho}_{0} } & \int dM  \frac{dn}{dM} M \Bigg\{ \frac{2}{3\pi} G^2 m_{\nu}^4 M \int d\log k \, \frac{g(k)}{\left(1+\frac{k_{\textrm{fs}}}{k}\right)^3} \Bigg[|u(k|M)|^2 + \\& + \int dM' \frac{dn}{dM'} \frac{M'}{M} u(k|M)u(k|M') b(M) b(M') P_{\delta \delta}^L(k) \Bigg] \Bigg\} \,.
	\end{split}
\end{equation}
This can be interpreted as an average over halos, from which one can read off the expression for the inverse characteristic velocity decay time for a halo of mass $M$: 
\begin{equation}
	\label{eq:dfhm2}
	\tau^{-1}(M) = \frac{2}{3\pi} G^2 m_{\nu}^4 M \int d\log k \, \frac{g(k)}{\left(1+\frac{k_{\textrm{fs}}}{k}\right)^3} \Bigg[|u(k|M)|^2  + \int dM' \frac{dn}{dM'} \frac{M'}{M} u(k|M)u(k|M') b(M) b(M') P_{\delta \delta}^L(k) \Bigg] \,.
\end{equation}
A comparison of Eq.~(\ref{eq:dfhm2}) with Eq.~(\ref{eq:timescale}), when ignoring the second term in the right-hand side of Eq.~(\ref{eq:dfhm2}) which is the 2-halo term, suggests the identification
\begin{equation}
	\label{eq:lambda}
	\log \Lambda(M) = \int d\log k \, \frac{g(k)}{\left(1+\frac{k_{\textrm{fs}}}{k}\right)^3} |u(k|M)|^2 \,.
\end{equation}
Let us stop for a moment to compare this expression with the estimate made in Sec. \ref{sec:1h}, i.e., $\Lambda \sim \lambda_{\textrm{coh}}/R$ with $ \lambda_{\textrm{coh}}$ the CDM velocity coherence scale and $R$ the halo radius. The normalized NFW profile is such that $u(k|M) \approx 1$ for $k\lesssim 1/R$ and drops to zero at smaller scales (see Fig.9 in \cite{cooray2002halo}). Based on Fig.~\ref{fig:f2plot}, it is also true that $g(k)/(1+k_{\textrm{fs}}/k)^3 \approx 1$ at scales $k\gtrsim k_{\textrm{coh}}$ and drops to zero at larger scales (assuming $k_{\textrm{coh}} \gg k_{\textrm{fs}}$, which is approximately true for the neutrino masses we consider). We then conclude that the integrand in Eq.~(\ref{eq:lambda}) is approximately a step function that assumes the value unity between the scales $k_{\textrm{coh}}$ and $1/R$ and drops to zero outside this range, when assuming the hierarchy of scales $k_{\textrm{fs}} \ll k_{\textrm{coh}}\ll 1/R$. In that case,
\begin{equation}
	\label{eq:estimate}
	\log \Lambda(M) = \int_{k_{\textrm{coh}}}^{1/R} \frac{dk}{k} \sim \log (1/k_{\textrm{coh}}R) \implies \Lambda \sim \lambda_{\textrm{coh}}/R \,,
\end{equation}
which is in agreement with our previous estimate. More generally Eq.~(\ref{eq:lambda}) can be well fit by a power law:
\begin{equation}
	\label{eq:plfit1}
	\Lambda(M) \approx K_{\Lambda} \left(\frac{10^{13}M_{\bigodot}}{M}\right)^{\chi} \,,
\end{equation}
with $K_{\Lambda} = \Lambda(10^{13}M_{\bigodot})$ a numerical coefficient that depends on the neutrino mass and redshift, and we find $\chi \approx 0.35$ to be robust against changes in the neutrino mass and redshift for the values we consider.  We present numerical values for the coefficient $K_{\Lambda}$ in Tab.~\ref{table:fitlambda}. The free-streaming length $2\pi/k_{\textrm{fs}}$, which is smaller for higher values of the neutrino mass and lower redshifts as can be seen in Fig.~\ref{fig:scales}, acts as a cutoff to the integral in Eq.~(\ref{eq:lambda}). This (mostly) sets the dependence of $K_{\Lambda}$ on the neutrino mass and redshift. 
\begin{table}[]
	\centering
	\begin{tabular}{|c|c|c|c|}
		\hline
		$K_{\Lambda}=\Lambda(10^{13}M_{\bigodot})$ & $m_{\nu}=0.05$eV & $m_{\nu}=0.1$eV & $m_{\nu}=0.15$eV \\ \hline
		$z=0$ & 60  & 28  & 18 \\ \hline
		$z=1$ & 79  & 38 & 24 \\ \hline
	\end{tabular}
	\caption{Numerical values for $K_{\Lambda}=\Lambda(10^{13}M_{\bigodot})$ involved in the power-law fit of Eq.~(\ref{eq:plfit1}) to the 1-halo cutoff $\Lambda$. }
	\label{table:fitlambda}
\end{table} 

After having carefully investigated the 1-halo term, we now proceed to study the contribution to Eq.~(\ref{eq:dfhm2}) from the clustering of nearby halos which can be parameterized by 
\begin{equation}
	\label{eq:2hterm}
	\begin{split}
		\Sigma(M) &= \frac{1}{\log \Lambda(M)} \int d\log k \, \frac{g(k)}{\left(1+\frac{k_{\textrm{fs}}}{k}\right)^3} \int dM' \frac{dn}{dM'} \frac{M'}{M} u(k|M)u(k|M') b(M') P_{\delta \delta}^L(k) \\ & = \frac{1}{\log \Lambda(M)} \int d\log k \, I(k|M) \Gamma(k) \,,
	\end{split}
\end{equation}
where:
\begin{equation}
	\label{eq:auxfunc2}
	\begin{split}
		& I(k|M) = \frac{\bar{\rho}_{0}}{M} \frac{g(k)}{\left(1+\frac{k_{\textrm{fs}}}{k}\right)^3} u(k|M) P^L_{\delta \delta}(k) \,, \\ & \Gamma(k) = \frac{1}{\bar{\rho}_{0}} \int dM \frac{dn}{dM} M b(M) u(k|M) \,, 
	\end{split}
\end{equation}
are dimensionless functions. Since $u(k|M) \approx 1$ at scales $k\lesssim 1/R$ while dropping to zero as $k\gg 1/R$, we conclude that $\Gamma(k)$ is unity at sufficiently large scales and drops to zero at small scales, using Eq.~(\ref{eq:unitymeanbias}). The transition between the two regimes happen at the scale $k_{\textrm{tr}} \sim 2\pi/R_{\textrm{peak}}$, where $R_{\textrm{peak}}$ is the radius of the halo whose mass $M_{\textrm{peak}}$ gives the largest contribution to the integral that defines the function $\Gamma(k)$, which is in general dominated by larger halos so we choose $M_{\textrm{peak}} \sim 10^{15}M_{\bigodot}$ with the associated $k_{\textrm{tr}} \sim 2$Mpc$^{-1}$. On the other hand, the function $I(k|M)$ peaks at scales much larger than this transition scale and hence it suffices to set $\Gamma(k) \approx 1$ in Eq.~(\ref{eq:2hterm}).\footnote{We verified this explicitly with the halo mass function from \cite{tinker2008toward}, and halo bias from \cite{tinker2010large}. Note that this approximation is analogous to setting $P_{\delta \delta}^{\textrm{2h}}(k) = P_{\delta \delta}^{L}(k)$ in the halo model and is made in, e.g., the halo reaction approach for the nonlinear power spectrum \cite{cataneo2019road} in beyond $\Lambda$CDM cosmologies.} This greatly simplifies the calculation (note that it is unnecessary to introduce the halo mass function or halo bias), and it shows that dynamical friction is mostly insensitive to the specifics of the halo model parameters. We arrive at
\begin{equation}
	\label{eq:2hterm2}
	\begin{split}
		\Sigma(M) &= \frac{1}{\log \Lambda(M)} \frac{\bar{\rho}_{0}}{M} \int d\log k \,  \frac{g(k)}{\left(1+\frac{k_{\textrm{fs}}}{k}\right)^3} u(k|M)P^{L}_{\delta \delta}(k)\,,  \\ & \approx \frac{1}{\log \Lambda(M)} \frac{\bar{\rho}_{0}}{M} \int d\log k \, \frac{g(k)}{\left(1+\frac{k_{\textrm{fs}}}{k}\right)^3} P^{L}_{\delta \delta}(k) \implies \Sigma(M) \approx \frac{K_{\Sigma}}{\log \Lambda(M)} \frac{10^{13}M_{\bigodot}}{M} \,,
	\end{split}
\end{equation}
where,
\begin{equation}
	\label{eq:fit2}
	K_{\Sigma} = \frac{\bar{\rho}_{0}}{10^{13}M_{\bigodot}} \int d\log k \, \frac{g(k)}{\left(1+\frac{k_{\textrm{fs}}}{k}\right)^3} P^{L}_{\delta \delta}(k) \,,
\end{equation}
is a dimensionless number, and in order to go from the first to second line in Eq.~(\ref{eq:2hterm2}) we used the fact that $u(k|M) \approx 1$ at scales that contribute to the integral for the range of halo masses we are considering, i.e., $(10^{9}-10^{16})M_{\bigodot}$. In Tab.~\ref{table:fitsigma} we present numerical values for the coefficient $K_{\Sigma}$. It decreases with neutrino mass due to the cutoff at the free-streaming scale, and it decreases with redshift due to the linear growth of matter power.  
\begin{table}[]
	\centering
	\begin{tabular}{|c|c|c|c|}
		\hline
		$K_{\Sigma}$ & $m_{\nu}=0.05$eV & $m_{\nu}=0.1$eV & $m_{\nu}=0.15$eV \\ \hline
		$z=0$ & 60  & 28  & 17 \\ \hline
		$z=1$ & 30  & 15 & 10  \\ \hline
	\end{tabular}
	\caption{Numerical values for the coefficient $K_{\Sigma}$ defined in Eq.~(\ref{eq:fit2}), involved in the contribution to the dynamical friction effect from the clustering of halos.}
	\label{table:fitsigma}
\end{table}

We can finally substitute Eqs.~(\ref{eq:lambda})-(\ref{eq:fit2}) into Eq.~(\ref{eq:dfhm2}) to obtain
\begin{equation}
	\label{eq:invtauhm}
	\tau^{-1}(M) = \frac{2}{3\pi} \log \Lambda(M) G^2 M m_{\nu}^4  [1+b(M)\Sigma(M)] \,.
\end{equation} 
The function $\Sigma(M)$ gives the ratio of two to one halo contributions to the inverse characteristic time, up to the halo bias $b(M)$. From Eqs.~(\ref{eq:plfit1}), (\ref{eq:2hterm2}) and the numerical values in Tables \ref{table:fitlambda} and \ref{table:fitsigma}, we conclude that the 2-halo term dominates for halo masses $M \lesssim 10^{14} M_{\odot}$ while the 1-halo term dominates for halo masses $M \gtrsim 10^{14} M_{\odot}$ [when assuming $b(M)\approx1$]. 

The inverse characteristic time in Eq.~(\ref{eq:invtauhm}) is plotted in Fig.~\ref{fig:invtau} under the assumption that $b(M)=1$ for simplicity. It asymptotically approaches a constant for small halo masses, as determined by the numerical values in Eqs.~(\ref{eq:invtau}) and (\ref{eq:powerlaw}). This follows from a comparison between Eqs.~(\ref{eq:2hterm2}-(\ref{eq:invtauhm}) and (\ref{eq:invtau}), in the regime where the 2-halo term dominates. This result can be interpreted as follows:  Halos of sufficiently low mass can be thought of as surfing on the neutrino wakes produced by larger halos. On the other extreme of Fig.~\ref{fig:invtau}, the dynamical friction experienced by larger halos is due to their own (strong) wakes leading to a more pronounced effect. This, in combination with the fact that the halo bias $b(M)$ increases at the high mass end (and can reach values as high as  $\sim 10$) indicates that neutrino wakes can lead to a percent-level decrease in halo velocities.\footnote{However, such high mass halos are rare and hence have limited statistics.}

\begin{figure}
	\centering	\includegraphics[width=0.75\textwidth]{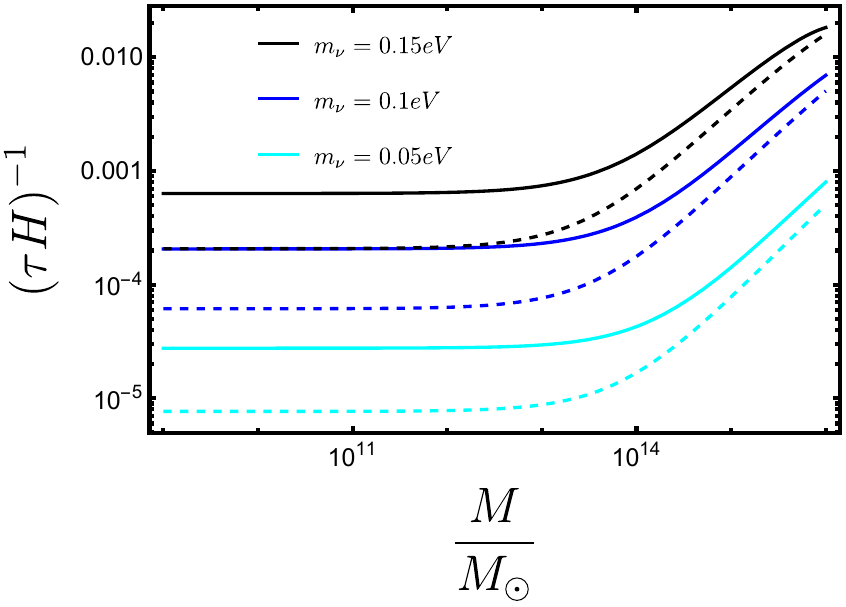}
	\caption{The inverse characteristic velocity decay time, due to neutrino dynamical friction, in units of the Hubble expansion rate, as given by Eq.~(\ref{eq:invtauhm}). This also corresponds to the fractional decrease in the velocity of halos due to the dynamical friction over the age of the universe. The individual neutrino mass varies from $m_{\nu}=0.15$eV to $m_{\nu}=0.05$eV from top to bottom. Solid curves correspond to $z=0$, and dashed curves to $z=1$.}
	\label{fig:invtau}
\end{figure} 

We have accomplished our goal of comparing our simplified framework in Sec.~\ref{sec:1h}, culminating on Eq.~(\ref{eq:timescale}), with the more general approach of Sec.~\ref{sec:dfolss} leading to Eq.~(\ref{eq:invtauhm}), that solves both issues with the Chandrasekhar dynamical friction formula as previously pointed out: arbitrariness in the choice of a cutoff $\Lambda$, and lack of halo clustering effects.

\subsection{Distortions to the neutrino-CDM cross power spectrum}
\label{ssec:twopointcc}
In the previous section we have explicitly checked that the distortion in the neutrino density field due to the peculiar motion of halos, as given by the second term in the right-hand side of Eq.~(\ref{eq:centraleq}), is connected to the dynamical friction effect. We are now finally read to apply SPT to explore its potential signatures on the large scale structure, offering new opportunities for a cosmological measurement of the neutrino masses.

We start with the CDM-$\nu$ two-point cross-correlation, or the cross-power spectrum in Fourier space. We repeat Eq.~(\ref{eq:crosspower}) for convenience,
\begin{equation}
	\label{eq:crosspoweragain}
	P_{\delta_{\nu} \delta}(k) = \frac{\left(\frac{k_{\textrm{fs}}}{k}\right)^2}{(1+\frac{k_{\textrm{fs}}}{k})^2} \left[ P_{\delta \delta}(k) + \frac{ \gamma}{1+\frac{k_{\textrm{fs}}}{k}} \frac{P_{p\delta}(k)}{\sigma_{\nu}} \right] \,,
\end{equation}
and Eq.~(\ref{eq:nmf}) for the nonlinear contribution to the CDM momentum field, 
\begin{equation}
	\label{eq:nmfagain}
	p(\vec{k}) =  \sigma \frac{k_{\textrm{NL}}}{k} \int \frac{d^3\vec{k}_{1}}{(2\pi)^3} \int \frac{d^3\vec{k}_{2}}{(2\pi)^3}(2\pi)^3 \delta^{(3)}(\vec{k}-\vec{k}_{12}) \alpha(\vec{k}_{1},\vec{k}_{2}) \delta(\vec{k}_{1}) \frac{\theta(\vec{k}_{2})}{aH} \,.
\end{equation}
To leading order in perturbation theory \cite{bernardeau2002large},
\begin{equation}
	\label{eq:ptlss}
	\begin{split}
		& \delta (\vec{k}) = \delta_{L} (\vec{k}) + \int \frac{d^3\vec{k}_{1}}{(2\pi)^3} \int \frac{d^3\vec{k}_{2}}{(2\pi)^3}(2\pi)^3 \delta^{(3)}(\vec{k}-\vec{k}_{12}) F_{2}(\vec{k}_{1},\vec{k}_{2}) \delta_{L} (\vec{k}_{1}) \delta_{L} (\vec{k}_{2}) \,, \\ & \theta (\vec{k}) = -faH\left[\delta_{L} (\vec{k}) + \int \frac{d^3\vec{k}_{1}}{(2\pi)^3} \int \frac{d^3\vec{k}_{2}}{(2\pi)^3}(2\pi)^3 \delta^{(3)}(\vec{k}-\vec{k}_{12}) G_{2}(\vec{k}_{1},\vec{k}_{2}) \delta_{L} (\vec{k}_{1}) \delta_{L} (\vec{k}_{2}) \right] \,,
	\end{split}
\end{equation}
with $\delta_{L} (\vec{k})$ the CDM linear fluctuation field, and perturbation theory kernels
\begin{equation}
	\label{eq:ptlsscoe}
	\begin{split}
		& F_{2}(\vec{k}_{1},\vec{k}_{2}) = \frac{5}{7} + \frac{1}{2} \frac{\vec{k}_{1} \cdot \vec{k}_{2}}{k_1 k_2} \left(\frac{k_1}{k_2}+\frac{k_2}{k_1}\right) + \frac{2}{7} \frac{(\vec{k}_{1} \cdot \vec{k}_{2})^2}{k_1^2 k_2^2} \,, \\ & G_{2}(\vec{k}_{1},\vec{k}_{2}) = \frac{3}{7} + \frac{1}{2} \frac{\vec{k}_{1} \cdot \vec{k}_{2}}{k_1 k_2} \left(\frac{k_1}{k_2}+\frac{k_2}{k_1}\right) + \frac{4}{7} \frac{(\vec{k}_{1} \cdot \vec{k}_{2})^2}{k_1^2 k_2^2} \,.
	\end{split}
\end{equation}
Combining Eqs.~(\ref{eq:crosspoweragain})-(\ref{eq:ptlsscoe}), we arrive at the one-loop expression to the  CDM-$\nu$ cross-power spectrum\footnote{The one-loop power spectrum requires a next to leading order calculation in perturbation theory, i.e., we will also need the kernel $F_{3}(\vec{k_{1}},\vec{k_{2}},\vec{k_{3}})$. An expression for it can be found in \cite{bernardeau2002large}.}
\begin{equation}
	\label{eq:crossps}
	P_{\delta \delta_{\nu}}^{\textrm{1-loop}}(k) = P_{\delta \delta_{\nu}}^{L}(k)\left[1 + \frac{\Delta P_{\delta \delta_{\nu}}^{\textrm{1-loop}}(k) }{P_{\delta \delta_{\nu}}^{L}(k)}\Bigg|_{\textrm{SPT}} +  \frac{\Delta P_{\delta \delta_{\nu}}^{\textrm{1-loop}}(k) }{P_{\delta \delta_{\nu}}^{L}(k)}\Bigg|_{\textrm{Dist}}\right] \,,
\end{equation}
with,
\begin{equation}
\label{eq:linearcross}
	P_{\delta \delta_{\nu}}^{L}(k) = \frac{\left(\frac{k_{\textrm{fs}}}{k}\right)^2}{(1+\frac{k_{\textrm{fs}}}{k})^2} P_{\delta \delta}^{L}(k) \,,
\end{equation}
the linear theory cross power spectrum and
\begin{equation}
	\label{eq:lambdas}
	\begin{split}
		&\frac{\Delta P_{\delta \delta_{\nu}}^{\textrm{1-loop}}(k) }{P_{\delta \delta_{\nu}}^{L}(k)}\Bigg|_{\textrm{SPT}} = 2 \int \frac{d^3 \vec{k}'}{(2\pi)^3} \left\{ \left[F_{2}(\vec{k}',\vec{k}-\vec{k}')\right]^2 \frac{P^L_{\delta \delta}(|\vec{k}-\vec{k}'|)}{P^L_{\delta \delta}(k)} + 3 F_{3}(\vec{k},\vec{k}',-\vec{k}')  \right\} P^L_{\delta \delta}(k') \,, \\ & \frac{\Delta P_{\delta \delta_{\nu}}^{\textrm{1-loop}}(k) }{P_{\delta \delta_{\nu}}^{L}(k)}\Bigg|_{\textrm{Distortion}} =  -f \frac{\sigma}{\sigma_{\nu}} \frac{ \gamma}{1+\frac{k_{\textrm{fs}}}{k}} \int \frac{d^3 \vec{k}'}{(2\pi)^3} \frac{k_{\textrm{NL}}}{k'} (\hat{k} \cdot \hat{k}') \times \\ & \times \left[ F_{2}(\vec{k}',-\vec{k})P^L_{\delta \delta}(k') + G_{2}(\vec{k}'-\vec{k},\vec{k}) P^L_{\delta \delta}(|\vec{k}-\vec{k}'|)  + F_{2}(\vec{k}-\vec{k}',\vec{k}') \frac{P^L_{\delta \delta}(|\vec{k}-\vec{k}'|) P^L_{\delta \delta}(k')}{P^L_{\delta \delta}(k)} \right] \,.
	\end{split}
\end{equation}
In Eq.~(\ref{eq:crossps}), the second term in the right hand side gives the SPT one-loop correction to the power spectrum, while the third term represents the contribution from the distortion in neutrino density field due to the peculiar motion of halos, whose effect is to slightly suppress the cross power. In Fig.~\ref{fig:nucdm2pt} we plot the one-loop expression to the CDM-$\nu$ cross power spectrum, as given by Eq.~(\ref{eq:crossps}), for the reference values $m_{\nu}=0.1$eV and $z=0$. We can see that in principle one needs to properly account for the distortion effect in order to produce accurate predictions for the nonlinear corrections to the cross power spectrum at scales around $k_{\textrm{NL}} \approx 0.1$Mpc$^{-1}$, but the effect is quite small.
\begin{figure}
	\centering
	\includegraphics[width=0.75\textwidth]{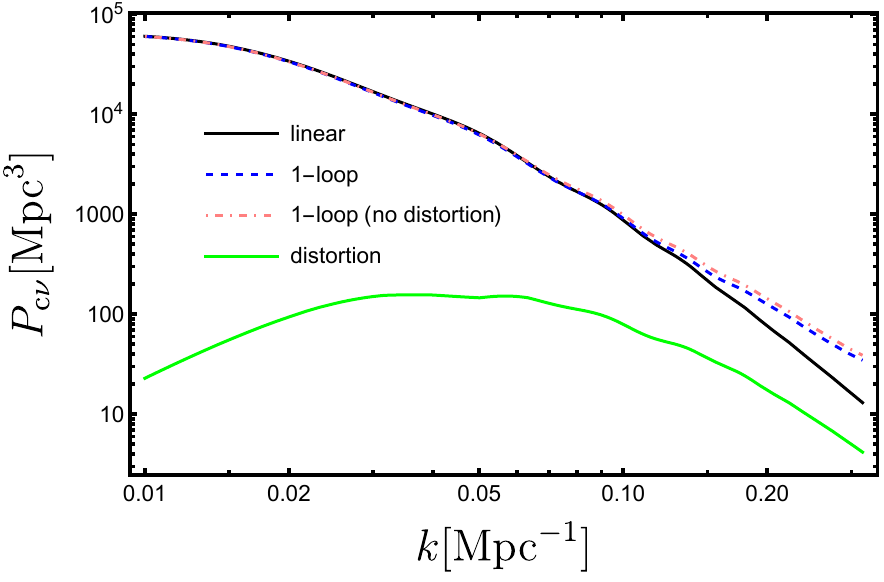}
	\caption{Contributions to the CDM-$\nu$ cross power spectrum, for the reference values $m_{\nu}=0.1$eV and $z=0$. The black solid curve is the linear spectrum produced by CLASS, while the blue dashed curve is the one-loop expression from Eq.~(\ref{eq:crossps}). In the pink dashed curve we also show the SPT one-loop result when not including the contribution from the distortion effect, i.e., dropping the third term in the right hand side of Eq.~(\ref{eq:crossps}). Finally, the solid green curve isolates the distortion contribution to the one-loop result in blue (in absolute value, since the contribution is negative).}
	\label{fig:nucdm2pt}
\end{figure} 

Note that, in principle, two-point cross correlations of CDM and $\nu$ are sensitive to the distortion effect above. However, this small signal is degenerate with the much larger contribution from standard structure formation, and hence analysis involving two-point correlations cannot isolate the distortion effect. However, we expect three-point cross correlations of CDM and $\nu$ to reveal the distortion effect. For instance, as we argued in Sec.~\ref{sec:dfolss}, in the presence of a long wavelength CDM velocity field, a dipole appears in the local CDM-$\nu$ cross power. In a different language, the presence of a specific nonzero squeezed limit bispectrum involving CDM and $\nu$ correlations would be a smoking gun of the distortion effect.

\subsection{Neutrino distortion to the bispectrum}
\label{ssec:threepointcc}
We are interested in three-point correlations of the form $\langle \delta_{\nu}(\vec{k}_1) \delta(\vec{k}_2) \delta(\vec{k}_3) \rangle$. Our starting point is Eq.~(\ref{eq:bispectrum}) for the bispectrum, which we repeat here for convenience:
\begin{equation}
	\label{eq:bispectrumagain}
	B(k_1,k_2,k_3) = \frac{\left(\frac{k_{\textrm{fs}}}{k_1}\right)^2}{(1+\frac{k_{\textrm{fs}}}{k_1})^2} \left[ \langle \delta(\vec{k}_1) \delta(\vec{k}_2) \delta(\vec{k}_3) \rangle ' + \frac{ \gamma}{1+\frac{k_{\textrm{fs}}}{k_1}} \langle \frac{p(\vec{k}_1)}{\sigma_{\nu}} \delta(\vec{k}_2) \delta(\vec{k}_3) \rangle ' \right] \,.
\end{equation}
This is written in terms of CDM-only bispectra that can be readily evaluated to tree level in SPT using Eqs.~(\ref{eq:nmfagain})-(\ref{eq:ptlsscoe})
\begin{equation}
\label{eq:3pt1}
		 \langle \delta(\vec{k}_1) \delta(\vec{k}_2) \delta(\vec{k}_3) \rangle ' = 2\left[ F_{2}(\vec{k}_2,\vec{k}_3) P^L_{\delta \delta}(k_2) P^L_{\delta \delta}(k_3) + F_{2}(\vec{k}_1,\vec{k}_3) P^L_{\delta \delta}(k_1) P^L_{\delta \delta}(k_3) + F_{2}(\vec{k}_1,\vec{k}_2) P^L_{\delta \delta}(k_1) P^L_{\delta \delta}(k_2) \right] \,,
\end{equation}
and,
\begin{equation}
\label{eq:3pt2}
	\langle \frac{q(\vec{k}_1)}{\sigma_{\nu}} \delta(\vec{k}_2) \delta(\vec{k}_3) \rangle ' = -f \frac{\sigma}{\sigma_{\nu}} \frac{k_{\textrm{NL}}}{k_1} \left[\alpha(\vec{k}_2,\vec{k}_3) + \alpha(\vec{k}_3,\vec{k}_2)  \right]  P^L_{\delta \delta}(k_2) P^L_{\delta \delta}(k_3) \,.
\end{equation}
We can then write Eq.~(\ref{eq:bispectrumagain}) as a sum of SPT and distortion (dist) contributions,
\begin{equation}
\label{eq:bissplit}
	B(k_1,k_2,k_3) = B_{\textrm{SPT}}(k_1,k_2,k_3) + B_{\textrm{dist}}(k_1,k_2,k_3) \,,
\end{equation}
with,
\begin{equation}
\label{eq:bissplit2}
\begin{split}
	B_{\textrm{SPT}}(k_1,k_2,k_3) &= 2 \frac{\left(\frac{k_{\textrm{fs}}}{k_1}\right)^2}{(1+\frac{k_{\textrm{fs}}}{k_1})^2} \left[ F_{2}(\vec{k}_2,\vec{k}_3) P^L_{\delta \delta}(k_2) P^L_{\delta \delta}(k_3) + F_{2}(\vec{k}_1,\vec{k}_3) P^L_{\delta \delta}(k_1) P^L_{\delta \delta}(k_3) + F_{2}(\vec{k}_1,\vec{k}_2) P^L_{\delta \delta}(k_1) P^L_{\delta \delta}(k_2) \right] \,, \\ B_{\textrm{dist}}(k_1,k_2,k_3) &= \gamma f \frac{\sigma}{\sigma_{\nu}} \frac{\left(\frac{k_{\textrm{fs}}}{k_1}\right)^2}{(1+\frac{k_{\textrm{fs}}}{k_1})^3} \left( \frac{k_{\textrm{NL}}}{k_3}\mu_{13} + \frac{k_{\textrm{NL}}}{k_2}\mu_{12}\right) P^L_{\delta \delta}(k_2) P^L_{\delta \delta}(k_3) \,,
\end{split}
\end{equation}
where we used the relations $\alpha(\vec{k}_1,\vec{k}_2) = (\vec{k}_{12}\cdot \vec{k}_2)/k_2^2$ and $\vec{k}_1+\vec{k}_2+\vec{k}_3=0$, and introduce the notation $\mu_{ij} = \hat{k}_i \cdot \hat{k}_j$. 

We are interested in the contribution from the distortion effect, captured by $B_{\textrm{dist}}$, which we will see can be isolated from $B_{SPT}$. The distortion term, $B_{\textrm{dist}}(k_1,k_2,k_3)$, is symmetric under the exchange $k_2 \leftrightarrow k_3$, so we can assume without loss of generality that $k_3 \leq k_2$ and parameterize different triangle configurations by the values of $k_2$, $x_3=k_3/k_2 \leq 1$ and $x_1=k_1/k_2$. From the requirement that three wavevectors must fit into a triangle configuration, it follows that $1-x_3 \leq x_1 \leq 1+x_3$. In Fig.~\ref{fig:bisshape} we plot the triangle shape dependence of the bispectrum for a choice of $k_2=0.05$Mpc$^{-1}$ and our fiducial values $m_{\nu}=0.1$eV and $z=0$. The bispectrum peaks at squeezed configurations for which $k_1 \approx k_2 \gg k_3$. The difference in sign between the $x_1=1+x_3$ and $x_1=1-x_3$ cases is due to the fact that the bispectrum is proportional to the angle between wavevectors and hence depends on their relative orientations. In Fig.~\ref{fig:biselong} we plot the same bispectrum along elongated triangle configurations for which $x_1=1+x_3$. 
\begin{figure}
	\centering
	\includegraphics[width=0.75\textwidth]{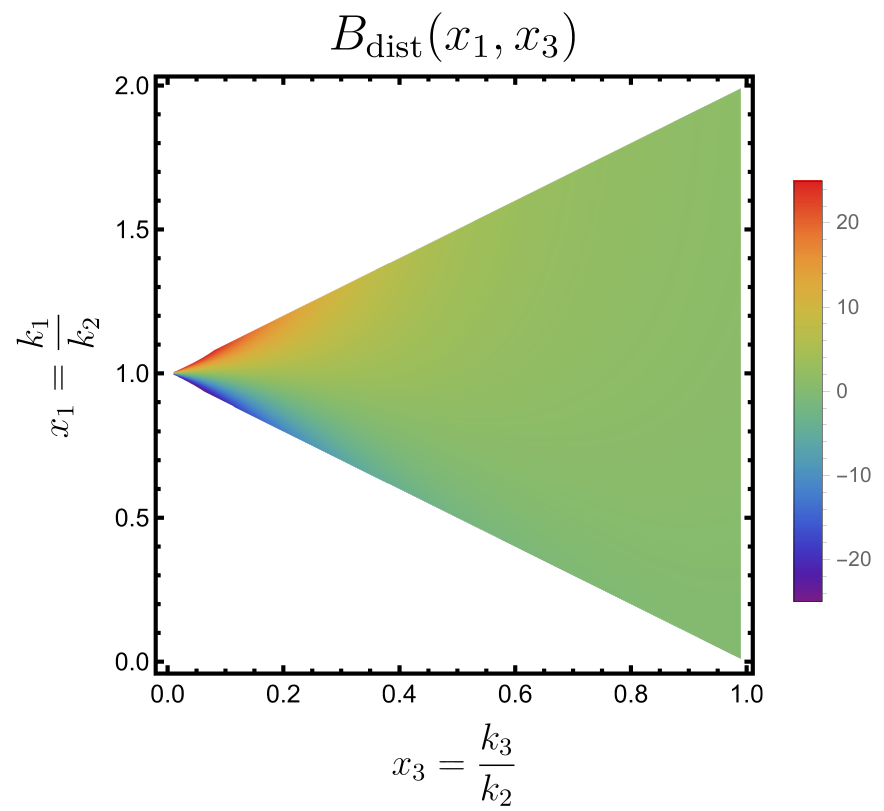}
	\caption{Contribution to the bispectrum from the neutrino distortion effect as given by the second line in Eq.~(\ref{eq:bissplit2}), normalized by its value at the equilateral triangle configuration $B_{\textrm{dist}} (x_1=1,x_3=1)$, as a function of the triangle shape parameterized by $x_1=k_1/k_2$ and $x_3=k_3/k_2$. We fix $k_2=0.05$Mpc$^{-1}$, and assume our fiducial values of a single neutrino mass state with $m_{\nu}=0.1$eV and $z=0$. The left corner of the triangle corresponds to the squeezed limit $k_3\ll k_1, k_2$. The upper boundary of the triangle corresponds to the ``flattened" triangle configuration with $\vec{k}_3$ aligned with $\vec{k}_2$, and $\vec{k}_1$ pointing in the exact opposite direction of $\vec{k}_2$, $\vec{k}_3$. The lower boundary of the triangle corresponds to the opposite limit of $\vec{k}_1$ parallel with $\vec{k}_3$ and $\vec{k}_2$ antiparallel to $\vec{k}_1$, $\vec{k}_3$. The change in sign of $B_{\rm dist}$ along these flattened configurations in the limit $k_3\ll k_1,k_2$ illustrates the anisotropy in the bispectrum under the exchange of $\vec{k}_1$ and $\vec{k}_2$ or equivalently, under exchanging CDM and $\nu$ density fields in $\langle \delta_\nu(\vec{k}_1) \delta(\vec{k}_2)\delta(\vec{k}_3)\rangle$.}
	\label{fig:bisshape}
\end{figure} 
  
\begin{figure}
	\centering
	\includegraphics[width=0.75\textwidth]{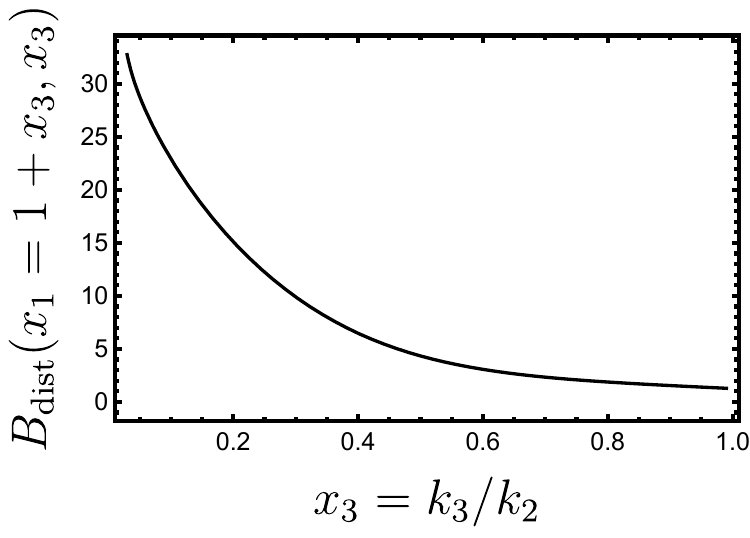}
	\caption{Contribution to the bispectrum from the neutrino distortion effect as given by the second line in Eq.~(\ref{eq:bissplit2}), normalized by its value at the equilateral triangle configuration $B_{\textrm{dist}} (x_1=1,x_3=1)$, as a function of $x_3=k_3/k_2$ and for elongated configurations for which $x_1=1+x_3$. We fix $k_2=0.05$Mpc$^{-1}$, and assume our fiducial values of a single neutrino mass state $m_{\nu}=0.1$eV and $z=0$. }
	\label{fig:biselong}
\end{figure} 

We can provide a physical interpretation for this bispectrum. For this consider the squeezed limit with $k_1 \approx k_2 \gg k_3$ in the second line of Eq.~(\ref{eq:bissplit2}), and use Eq.~(\ref{eq:linearcross}). We arrive at
\begin{equation}
	\label{eq:squeezed}
	B_{\textrm{dist}}(k_1,k_2,k_3) \approx \gamma f \frac{\sigma}{\sigma_{\nu}}  \frac{k_{\textrm{NL}}}{k_3}\mu_{13} \frac{P^L_{\delta \delta_{\nu}}(k_1)}{1+\frac{k_{\textrm{fs}}}{k_1}} P^L_{\delta \delta}(k_3) \,.
\end{equation}
In order to interpret Eq.~(\ref{eq:squeezed}), note that:
\begin{equation}
	\frac{\vec{v}(\vec{k}_3)}{\sigma_{\nu}} = -\frac{i\hat{k}_3}{k_3} \frac{\theta(\vec{k}_3)}{\sigma_{\nu}}= i\hat{k}_3 f \frac{\sigma}{\sigma_{\nu}} \frac{k_{\textrm{NL}}}{k_3} \delta(\vec{k}_3) \,,
\end{equation}
where we used the relation $\theta(\vec{k}_3) = -(faH)\delta(\vec{k}_3)$ and insert the identity $\sigma k_{\textrm{NL}}/aH=1$. We may then write,
\begin{equation}
	B_{\textrm{dist}}(k_1,k_2,k_3) \approx \left\langle \langle\delta_{\nu}(\vec{k}_1) \delta(\vec{k}_2)\rangle_{\phi(\vec{K})} \delta(\vec{k}_3) \right\rangle \,,
\end{equation}
with $\vec{K}=-\vec{k}_3= \vec{k}_1+\vec{k}_2$ and,
\begin{equation}
	\label{eq:dipoleagain}
	\langle \delta_{\nu}(\vec{k}_1) \delta(\vec{k}_2)\rangle_{\phi(\vec{K})} = \frac{\vec{k}_1\cdot \vec{K}}{k_1}  \frac{P^L_{\delta \delta_{\nu}}(k_1)}{1+\frac{k_{\textrm{fs}}}{k_1}} \phi(\vec{K}) \,,
\end{equation}
is a dipole contribution to the local CDM-$\nu$ cross correlation in the presence of a long wavelength velocity potential $\phi(\vec{K})$ defined as, in position space,
\begin{equation}
	\label{eq:velpotential}
\gamma \frac{\vec{v}}{\sigma_{\nu}} = \vec{\nabla} \phi \,.
\end{equation}
We expect Eq.~(\ref{eq:dipoleagain}) to be robust against higher order perturbation theory corrections, provided one applies the replacement $P^L_{\delta \delta_{\nu}}(k_1) \to P^{\textrm{MON}}_{\delta \delta_{\nu}}(k_1) = ((\frac{k_{\textrm{fs}}}{k})^2/(1+\frac{k_{\textrm{fs}}}{k})^2) P_{\delta \delta}(k_1)$ to the monopole contribution to the local CDM-$\nu$ cross power spectrum, where $P_{\delta \delta}(k_1)$ is the nonlinear CDM power spectrum. 

Note from Eq.~(\ref{eq:dipoleagain}) that the dipole is proportional to both the large-scale displacement field $\phi(\vec{K})$ and the short scale monopole contribution to the cross power spectrum $P^{\textrm{MON}}_{\delta \delta_{\nu}}(k_1)$. A measurement of the squeezed limit bispectrum can then be phrased as a reconstruction of the displacement field $\phi(\vec{K})$, independent of the specific realization of the small scale CDM power. As a consequence, such a measurement is not limited by cosmic variance but rather by the finite number of galaxies in the survey as we will see in the next subsection.

Equation (\ref{eq:dipoleagain}) is in qualitative agreement with the results obtained in \cite{zhu2020measuring} in a simplified model where the authors introduce an effective relative displacement between CDM and neutrino fluid elements to show that shift nonlinearities arise in the neutrino density field, which is exactly what we find in a explicit investigation of the solution to the Boltzmann equation for the distribution function of relic neutrinos around nonlinear CDM structure, i.e., from first principles. This novel approach provides some additional insight into the distortion effect and its potential observational signatures; it is now clear how the neutrino thermal velocity enters the picture, and that the dipole distortion is indeed directly connected to the dynamical friction effect. Additionally, and from a more practical point of view, we now have an explicit expression for the effective displacement potential as given by Eq.~(\ref{eq:velpotential}). We also have an expression for the bispectrum that goes beyond the squeezed limit as given by the second line in Eq.~(\ref{eq:bissplit2}), and that is important for two reasons. First, it opens up the opportunity to check if the squeezed limit indeed dominates the contribution to the signal-to-noise and to account for additional information in other triangle configurations. Second, it allows us to quantitatively assess the degeneracy between the distortion contribution and standard nonlinear structure formation. 

To elaborate further in the latter point, we now argue that the distortion contribution can in principle be  separated from the SPT contribution. For this, the key observation is that the expression in Eq.~(\ref{eq:3pt1}) is symmetric under the exchange $\vec{k}_1 \xleftrightarrow{} \vec{k}_2$, while the expression in Eq.~(\ref{eq:3pt2}), i.e., the contribution from the distortion effect, does not share the same symmetry. In order to exploit this, we consider (where the superscript $A$ denotes an antisymmetrization),
\begin{equation}
\label{eq:obs}
\begin{split}
	B^{A}(k_1,k_2,k_3) & = \frac{1}{2}\left[B(k_1,k_2,k_3) - B(k_2,k_1,k_3)\right]\,,\\ &  = B^{A}_{\textrm{SPT}}(k_1,k_2,k_3) + B^{A}_{\textrm{dist}}(k_1,k_2,k_3) \,,
\end{split}
\end{equation}
which once again can be written as a sum of SPT and distortion (dist) contributions. The first reads,
\begin{equation}
\label{eq:bispspt}
	\begin{split}
	 B^{A}_{\textrm{SPT}}(k_1,k_2,k_3) = \left[ \frac{\left(\frac{k_{\textrm{fs}}}{k_1}\right)^2}{(1+\frac{k_{\textrm{fs}}}{k_1})^2} - \frac{\left(\frac{k_{\textrm{fs}}}{k_2}\right)^2}{(1+\frac{k_{\textrm{fs}}}{k_2})^2}\right] \Bigg[  F_{2}(\vec{k}_2,\vec{k}_3) & P^L_{\delta \delta}(k_2) P^L_{\delta \delta}(k_3) + F_{2}(\vec{k}_1,\vec{k}_3) P^L_{\delta \delta}(k_1) P^L_{\delta \delta}(k_3) + \\ & F_{2}(\vec{k}_1,\vec{k}_2) P^L_{\delta \delta}(k_1) P^L_{\delta \delta}(k_2) \Bigg] \,,
	\end{split}		
\end{equation}
and the second, 
\begin{equation}
\label{eq:bispdist}
\begin{split}
		 B^{A}_{\textrm{dist}}(k_1,k_2,k_3) = \frac{\gamma}{2} f \frac{\sigma}{\sigma_{\nu}} \Bigg\{ \frac{\left(\frac{k_{\textrm{fs}}}{k_1}\right)^2}{(1+\frac{k_{\textrm{fs}}}{k_1})^3}  & \left(\frac{k_{\textrm{NL}}}{k_3}\mu_{13} + \frac{k_{\textrm{NL}}}{k_2} \mu_{12} \right) P^L_{\delta \delta}(k_2) P^L_{\delta \delta}(k_3)- \\ & \frac{\left(\frac{k_{\textrm{fs}}}{k_2}\right)^2}{(1+\frac{k_{\textrm{fs}}}{k_2})^3} \left(\frac{k_{\textrm{NL}}}{k_3}\mu_{23} + \frac{k_{\textrm{NL}}}{k_1} \mu_{21} \right) P^L_{\delta \delta}(k_1) P^L_{\delta \delta}(k_3)\Bigg\} \,,	
\end{split}
\end{equation}
where we combined Eqs.~(\ref{eq:bissplit2}) and (\ref{eq:obs}). In the squeezed limit where $k_1 \approx k_2 \gg k_3$ we then obtain $B^{A}_{\textrm{SPT}}(k_1,k_2,k_3) \approx 0$ and $B^{A}_{\textrm{dist}}(k_1,k_2,k_3) \approx B_{\textrm{dist}}(k_1,k_2,k_3)$ such that the antisymmetrization procedure cleans out the SPT  contribution and hence provides a smoking gun for the distortion effect. This argument assumes that all the information from the distortion effect in the bispectrum comes from the squeezed limit, which at this point is a reasonable assumption based on  Figs.~\ref{fig:bisshape} and \ref{fig:biselong}. 

In the next section we extend our analysis with a calculation of the signal-to-noise ratio based on a multi-tracer approach. In order to build intuition on the size of the distortion effect, let us first compare our squeezed limit bispectrum in Eq.~(\ref{eq:squeezed}), which we repeat here in simpler notation for convenience,
\begin{equation}
\label{eq:squeezedagain}
	B_{\textrm{dist}} = f_{\nu} \gamma f \frac{\sigma}{\sigma_{\nu}}  \frac{k_{\textrm{NL}}}{k_L}\mu \frac{\left(\frac{k_{\textrm{fs}}}{k_S}\right)^2}{\left(1+\frac{k_{\textrm{fs}}}{k_S}\right)^3} P_{\delta \delta}(k_S) P_{\delta \delta}(k_L) \,,
\end{equation}
with the familiar squeezed limit bispectrum (of the late-time matter density contrast) generated by the presence of local primordial non-Gaussianity \cite{komatsu2001acoustic},
\begin{equation}
\label{eq:localshape}
 B_{\textrm{local}} = 	\frac{6\Omega_{\textrm{m}}(a)}{g(a)}(aH)^2 f_{\textrm{NL}} \frac{1}{k_L^2T(k_L)}  P_{\delta \delta}(k_S) P_{\delta \delta}(k_L) \,.
\end{equation}
In Eq.~(\ref{eq:squeezedagain}), $k_S= k_1 \approx k_2$, $k_L = k_3$, $\mu$ is the cosine of the angle between short and long wave vectors and we used Eq.~(\ref{eq:glue1}). We also inserted a factor of the fractional contribution of neutrinos to the matter density $f_{\nu}$ which always multiplies the neutrino density contrast $\delta_{\nu}$ extracted from a measurement of the total mass fluctuation $\delta_m = (1-f_\nu)\delta+f_\nu\delta_\nu$. In Eq.~(\ref{eq:localshape}), $\Omega_{\textrm{m}}(a)$ is the fractional contribution of matter to the critical density, $g(a)= D_{\textrm{L}}(a)/a$ with $D_{\textrm{L}}(a)$ the linear growth factor and $T(k)$ is the matter transfer function. The shape and scale dependence of the two bispectra are different, but both peak at squeezed triangle configurations and we can formally compare Eqs.~(\ref{eq:squeezedagain}) and (\ref{eq:localshape}) to define an effective non-Gaussian parameter to the distortion bispectrum:  
\begin{equation}
\label{eq:fnleff}
	f_{\textrm{NL}}^{\textrm{eff}} = \frac{f_{\nu}}{\sigma_{\nu}^2} \gamma \frac{fg}{6\Omega_{\textrm{m}}} \sqrt{\frac{3}{2} \Omega_{\textrm{m}}} \left(\frac{k_L}{k_S}\right) \mu T(k_L) \frac{\left(\frac{k_{\textrm{fs}}}{k_S}\right)^2}{\left(1+\frac{k_{\textrm{fs}}}{k_S}\right)^3} \,,
\end{equation}
where we used Eqs.~(\ref{eq:scalenl}) and (\ref{eq:fstreaming}). For a numerical estimate we evaluate this quantity at redshift $z=0$, with $\Omega_{\textrm{m}}=0.32$, $f = \Omega_{\textrm{m}}^{4/7}$ and $g=1$. We also set $k_S = k_{\textrm{fs}}$, $k_L/k_S=0.1$, $\mu =1$ and $T(k_L) \approx 1$. We the obtain $f_{\textrm{NL}}^{\textrm{eff}} \approx 0.5$ for an individual neutrino mass $m_{\nu}=0.05$eV, $f_{\textrm{NL}}^{\textrm{eff}} \approx 4$ for $m_{\nu} = 0.1$eV and $f_{\textrm{NL}}^{\textrm{eff}} \approx 14$ for $m_{\nu} = 0.15$eV. In the case of three neutrino mass eigenstates, the net effective non-Gaussian parameter is given by the sum of the individual contributions to $f_{\textrm{NL}}^{\textrm{eff}} $. For comparison, the current best constraints on local non-Gaussianity from galaxy surveys are $f_{\rm NL} = -33\pm28$ at $95\%$ confidence, though this comparison is not straightforward since those measurements come from joint analysis of the power spectrum and bispectrum, including scale-dependent bias and loop-corrections. Yet, near-term surveys target  $\sigma(f_{\textrm{NL}}) = 0.2-0.5$, with significant constraining power coming from bispectrum measurements analogous to those proposed here \cite{dore2014cosmology, schlegel2022spectroscopic}. Of course, the signal shapes and scalings of neutrinos and $f_{\rm NL}$ are different but this comparison nevertheless provides a rough estimate of the survey needs to use the bispectrum to study neutrino wakes.

\subsection{Multi-tracer bispectrum forecast}  

In the previous section we studied the contribution to the bispectrum from the distortion effect, and used the fact that the signal peaks at squeezed triangle configurations to argue that it can be separated from the contributions to the bispectrum from standard nonlinear structure formation [the SPT term in Eq.~(\ref{eq:bissplit2})]. There are two more ingredients we need to account for: The intrinsic noise from cosmic variance, and the number of modes for a given triangle configuration. This will enable us to forecast the observability of the distortion effect in future surveys and to quantitatively assess how degenerate the distortion signal is with standard nonlinear structure formation.  

In order to isolate the distortion contribution to the bispectrum in Eq.~(\ref{eq:bissplit2}) we adopt the antissymetrization strategy of Eq.~(\ref{eq:obs}), as argued in the previous section. For this we need a probe of the total matter field $\delta_{\textrm{m}} = (1-f_{\nu})\delta +f_{\nu}\delta_{\nu}$ and also a galaxy data, $\delta_{g} = b\delta$, with $b$ the linear bias, $f_{\nu}=\bar{\rho}_{\nu}/\bar{\rho}$ the fractional contribution of neutrinos to the matter budget. Note that we follow the standard practice of assuming the galaxy field to trace CDM and baryons, excluding neutrinos \cite{LoVerde:2014pxa, vagnozzi2018bias, raccanelli2019biases, villaescusa2018imprint, costanzi2013cosmology, castorina2014cosmology, villaescusa2014cosmology, castorina2015demnuni}.

It follows that, schematically 
\begin{equation}
\label{eq:signal}
		 B^{A}_{\textrm{obs}}= \frac{1}{2}\langle (\delta_{\textrm{m}}\delta_{\textrm{g}} -\delta_{\textrm{g}}\delta_{\textrm{m}})\delta_{\textrm{m}}\rangle \  \approx \frac{1}{2} f_{\nu}b \langle (\delta_{\nu}\delta -\delta\delta_{\nu})\delta \rangle \,,
\end{equation}
is the bispectrum we wish to extract from the surveys, when combined with Eqs.~(\ref{eq:obs})-(\ref{eq:bispdist}), where we work to leading order in the neutrino fraction $f_{\nu} \ll 1$ \footnote{One could also choose $B^{A}_{\textrm{obs}}= \frac{1}{2}\langle (\delta_{\textrm{m}}\delta_{\textrm{g}} -\delta_{\textrm{g}}\delta_{\textrm{m}})\delta_{\textrm{g}}\rangle$, with a galaxy density contrast on the long-wavelength mode as opposed to a matter density contrast. To leading order in $f_{\nu}$ this produces the same observable (up to a factor of the linear bias $b$), with some additional (shot) noise due to the finite sampling of the galaxy density field.}. This includes our signal of the distortion in the neutrino density field due to the peculiar motion of halos but also a background contribution from standard nonlinear structure formation. We are then left with the problem of forecasting the observability of our effect in future surveys, when also accounting for potential degeneracies with the physics of standard small scale structure formation.

In Appendix \ref{sec:stn} we derive an optimal estimator for the bispectrum in Eq.~(\ref{eq:signal}).  The following formula for the cumulative signal-to-noise ratio per galaxy (or the fisher matrix $F$) is obtained for our signal, which corresponds to the contribution to the bispectrum from the distortion effect as given by Eq.~(\ref{eq:bispdist}):
\begin{equation}
\label{eq:stnformula}
\frac{\textrm{SNR}^2}{N_{g}} =	F_{B_{\textrm{dist}} B_{\textrm{dist}}} = 2f_{\nu}^2 b^2 	\int \frac{d^3 \vec{k}_1}{(2\pi)^3} \int \frac{d^3 \vec{k}_2}{(2\pi)^3} \int \frac{d^3 \vec{k}_3}{(2\pi)^3} (2\pi)^3\delta^{(3)}(\vec{k}_{123}) \frac{[B^{A}_{\textrm{dist}}(k_1,k_2,k_3)]^2}{P_{\delta \delta}(k_3)[P_{\delta \delta}(k_1)+P_{\delta \delta}(k_2)+P_{\delta \delta}(k_3)]} \,,
\end{equation}
where $\textrm{SNR}^2$ is the cumulative signal-to-noise ratio squared, $N_{\textrm{g}}$ is the total number of galaxies in the galaxy survey, and $P_{\delta \delta}(k)$ stands for the (linear) power spectrum.  We can now use Eq.~(\ref{eq:obs}) to rewrite this expression as follows:
\begin{equation}
\label{eq:stnr}
\begin{split}
		F_{B_{\textrm{dist}}  B_{\textrm{dist}}}& = \frac{f_{\nu}^2b^2}{2}  \int \frac{d^3 \vec{k}_1}{(2\pi)^3} \int \frac{d^3 \vec{k}_2}{(2\pi)^3} \int \frac{d^3 \vec{k}_3}{(2\pi)^3} (2\pi)^3\delta^{(3)}(\vec{k}_{123}) \frac{B_{\textrm{dist}}(k_1,k_2,k_3)}{P_{\delta \delta}(k_2)P_{\delta \delta}(k_3)[P_{\delta \delta}(k_1)+P_{\delta \delta}(k_2)+P_{\delta \delta}(k_3)]} \, \times \\ & \times \left\{\left[P_{\delta \delta}(k_2)+P_{\delta \delta}(k_3)\right]B_{\textrm{dist}}(k_1,k_2,k_3) - P_{\delta \delta}(k_2)B_{\textrm{dist}}(k_2,k_1,k_3) -  P_{\delta \delta}(k_3)B_{\textrm{dist}}(k_3,k_1,k_2) \right\} \,.
\end{split}
\end{equation}
Due to the Dirac delta enforcing the requirement that the wavevectors fit into a triangle configuration, we can immediately integrate over $\vec{k}_1$ such that we are left with integrals over $k_2$, $k_3$ and $\mu_{23}=\hat{k}_2 \cdot \hat{k}_3$. First, we replace the variable $k_3$ by $x_3=k_3/k_2$. Since the integrand in Eq.~(\ref{eq:stnr}) is symmetric under the exchange $k_2 \leftrightarrow k_3$, we can assume $x_3 \leq 1$ without loss of generality. Second, we replace $\mu_{23}$ with $x_1=k_1/k_2$ according to:
\begin{equation}
	\label{eq:angleagain}
	\mu_{23} = \hat{k}_2 \cdot \hat{k}_3 = \frac{x_1^2 -1-x_3^2}{2x_3} \implies d\mu_{23} = \frac{x_1}{x_3} dx_1 \,,
\end{equation}
and the range of integration is $1-x_3 \leq x_1 \leq 1+x_3$. Implementing these changes in the integration variables leads to:
\begin{equation}
	\label{eq:subs}
	\int \frac{d^3 \vec{k}_1}{(2\pi)^3} \int \frac{d^3 \vec{k}_2}{(2\pi)^3} \int \frac{d^3 \vec{k}_3}{(2\pi)^3} (2\pi)^3\delta^{(3)}(\vec{k}_{123}) = \frac{1}{4\pi^4} \int_{0}^{\infty} \frac{dk_2}{k_2} \int_{0}^{1} dx_3 \int_{1-x_3}^{1+x_3} dx_1 \, k_2^6 x_1x_3 \,,
\end{equation}
and hence to our final formula for the Fisher matrix,
\begin{equation}
	\label{eq:stnragain}
	\begin{split}
		F_{B_{\textrm{dist}}  B_{\textrm{dist}}}& = \frac{f_{\nu}^2b^2}{8\pi^4}  \int_{0}^{\infty} \frac{dk_2}{k_2} \int_{0}^{1} dx_3 \int_{1-x_3}^{1+x_3} dx_1 \, k_2^6 x_1x_3 \frac{B_{\textrm{dist}}(k_1,k_2,k_3)}{P_{\delta \delta}(k_2)P_{\delta \delta}(k_3)[P_{\delta \delta}(k_1)+P_{\delta \delta}(k_2)+P_{\delta \delta}(k_3)]} \, \times \\ & \times \left\{\left[P_{\delta \delta}(k_2)+P_{\delta \delta}(k_3)\right]B_{\textrm{dist}}(k_1,k_2,k_3) - P_{\delta \delta}(k_2)B_{\textrm{dist}}(k_2,k_1,k_3) -  P_{\delta \delta}(k_3)B_{\textrm{dist}}(k_3,k_1,k_2) \right\} \,.
	\end{split}
\end{equation}
We can integrate over $x_1$ and $x_3$ (the triangle shape), for a given fixed $k_2$, to investigate the scale dependence of the signal-to-noise. This is plotted in Fig.~\ref{fig:stnk} for our reference values $m_{\nu}=0.1$eV and $z=0$. It peaks at large quasi-linear scales, as set by the neutrino free-streaming scale, and hence one does not enhance the signal-to-noise by adding more modes. This means, in particular, that we are not limited by the precise modeling of complicated nonlinear and galaxy formation physics (and a linear halo bias should be sufficient). The measurement is also not degenerate with the optical depth to reionization since it is independent from the total matter power spectrum, and it is not cosmic variance limited either since the signal-to-noise can be made arbitrarily large by taking the limit $N_{g} \to \infty$. Indeed, cosmic variance cancellation is a common theme of multi-tracer approaches \cite{loverde2016neutrino,Schmittfull:2017ffw}. 

To elaborate on this, let us briefly make a comparison to the more familiar case of how cosmic variance cancellation applies to measuring a linear galaxy bias factor $b$ in $\delta_g = b\delta$ (e.g. \cite{Seljak:2008xr}). The linear bias factor can be measured without cosmic variance because it represents the linear response of the fluctuation in the galaxy density field $\delta_g$ to a fluctuation in the matter density field $\delta$ and one only needs a single realization of $\delta$ and $\delta_g$ to determine this response $b$. Moreover, the particular amplitude or shape of $P_{\delta\delta}$ is, in principle, unimportant for this so long as $P_{g\delta} = b P_{\delta\delta}$ and one measures both $P_{g\delta}$ and $P_{\delta\delta}$. In our case, the distortion to the neutrino density field is a linear response of the neutrino field $\delta_\nu$ to the local realization of the small-scale CDM momentum field $p = \delta v$. One similarly only needs a single realization of $ \delta v$ to determine this response [also see the discussion below Eq.~(\ref{eq:dipoleagainpre})].

Returning to Eq.~(\ref{eq:stnragain}), we can also integrate over $k_2$ and $x_1$, for a given fixed $x_3$. This is plotted in Fig.~\ref{fig:stnshape}, again for $m_{\nu}=0.1$eV and $z=0$. It peaks at a small but nonzero value of $x_3 \sim 0.1$ due to a combination of two effects: The bispectrum prefers the $x_3 \to 0$ limit as can be seen in Figs. \ref{fig:bisshape} and \ref{fig:biselong}, but the number of modes is suppressed in this limit. As a consequence, real surveys need to have a sufficiently large volume in order to probe the distortion bispectrum close to the squeezed limit. For our reference value $m_{\nu}=0.1$eV, we need $k_{\textrm{min}} \lesssim k_{\textrm{L}} \sim x_3 k_{\textrm{S}} \approx 0.005$Mpc$^{-1}$, from Figs.~\ref{fig:stnk} and \ref{fig:stnshape}.
  \begin{figure}
	\centering
	\includegraphics[width=0.75\textwidth]{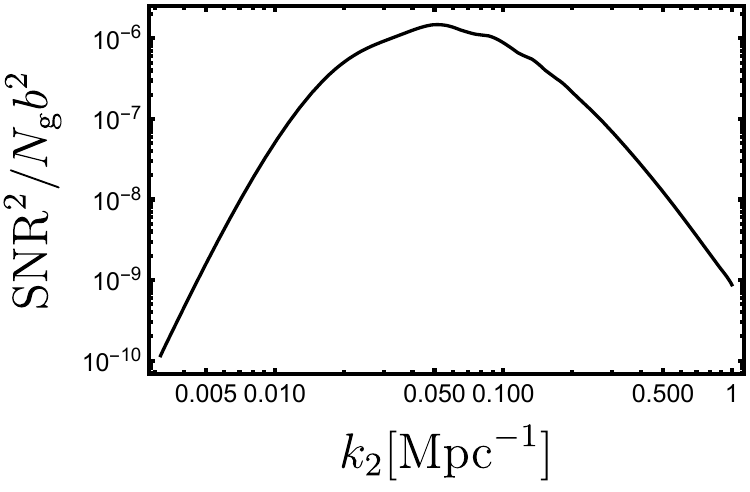}
	\caption{Cumulative distortion bispectrum signal-to-noise ratio squared per galaxy, divided by the galaxy bias squared, as a function of scale $k_2$ after integrating Eq.~(\ref{eq:stnragain}) over the triangle shape. We choose the reference values of a single neutrino mass state with $m_{\nu}=0.1$eV and measure the bispectrum at $z=0$. With this choice, consistency with oscillation data requires three neutrino mass states each with $m_{\nu i}\approx 0.1$eV and the net signal-to-noise squared would increase by roughly a factor of three.}
	\label{fig:stnk}
\end{figure}

\begin{figure}
	\centering
	\includegraphics[width=0.75\textwidth]{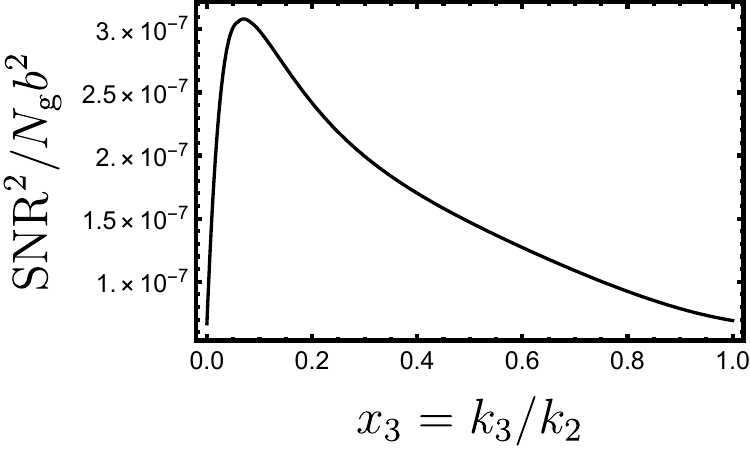}
	\caption{Cumulative distortion bispectrum signal-to-noise ratio squared per galaxy, divided by the galaxy bias squared, as a function of the long-wavelength mode $x_3=k_3/k_2$ after  integrating Eq.~(\ref{eq:stnragain}) over $k_2$ and $x_1=k_1/k_3$. We choose the reference values of a single neutrino mass state with $m_{\nu}=0.1$eV and measure the bispectrum at $z=0$. With this choice, consistency with oscillation data requires three neutrino mass states each with $m_{\nu i}\approx 0.1$eV and the net signal-to-noise squared would be increased by roughly a factor of three.}
	\label{fig:stnshape}
\end{figure}

We finally proceed to evaluate Eq.~(\ref{eq:stnragain}) numerically, for the individual neutrino masses and redshifts of interest, in order to forecast the observability of the effect in future surveys. The results are presented in Tab.~\ref{table:forecast}. At $z=0$ we find that $\textrm{SNR}^2/b^2 N_{g} \approx (2.5 \times 10^{-9}-1.6 \times 10^{-6})$ for individual neutrino masses in the range $m_{\nu}=(0.05-0.15)$eV. \footnote{The total signal-to-noise for all three neutrino mass eingenstates can then be obtained by summing the individual contributions in quadrature, i.e., $SNR_{\textrm{tot}}^2= \sum_{m_{\nu}} SNR_{\textrm{tot}}^2|_{m_{\nu}}$.} This corresponds to (setting $b = 1$ for the galaxy bias) a range of $N_{g} \gtrsim (4\times 10^{9}- 6\times 10^{6}$) for the minimal number of galaxies in the survey in order to reach a 3$\sigma$ detection of the distortion effect. The signal-to-noise increases with the neutrino mass and decreases with redshift, as expected from our intuition from the dynamical friction effect [see Eq(\ref{eq:powerlaw})]. However, we should keep in mind that we work under the assumption that $k_{\textrm{fs}}/k_{\textrm{NL}} \ll 1$ (see Fig.~\ref{fig:scales}). For a sufficiently large neutrino mass $k_{\textrm{fs}} \sim k_{\textrm{NL}}$, and hence we expect the signal-to-noise to eventually decrease due to the cutoff at the free-streaming scale.  

When accounting for the cosmological upper bound $\sum_{\nu} m_{\nu} \leq 0.12$eV, together with oscillation experiments, we are led to the optimistic scenario of two neutrino eigenstates with mass $m_{\nu}\approx 0.05$eV under the inverted hierarchy. This yields a total number of galaxies $N_{g} \approx 8 \times 10^{8}$ in order to reach a 2-$\sigma$ detection, which may be possible with future surveys. This specific choice of neutrino mass splittings produces the highest possible signal-to-noise ratio while saturating the current cosmological upper bound in the sum of mass eigenstates. 
\begin{table}[]
	\centering
	\begin{tabular}{|c|c|c|c|}
		\hline
		$\textrm{SNR}^2/b^2 N_{g}$ & $m_{\nu}=0.05$eV & $m_{\nu}=0.1$eV & $m_{\nu}=0.15$eV \\ \hline
		$z=0$ & $ 2.5 \times 10^{-9}$  & $1.6 \times 10^{-7}$  & $1.6 \times 10^{-6}$ \\ \hline
		$z=1$ & $8.6 \times 10^{-11}$  & $6.3 \times 10^{-9}$  &  $6.8 \times 10^{-8}$ \\ \hline
	\end{tabular}
	\caption{Numerical values for the cumulative distortion bispectrum signal-to-noise ratio squared per galaxy, divided by the galaxy bias squared, for the (individual) neutrino masses and redshifts of interest. For three massive neutrino states, the net signal-to-noise squared is roughly the sum of the signal-to-noise squared for each state.}
	\label{table:forecast}
\end{table}

Our  estimates for the signal-to-noise ratio in Tab.~\ref{table:forecast} are more pessimistic than previous estimates found in the literature for the observability of the CDM-$\nu$ relative flow effect in the large-scale structure \cite{zhu2020measuring, okoli2017dynamical, zhu2014measurement}. We attribute this to the fact that previous approaches to study the effect of CDM-$\nu$ dynamical friction/relative flow are based on the introduction of an ad-hoc effective displacement between CDM and $\nu$ fluids, and hence do not fully account for the cutoff at the velocity coherence and free-streaming scales. Our first-principles approach, based on an explicit extraction of the distortion effect from the solution to the Boltzmann equation for the neutrino distribution function in the background of the nonlinear CDM structure, does not suffer from this problem. The cutoffs at both the velocity coherence and free-streaming scale are naturally accounted for. Yet, the presence of these cutoffs reduces the amplitude of the signal.

We previously made the observation that standard nonlinear structure formation also gives a contribution to the bispectrum of interest, as given by Eq.~(\ref{eq:bispspt}). We also argued that the antisymmetrization effectively separates it from the distortion contribution, provided that all the information comes from the squeezed limit $x_{3} \to 0 $. However, the signal-to-noise ratio peaks at some small but nonzero value of $x_{3}$, as indicated in Fig.~\ref{fig:stnshape}. One may then worry that this can lead to a significant degeneracy between the distortion and SPT contributions to the bispectrum, effectively erasing our signal. To quantitatively investigate this possibility, we first proceed to define a Fisher matrix $F_{B_{\textrm{SPT}}B_{\textrm{SPT}}}$ exactly as before in Eq.~(\ref{eq:stnformula}), by simply replacing $B_{\textrm{dist}} \to B_{\textrm{SPT}}$. The next step is to introduce:
\begin{equation}
\label{eq:fisher}
	F_{B_{\textrm{dist}} B_{\textrm{SPT}}} = 2f_{\nu}^2 b^2 	\int \frac{d^3 \vec{k}_1}{(2\pi)^3} \int \frac{d^3 \vec{k}_2}{(2\pi)^3} \int \frac{d^3 \vec{k}_3}{(2\pi)^3} (2\pi)^3\delta^{(3)}(\vec{k}_{123}) \frac{B_{\textrm{dist}}(k_1,k_2,k_3)B_{\textrm{SPT}}(k_1,k_2,k_3)}{P_{\delta \delta}(k_3)[P_{\delta \delta}(k_1)+P_{\delta \delta}(k_2)+P_{\delta \delta}(k_3)]} \,,
\end{equation}
in terms of which we can define the cosine of an angle between the two bispectra as follows,
\begin{equation}
\label{eq:angle}
	\mu = \frac{F_{B_{\textrm{dist}} B_{\textrm{SPT}}}}{\sqrt{F_{B_{\textrm{dist}} B_{\textrm{dist}}}F_{B_{\textrm{SPT}} B_{\textrm{SPT}}}}}
\end{equation}  
with $|\mu| \leq 1$, $|\mu|=1$ corresponding to the case of perfect degeneracy and $|\mu|=0$ to no degeneracy between the two bispectra. For all individual neutrino masses and redshifts of interest, we find $|\mu| \leq 0.23$ when choosing $k_{\max}=0.3$Mpc$^{-1}$, further reducing to $|\mu| \leq 0.13$ for $k_{\max}=1$Mpc$^{-1}$. Of course, such a choice of maximum wavenumber actually requires a better handle into nonlinear structure formation and galaxy formation physics in order to accurately model the nonlinear dynamics. However, a naive application of SPT combined with a simple linear galaxy bias already illustrates the point that one can best disentangle the signal from the background by pushing to smaller scales. This is to be expected since these two effects mostly operate at different scales: The distortion contribution peaks at large quasi-linear scales (see Fig.~\ref{fig:stnk}), while we expect the contributions from standard nonlinear structure formation to show up at smaller nonlinear scales. However, even for a choice of $k_{\max}=0.3$Mpc$^{-1}$ the degeneracy is already small. This is because the distortion signal is dominated by triangle configurations that are close to the squeezed limit (see Fig.~\ref{fig:stnshape}), while we expect the contribution from standard nonlinear structure formation to be suppressed in the squeezed limit \cite{jeong2009primordial}. We expect similar arguments to also hold for the contributions to the bispectrum from local galaxy formation physics, or, for instance higher-order terms in a bias expansion (e.g. \cite{Desjacques:2016bnm}), which would justify our use of a simple linear bias. On the other hand, these terms would add additional sources of stochasticity to the noise, analogous to higher-order terms we have already dropped in the calculation of the covariance matrix in Appendix \ref{sec:stn}. We leave detailed investigations of these considerations to future work. 

\section{Conclusion}
\label{sec:conc}

At late times relic neutrinos become nonrelativistic and cluster anisotropically behind moving cold dark matter structures, which generates a distortion in the neutrino density field. This effect gets imprinted into cosmological observables and can potentially be detected with upcoming surveys via its signature on three-point cross correlations of matter and galaxies. 

In Sec.~\ref{sec:1h} we first considered a CDM distribution consisting of a single moving point mass halo, which leads to the anisotropic clustering of neutrinos as shown in Fig.~\ref{fig:2dplot}. This in turn produces a dynamical friction effect that slows the halo down according to Chandrasekhar's formula,  Eq.~(\ref{eq:cdff}). 

In Sec.~\ref{sec:dfolss} we moved on to a more general (and hence realistic) nonlinear cold dark matter (CDM) distribution, relying solely on the validity of the continuity and Poisson's equation, and hence remaining agnostic about the nonlinear gravitational evolution. We showed how to extract the distortion in the neutrino density field produced by the peculiar motion of halos from the solution to the Boltzmann equation for the neutrino distribution function, as given by Eqs.~(\ref{eq:combine}) and (\ref{eq:centraleq}). This enabled us to greatly improve on, and solve a few problems with, our previous framework based on a single moving point mass halo. The consistency of our approach required a hierarchy of scales between the neutrino free-streaming scale $k_{\textrm{fs}}$, and the scale of nonlinearities  $k_{\textrm{NL}}$, i.e., $k_{\textrm{fs}} \ll k_{\textrm{NL}}$, which is consistent with current cosmological upper bounds on the neutrino mass scale as illustrated in Fig.~\ref{fig:scales}. 

We determined that the average decrease of halo peculiar velocities due to the dynamical friction effect is small and hence future experiments based on measurements of galaxy velocities will likely remain insensitive to the distortion effect in the foreseeable future [see Eq.~(\ref{eq:powerlaw})]. However, the distortion effect also gets imprinted in the large scale structure, and one has to account for it in order to accurately model the nonlinear corrections to the CDM-$\nu$ cross power spectrum, see Eq.~(\ref{eq:crosspower}). We also argued that a clean probe of the distortion effect is to be found in three-point cross correlations of CDM and neutrinos, or its Fourier transform the bispectrum, as given by Eq.~(\ref{eq:bispectrum}). We emphasize that Eqs.~(\ref{eq:crosspower}) and (\ref{eq:bispectrum}) are agnostic about the nonlinear dynamics of CDM and can be evaluated with any given model for the CDM power spectrum and bispectrum. This fact can also be used to make inferences about neutrino wakes from data that are also agnostic about the nonlinear dynamics of CDM, that is, they will depend only on the snapshot of CDM field in a given region.

In Sec.~\ref{sec:calc} we considered simplified models for the nonlinear gravitational evolution in order to make some numerical calculations, building upon the framework developed in Sec.~\ref{sec:dfolss}. The halo model is applied to extend the results of Sec.~\ref{sec:1h}  based on a single moving halo, now accounting for the clustering of nearby halos and its contribution to the dynamical friction effect. We calculate the decrease in halo peculiar velocities as a function of halo mass, see Fig.~\ref{fig:invtau}. Next we introduce 1-loop standard perturbation theory (SPT) to extract the distortion contribution to the CDM-$\nu$ cross power spectrum, as shown in Fig.~\ref{fig:nucdm2pt}. We find it to be a small decrease in power, which is also degenerate with standard nonlinear structure formation. Finally, we compute the bispectrum to tree-level in SPT. The contribution from the distortion effect peaks at squeezed triangle configurations, see Figs.~\ref{fig:bisshape} and \ref{fig:biselong}. This has a simple physical interpretation: The local CDM-$\nu$ cross power acquires a dipole in the presence of a long-wavelength displacement potential, as given by Eq.~(\ref{eq:dipoleagain}). We show that this signal can be extracted using three-point cross correlations of matter and galaxies, i.e., in a multi-tracer approach. 

In Eq.~(\ref{eq:dipoleagain}) the dipole is proportional to the monopole and hence one can extract the displacement potential from the ratio between the bispectrum and the power spectrum \cite{zhu2020measuring}, independently of the specific realization of the small scale CDM power. This implies that the signal is not limited by cosmic variance or potential inaccuracies in the modeling of complicated nonlinear structure and galaxy formation physics (this can also be seen from the fact that the signal-to-noise ratio peaks at large quasi-linear scales as illustrated in Fig.~\ref{fig:stnk}). This signature of neutrino masses on the large-scale structure has some other desirable features; it is not degenerate with the optical depth to reionization, and is also model independent in the sense that a dynamical dark energy component cannot reproduce the effect.\footnote{Here we assume a standard dynamical dark energy component that changes the expansion history, and hence the linear growth of structure, but does not cluster.} Finally, it is not directly proportional to the neutrino mass.  The dynamical friction effect and distortion contributions to the bispectrum have an approximate scaling of $\sim m_{\nu}^4$ and $\sim m_{\nu}^3$ for sufficiently small neutrino masses,\footnote{These approximate scalings assume the hierarchy $k_{\textrm{fs}} \ll k_{\textrm{NL}}$. For a neutrino mass sufficiently large such that $k_{\textrm{fs}} \sim k_{\textrm{NL}}$, the free-streaming scales starts acting as a cutoff which reduces the amplitude of distortion effects, see Eq.~(\ref{eq:powerlaw}) and comments below.} respectively. These effects are hence dominated by the most massive neutrino eigenstate (see Eq.~(\ref{eq:powerlaw}) and Tab.~\ref{table:forecast}). This could, in principle, provide a cosmological window into the neutrino mass splittings \cite{wagner2012effects, archidiacono2020will, brinckmann2019promising}.

In practice, a detection of the distortion in the neutrino density field due the peculiar motion of halos will be challenging, and ultimately depends on the neutrino mass scale that is actually realized in nature. We estimate $N_{g} \gtrsim  (4\times 10^{9}-6\times 10^{6})$ [for individual neutrino masses in the range $m_{\nu} = (0.05-0.15)$eV] for the minimum number of galaxies in the survey in order to reach a 3$\sigma$ detection (in the optimistic scenario where all the galaxies sit at low redshifts $z\approx 0$, see Tab.~\ref{table:forecast}). In the optimistic 
scenario of two neutrino eigenstates with mass $m_{\nu} \approx 0.05$eV under the inverted hierarchy, which saturates the cosmological upper bound $\sum_{\nu} m_{\nu} \leq 0.12$eV and is consistent with oscillation experiments, a total number of galaxies $N_{g} \approx 8 \times 10^{8}$ is required in order to reach a 2-$\sigma$ detection. For reference, the Baryon Oscillation spectroscopic survey (BOSS) completed spectroscopy on $N_{g} \sim 10^6$ galaxies \cite{alam2021completed}, and the ongoing Dark Energy Spectroscopic Instrument (DESI) will reach $N_{g} \sim 4 \times 10^7$ \cite{abareshi2022overview}. A futuristic spectroscopy survey may reach $N_{g} \sim 10^9$ \cite{dodelson2016cosmic}. Also, a more realistic forecast will have to take into account that the matter field is only probed indirectly via the gravitational deflection it causes on the trajectory of light rays, i.e., via weak lensing. Hence, what we actually probe is the projected matter field along the line-of-sight. We can then consider photometric surveys since the exact knowledge about the redshift of galaxies is no longer essential. In one hand we expect this leads to a degradation on the signal due to the mixing of scales in projection, while on the other hand it allows for a much larger number of galaxies. For example, the Dark Energy Survey catalogs include hundreds of millions of galaxies \cite{DES:2020aks, DES:2018gui} while the upcoming Vera C. Rubin Observatory (LSST) will extract photometric redshifts of $N_{g} \sim 4 \times 10^{9}$ galaxies at $z\lesssim 1$ \cite{mandelbaum2018lsst}. It would be interesting to do a detailed exploration of prospects for constraining this signal with photometric surveys, we leave this for future work. In summary, it appears that future cosmological surveys have the statistical power to detect the distortion effect and further investigation is warranted.

It would be interesting to directly extract the distortion signal in the bispectrum from N-body simulations with massive neutrinos, and compare to our calculations \cite{inman2017simulating}. In the theory side, we plan on improving our modeling to relax the necessity of a hierarchy of scales, $k_{\textrm{fs}} \ll k_{\textrm{NL}}$. This requires one to assume a specific model for the nonlinear dynamics of CDM, and we speculate that an approach along the lines of the Renormalized Perturbation Theory (RPT) \cite{crocce2006renormalized} can provide the necessary ingredients to account for the effects of CDM bulk flows in the neutrino field to all orders in perturbation theory. Also, it seems likely that  our analysis can be extended to a more general warm dark matter component, as all we had to assume is that the background distribution functions peaks at some finite characteristic momentum (the temperature). For instance, it has been previously suggested that the CDM-baryon relative velocity leads to a unique signature in the galaxy bispectrum  \cite{slepian2017large, slepian2015signature, yoo2011supersonic}. We leave an investigation of the connection between these two effects to later work.

\acknowledgments
This work is supported by the Department of Physics and the College of Arts and Sciences at the University of Washington and by the Department of Energy grant DE-SC0023183. We thank John Franklin Crenshaw for helpful discussions, Zachary Weiner and David Cyncynates for valuable feedback on a draft, and Tongyan Lin and Benjamin Safdi for collaboration in early stages of this project. ML thanks Matias Zaldarriaga for (many years ago) asking a question about halo velocities and escape velocities that eventually led to thinking about this topic. 
\appendix 

\section{Signal-to-noise derivation}
\label{sec:stn}

We carry out the derivation of the optimal estimator for the bispectrum in Eq.~(\ref{eq:signal}), 
\begin{equation}
\label{eq:bis}
	\frac{1}{2}\langle [\delta_{\textrm{m}}(\vec{k}_1)\delta_{\textrm{g}}(\vec{k}_2) -\delta_{\textrm{g}}(\vec{k}_1)\delta_{\textrm{m}}(\vec{k}_2)]\delta_{\textrm{m}}(\vec{k}_3) \rangle = (2\pi)^3 \delta^{(3)}(\vec{k}_{1}+\vec{k}_{2}+\vec{k}_{3}) B_{\textrm{obs}}^{A}(k_1,k_2,k_3) \,,
\end{equation}
following the approach in \cite{smith2018ksz}. Here $\delta^{(3)}(\vec{k})$ stands for the Dirac delta. Let us start with the most general three-point estimator for the bispectrum amplitude: 
\begin{equation}
\label{eq:cubicest}
\begin{split}
	\hat{\mathcal{E}} & = \frac{1}{2VF} \int \frac{d^3 \vec{k}_1}{(2\pi)^3} \int \frac{d^3 \vec{k}_2}{(2\pi)^3} \int \frac{d^3 \vec{k}_3}{(2\pi)^3} (2\pi)^3\delta^{(3)}(\vec{k}_{123}) W(k_{1},k_{2},k_{3}) \left[\delta_{\textrm{m}}(\vec{k_{1}}) \delta_{\textrm{g}}(\vec{k_{2}}) - \delta_{\textrm{g}}(\vec{k_{1}}) \delta_{\textrm{m}}(\vec{k_{2}})\right] \delta_{\textrm{m}}(\vec{k}_{3}) \\ & = \frac{1}{VF} \int_{\{\vec{k}\}} \delta(\vec{k}_{123}) W(k_1,k_2,k_3) \delta_{\textrm{m}}(\vec{k_{1}}) \delta_{\textrm{g}}(\vec{k_{2}}) \delta_{\textrm{m}}(\vec{k}_{3}) \,,
\end{split}
\end{equation}
with $V$ the survey volume, and $W(k_{1},k_{2},k_{3})$ a weight function to be found in such a way as to minimize the noise. We take it to satisfy the condition $W(k_{1},k_{2},k_{3}) = -W(k_{2},k_{1},k_{3})$ without loss of generality, since a symmetric part would give a vanishing contribution to the estimator. We also introduced, for simplicity of notation:
\begin{equation}
\label{eq:shorthand}
	 \int_{\{\vec{k}\}} \delta(\vec{k}_{123})  \equiv \int \frac{d^3 \vec{k}_1}{(2\pi)^3} \int \frac{d^3 \vec{k}_2}{(2\pi)^3} \int \frac{d^3 \vec{k}_3}{(2\pi)^3} (2\pi)^3\delta^{(3)}(\vec{k}_{123}) \,.
\end{equation}
Finally, the Fisher matrix $F$ is a normalization coefficient that we fix by requiring that $\langle \hat{\mathcal{E}} \rangle =1$, which implies
\begin{equation}
\label{eq:fishernorm}
	F = \int_{\{\vec{k}\}} \delta(\vec{k}_{123}) W(k_1,k_2,k_3) B^{A}_{\textrm{obs}}(k_1,k_2,k_3) \,,
\end{equation}
after taking the expectation value of Eq.~(\ref{eq:cubicest}), using Eq.~(\ref{eq:bis}), and letting $(2\pi)^{3}\delta^{(3)}(0) \to V$ as usual. We now proceed to compute the leading order Gaussian contribution to the noise $\langle \hat{\mathcal{E}}^2 \rangle $, which effectively means only keeping terms in the resulting six-point function that can be written as a product of three two-point functions, i.e., the $PPP$ contributions. We obtain after a straightforward calculation:
\begin{equation}
\label{eq:noise}
\begin{split}
	\langle \hat{\mathcal{E}}^2 \rangle = \frac{1}{VF^2} \int_{\{\vec{k}\}} & \delta(\vec{k}_{123}) W(k_1,k_2,k_3) \Bigg\{ W(k_1,k_2,k_3) \left[P_{\textrm{m}\textrm{m}}(k_1)P_{\textrm{g}\textrm{g}}(k_2)- P_{\textrm{m}\textrm{g}}(k_1)P_{\textrm{m}\textrm{g}}(k_2)\right]P_{\textrm{m}\textrm{m}}(k_3) + \\ & + W(k_1,k_3,k_2) \left[P_{\textrm{m}\textrm{m}}(k_1)P_{\textrm{m}\textrm{g}}(k_3)- P_{\textrm{m}\textrm{g}}(k_1)P_{\textrm{m}\textrm{m}}(k_3)\right]P_{\textrm{m}\textrm{g}}(k_2) +  \\ & + W(k_3,k_2,k_1) \left[P_{\textrm{g}\textrm{g}}(k_2)P_{\textrm{m}\textrm{m}}(k_3)- P_{\textrm{m}\textrm{g}}(k_2)P_{\textrm{m}\textrm{g}}(k_3)\right]P_{\textrm{m}\textrm{m}}(k_1)\Bigg\} \,.
\end{split}
\end{equation}
We now make use of,
\begin{equation}
\label{eq:tracers}
	\begin{split}
		& P_{\textrm{m}\textrm{m}}(k) \approx P_{\delta \delta}(k) \,, \\ & P_{\textrm{g}\textrm{g}}(k) \approx b^2 P_{\delta \delta}(k) + \frac{1}{n_{\textrm{g}}} \,, \\ & P_{\textrm{m}\textrm{g}}(k) \approx bP_{\delta \delta}(k) \,,
	\end{split}
\end{equation}
where we work to leading order in the fractional contribution of neutrinos to the energy density $f_{\nu} \ll 1$, $P_{\delta \delta}(k)$ the power spectrum including a shot-noise contribution to the power spectrum of galaxies, $1/n_{\textrm{g}}$, with $n_{\textrm{g}}$ the number density of galaxies in the survey, and for simplicity we assume a constant bias factor $b$ \footnote{This assumption can be relaxed to accommodate a scale dependent bias which is more realistic at the scales we are considering, and it does not lead to qualitatively different results. We stick to a constant bias for simplicity.}. We arrive at
\begin{equation}
\label{eq:noiseagain}
\begin{split}
	\langle \hat{\mathcal{E}}^2 \rangle  = \frac{1}{2N_{g}F^2}  & \int_{\{\vec{k}\}}  \delta(\vec{k}_{123}) W(k_1,k_2,k_3) \Bigg\{ \left[W(k_1,k_2,k_3)+W(k_3,k_2,k_1)\right]P_{\delta \delta}(k_1)P_{\delta \delta}(k_3) - \\ & - \left[W(k_2,k_1,k_3)+W(k_3,k_1,k_2)\right]P_{\delta \delta}(k_2)P_{\delta \delta}(k_3) \Bigg\} \,,
\end{split}
\end{equation}
where $N_{\textrm{g}} = n_{\textrm{g}} V$ the total number of galaxies in the survey volume $V$. The Eq.~(\ref{eq:noiseagain}) can be conveniently written as follows
\begin{equation}
\label{eq:covariance}
	\langle \hat{\mathcal{E}}^2 \rangle = \frac{1}{N_{\textrm{g}}V F^2} \int_{\{\vec{k}\}}  \delta(\vec{k}_{123}) \int_{\{\vec{k'}\}}  \delta(\vec{k'}_{123}) W(k_1,k_2,k_3)\textrm{Cov}(\vec{k}_1,\vec{k}_2,\vec{k}_3;\vec{k}'_1,\vec{k}'_2,\vec{k}'_3) W(k'_1,k'_2,k'_3) \,,
\end{equation}
in terms of a covariance matrix:
\begin{equation}
\label{eq:covariancematrix}
\begin{split}
	& \textrm{Cov}(\vec{k}_1,\vec{k}_2,\vec{k}_3;\vec{k}'_1,\vec{k}'_2,\vec{k}'_3)  = \frac{1}{2} \Big[ \delta(\vec{k}'_1-\vec{k}_1) \delta(\vec{k}'_2-\vec{k}_2) \delta(\vec{k}'_3-\vec{k}_3) P_{\delta \delta}(k_1)P_{\delta \delta}(k_3) + \\ & + \delta(\vec{k}'_1-\vec{k}_3) \delta(\vec{k}'_2-\vec{k}_2) \delta(\vec{k}'_3-\vec{k}_1) P_{\delta \delta}(k_1)P_{\delta \delta}(k_3) - \delta(\vec{k}'_1-\vec{k}_2) \delta(\vec{k}'_2-\vec{k}_1) \delta(\vec{k}'_3-\vec{k}_3) P_{\delta \delta}(k_2)P_{\delta \delta}(k_3) - \\ & - \delta(\vec{k}'_1-\vec{k}_3) \delta(\vec{k}'_2-\vec{k}_1) \delta(\vec{k}'_3-\vec{k}_2) P_{\delta \delta}(k_2)P_{\delta \delta}(k_3) \Big] \,.
\end{split}
\end{equation}
We now have an optimization problem, where we would like to minimize the noise $\langle \hat{\mathcal{E}}^2 \rangle$, as given by Eq.~(\ref{eq:covariance}), when subject to the constraint Eq.~(\ref{eq:fishernorm}). By taking the variation of the noise with respect to the weight function and setting it to zero, we obtain the following equation:
\begin{equation}
\label{eq:sol}
	\int_{\{\vec{k'}\}}  \delta(\vec{k'}_{123}) \textrm{Cov}(\vec{k}_1,\vec{k}_2,\vec{k}_3;\vec{k}'_1,\vec{k}'_2,\vec{k}'_3) W(k'_1,k'_2,k'_3) = \delta(\vec{k}_{123}) B_{\textrm{obs}}(k_1,k_2,k_3) \,,
\end{equation}  
up to an arbitrary normalization constant. The solution to Eq.~(\ref{eq:sol}) can be parameterized as follows,
\begin{equation}
\label{eq:prec}
	W(k_1,k_2,k_3) = 	\int_{\{\vec{k'}\}}  \delta(\vec{k'}_{123}) \Sigma(\vec{k}_1,\vec{k}_2,\vec{k}_3;\vec{k}'_1,\vec{k}'_2,\vec{k}'_3) B_{\textrm{obs}}(k'_1,k'_2,k'_3) \,,
\end{equation}
in terms of a precision matrix $\Sigma(\vec{k}_1,\vec{k}_2,\vec{k}_3;\vec{k}'_1,\vec{k}'_2,\vec{k}'_3)$. The substitution of Eq.~(\ref{eq:prec}) into Eq.~(\ref{eq:sol}) now yields
\begin{equation}
\label{eq:solprecision}
	\int_{\{\vec{k'}\}}  \delta(\vec{k'}_{123}) \textrm{Cov}(\vec{k}_1,\vec{k}_2,\vec{k}_3;\vec{k}'_1,\vec{k}'_2,\vec{k}'_3)\Sigma(\vec{k}'_1,\vec{k}'_2,\vec{k}'_3;\vec{k}''_1,\vec{k}''_2,\vec{k}''_3) = \delta(\vec{k}''_1-\vec{k}_1) \delta(\vec{k}''_2-\vec{k}_2) \delta(\vec{k}''_3-\vec{k}_3) \,,
\end{equation}
such that the precision matrix is the inverse of the covariance matrix of Eq.~(\ref{eq:covariancematrix}). As we will see, in practice it is more convenient to directly solve Eq.~(\ref{eq:sol}) for the weight function instead of solving for the precision matrix. The substitution of Eq.~(\ref{eq:covariancematrix}) into Eq.~(\ref{eq:sol}) leads to:
\begin{equation}
\label{eq:solfilter}
\begin{split}
	 \left[P_{\delta \delta}(k_1) + P_{\delta \delta}(k_2)\right] &P_{\delta \delta}(k_3)W(k_1,k_2,k_3)  + P_{\delta \delta}(k_1)P_{\delta \delta}(k_3) W(k_3,k_2,k_1) - \\ & -  P_{\delta \delta}(k_2)P_{\delta \delta}(k_3) W(k_3,k_1,k_2) = 2 B_{\textrm{obs}}(k_1,k_2,k_3) \,.
\end{split}
\end{equation}
It is now straightforward to check that
\begin{equation}
\label{eq:solfilteragain}
	W(k_1,k_2,k_3) = \frac{2B_{\textrm{obs}}(k_1,k_2,k_3)}{P_{\delta \delta}(k_3)\left[P_{\delta \delta}(k_1)+P_{\delta \delta}(k_2)+P_{\delta \delta}(k_3)\right]} \,,
\end{equation} 
is the solution we are looking for, after using the identity
\begin{equation}
\label{eq:identity}
	B_{\textrm{obs}}(k_1,k_2,k_3) + B_{\textrm{obs}}(k_3,k_1,k_2) + B_{\textrm{obs}}(k_2,k_3,k_1) = 0
\end{equation}
which follows from Eq.~(\ref{eq:bis}). Alternatively, the Eq.~(\ref{eq:solfilteragain}) can be derived after considering the Eq.~(\ref{eq:solfilter}) in combination with the same equation when written in terms of a different permutation of the wavenumbers. We now have all the ingredients necessary to determine the cumulative signal-to-noise ratio. First note from Eqs.~(\ref{eq:fishernorm}),(\ref{eq:covariance}) and (\ref{eq:sol}) that
\begin{equation}
\label{eq:rel}
	\textrm{SNR}^2 = \frac{1}{	\langle \hat{\mathcal{E}}^2 \rangle } = N_{\textrm{g}} F \implies F = \frac{\textrm{SNR}^2}{N_{\textrm{g}}} \,,
\end{equation}
is given by, when substituting the Eq.~(\ref{eq:solfilteragain}) into Eq.~(\ref{eq:fishernorm}):
\begin{equation}
\label{eq:stnapp}
	F = 2 \int \frac{d^3 \vec{k}_1}{(2\pi)^3} \int \frac{d^3 \vec{k}_2}{(2\pi)^3} \int \frac{d^3 \vec{k}_3}{(2\pi)^3} (2\pi)^3\delta^{(3)}(\vec{k}_{123}) \frac{\left[B_{\textrm{obs}}(k_1,k_2,k_3)\right]^2}{P_{\delta \delta}(k_3)\left[P_{\delta \delta}(k_1)+P_{\delta \delta}(k_2)+P_{\delta \delta}(k_3)\right]} \,,
\end{equation}
when also writing the integration variables in full, according to Eq.~(\ref{eq:shorthand}). This, in combination with Eq.~(\ref{eq:rel}), is the formula we need for the cumulative signal-to-noise ratio.

\bibliographystyle{JHEP}
\bibliography{dynfriction.bib}

\end{document}